\documentclass[reprint,onecolumn]{revtex4-2}
\usepackage{amsmath}
\usepackage{mathtools}
\usepackage{graphicx}
\usepackage{amsfonts}
\usepackage{color}
\usepackage{microtype} %Improves TeX's choices of linebreaks, text justification, end-of-line hyphenation, kerning etc

\usepackage{verbatim} %for comments
\usepackage{braket}
\usepackage{physics}
\usepackage{bm}

\usepackage{float}

\newcommand{\tracenorm}[1]{\norm{#1}_1}
\newcommand{\accevent}{\Omega_{\text{acc}}}

\newcommand{\rhoMeas}{\rho_{AA'}^\text{prep}} % rho^meas real protocol state
\newcommand{\rhoMeasN}{\left(\rhoMeas\right)^{\otimes n}} % n copies of rho^meas
\newcommand{\tauMeas}{{\xi_{AA''}^\text{prep}}} % tau^meas virtual protocol state
\newcommand{\tauMeasN}{\left(\tauMeas\right)^{\otimes n}} % n copies of tau^meas
\newcommand{\demi}{\frac{1}{2}}
\newcommand{\protMapShield}[1]{\mathcal{E}^{(#1)}_\text{QKD-shield}} % Protocol map for shield system protocol
\newcommand{\protMapShieldSR}[1]{\mathcal{E}^{(#1)}_\text{QKD-S}} % Protocol map for source replacement + shield system protocol
\newcommand{\epsSec}{\varepsilon_{\text{sec}}} % Secrecy epsilon
\newcommand{\epsCor}{\varepsilon_{\text{cor}}} % correct epsilon
\newcommand{\epsAT}{\varepsilon_{\text{AT}}} % Acceptace test epsilon
\newcommand{\epsbar}{\bar{\varepsilon}} % Smoothing parameter
\newcommand{\epsPA}{\varepsilon_{\text{PA}}} % Privacy ampklification epsilon
\newcommand{\epsEV}{\varepsilon_{\text{EV}}} % epsilon EV
\newcommand{\epstilde}{\widetilde{\varepsilon}}
\newcommand{\ndecoy}{{N_{\text{ph}}}} % Photon number cutoff for decoy-state analysis
\newcommand{\nint}{n_{\text{int}}} % Number of intensities used for decoy-state analysis
\newcommand{\nmeas}{{n_{\text{meas}}}} % Number of Bob's POVM elements
\newcommand{\nFSS}{N} % Photon number cutoff for FSS

\DeclareMathOperator{\id}{\mathord{\rm id}} % Identity map

\newcommand{\protMap}[1]{\mathcal{E}^{\left(#1\right)}_\text{QKD}} % Protocol map
\newcommand{\protMapId}[1]{\mathcal{E}^{(#1),\text{ideal}}_\text{QKD}} % Ideal protocol map

\newcommand{\protMapVar}[1]{\mathcal{E}^{\left(#1_1,\dots,#1_M\right)}_\text{var-QKD}} % Protocol map
\newcommand{\protMapIdVar}[1]{\mathcal{E}^{(#1_1,\dots,#1_M),\text{ideal}}_\text{var-QKD}} %

\newcommand{\fullprotMap}[1]{\mathcal{M}^{\left(#1\right)}_\text{QKD}} % Protocol map
\newcommand{\fullprotMapId}[1]{\mathcal{M}^{(#1),\text{ideal}}_\text{QKD}} % Ideal protocol map
\newcommand{\protMapShieldId}[1]{\mathcal{E}^{(#1),\text{ideal}}_\text{QKD-shield}} % Ideal protocol map
\newcommand{\protMapShieldSRId}[1]{\mathcal{E}^{(#1),\text{ideal}}_\text{QKD-S}} % Ideal protocol map

\newcommand{\protMapSquash}[1]{\mathcal{E}^{\left(#1\right)}_\text{QKD-Sq}} % Protocol map fir squashed protocol
\newcommand{\protMapSquashId}[1]{\mathcal{E}^{(#1),\text{ideal}}_\text{QKD-Sq}} % Ideal protocol map for squashed protocol

\newcommand{\perm}{\pi} %Element of S_n
\newcommand{\permutation}[2]{{P^{#1,#2}_\perm}} % Representation of elements of S_n

\newcommand{\state}{\rho_{A^nB^n}}
\newcommand{\stateSquash}{\rho_{A^nQ^n}}
\newcommand{\promiseset}{S_{\hat{\sigma}} }
\newcommand{\promise}{\hat{\sigma}_A}
    \newcommand{\promiseShield}{\hat{\sigma}_{AA_S}}
\newcommand{\Sym}[1]{\text{Sym}^n\left(#1\right)}

\newcommand{\matrixset}[1]{\text{L}\left(#1\right)}
\newcommand{\posset}[1]{\text{Pos}\left(#1\right)}
\newcommand{\stateset}[1]{\text{S}_\circ\left(#1\right)}
\newcommand{\channelset}[1]{\mathcal{C}\left(#1\right)}
\newcommand{\mapset}[1]{\text{T} \left(#1\right)}
\newcommand{\boundedOpSet}[1]{\mathcal{B}\left(#1\right)}

\newcommand{\C}[1]{{\mathbb{C}^{#1}}}

\newcommand{\Cfull}{\widetilde{C}}
\newcommand{\Cat}{C_{AT}}
\newcommand{\cost}{g_{n,x}} % Cost of the lift
\newcommand{\dimSym}[1]{\dim\left(\Sym{\C{#1}} \right)} % Dimension of the symmetric subspace
\newcommand{\blockdim}[2]{{d_{#1}^{#2}}} % dimension of blocks

\newcommand{\FSScost}{f_\nFSS} % cost of flag-state squasher

\newcommand{\smoothmin}[1]{H^{#1}_{\text{min}}}

\newcommand{\lambdamin}[1]{\lambda_{\min}\left(#1\right)} % min eigenvalue of operator #1

\newcommand{\Fobs}{\mathbf{F}^\text{obs}}
\newcommand{\leak}[1]{\lambda^\text{EC}_{#1}}
\newcommand{\bstat}{b_\text{stat}}
\newcommand{\nkey}{n_K}

\makeindex
\usepackage{amsthm}
\usepackage{thmtools}
\theoremstyle{definition} 
\newtheorem{theorem}{Theorem}
\newtheorem{definition}{Definition}
\newtheorem{lemma}{Lemma}
\newtheorem{corollary}{Corollary}[theorem]
\newtheorem{remark}{Remark}

\usepackage{hyperref} %Package compatibility note: hyperref should be included after most other packages unless otherwise specified
\usepackage[capitalize]{cleveref} %To be included after hyperref
\hypersetup{
	colorlinks = true,
	linkcolor =blue,
	citecolor=blue, 
	urlcolor=blue 
}

\begin{document}
	\title{Postselection technique for optical Quantum Key Distribution with improved de Finetti reductions}

\author{Shlok Nahar$^1$, Devashish Tupkary$^1$, Yuming Zhao$^2$, Norbert L\"utkenhaus$^1$, Ernest Y.-Z. Tan$^1$}
\affiliation{$^1$Institute for Quantum Computing and Department of Physics and Astronomy, University of Waterloo, Waterloo, Ontario, Canada, N2L 3G1}
\affiliation{$^2$Institute for Quantum Computing and Department of Pure Mathematics, University of Waterloo, Waterloo, Ontario, Canada, N2L 3G1}
	
\begin{abstract}
    
    The postselection technique is an important proof technique for proving the security of quantum key distribution protocols against coherent attacks via the uplift of any security proof against IID-collective attacks.
    In this work, we go through multiple steps to rigorously apply the postselection technique to optical quantum key distribution protocols.
    First, we place the postselection technique on a rigorous mathematical foundation by fixing a technical flaw in the original postselection paper.
    Second, we extend the applicability of the postselection technique to prepare-and-measure protocols by using a de Finetti reduction with a fixed marginal. Third, we show how the postselection technique can be used for decoy-state protocols by tagging the source. Finally, we extend the applicability of the postselection technique to realistic optical setups by developing a new variant of the flag-state squasher.   
    We also improve existing de Finetti reductions, which reduce the effect of using the postselection technique on the key rate. These improvements can be more generally applied to other quantum information processing tasks.
    As an example to demonstrate the applicability of our work, we apply our results to the time-bin encoded three-state protocol.
    We observe that the postselection technique performs better than all other known proof techniques against coherent attacks.
    
\end{abstract}

\begin{titlepage}
    \maketitle     
\end{titlepage}
\setcounter{page}{1}

\section{Introduction}
\label{sec:introduction}

Proving the security of quantum key distribution (QKD) protocols against coherent attacks is a challenging task since the security proof must consider any arbitrary attack implemented by the adversary, as opposed to IID-collective attacks where the adversary attacks each round in an identical and independent manner. Despite this challenge, there has been tremendous progress \cite{koashi_simple_2009,tomamichel_tight_2012,dupuis_entropy_2020,metger_generalised_2022}. However, all known proof techniques have some drawbacks.

In particular, the phase-error rate \cite{koashi_simple_2009} and entropic uncertainty relation (EUR) \cite{tomamichel_largely_2017} based approaches cannot be used for practical detection setups with basis-dependent losses.
On the other hand, while the analysis of prepare-and-measure (PM) protocols are related to that of entangled-based (EB) protocols via the source-replacement scheme \cite{bennett_quantum_1992, curty_entanglement_2004,ferenczi_symmetries_2011} this introduces a constraint of fixed marginal on Alice's state (which we refer to as a fixed marginal constraint). For this reason, the entropy accumulation theorem \cite{dupuis_entropy_2020} and the existing postselection technique \cite{christandl_postselection_2009} are not directly applicable to PM protocols. Moreover, the postselection technique also gives pessimistic bounds on the key rate.
More recently, a generalised entropy accumulation theorem (GEAT) \cite{metger_generalised_2022} was developed that applies to PM protocols and has been shown to perform well for qubit protocols. However, it currently requires a condition that Eve's optimal attack satisfies a particular sequential property, and if that condition is enforced via limiting the repetition rate of the protocol \cite{sandfuchs_security_2023,curras-lorenzo_security_2023}, the resulting key rates for fiber-based protocols are low. Finally, Remark 4.3.3 of Renner's thesis \cite{renner_security_2005} can also be used with the exponential de Finetti theorem to accommodate PM protocols. However, finite-size key rates for PM QKD protocols using this technique have not been computed and studied rigorously in any work. Moreover, the proof technique only accommodates an approximate version of the fixed marginal constraint required for PM protocols, and thus would perform significantly worse than other proof techniques. Additionally, the `sacrifice' bits needed for Renner's exponential de Finetti theorem results in even lower key rates as evidenced by \cite[Figure 2]{sheridan2010finite}.

Note that in this work, ``the postselection technique'' refers to a generic QKD proof technique \cite{christandl_postselection_2009} that uplifts security proofs against IID-collective attacks to proofs against arbitrary attacks, and should not be confused with the selection of subsets of classical data as part of a quantum information application.
While the postselection technique is well-established, ours is the first \emph{full} proof based on the concept of the postselection technique. In particular, the following issues were not dealt with prior to our work:
\begin{enumerate}
    \item \underline{The infinite-dimensional optical systems:} Realistic models of Bob's detection setup are infinite-dimensional. Although this is well-studied in QKD \cite{fung_universal_2011, gittsovich_squashing_2014, zhang_security_2021, upadhyaya_dimension_2021}, it turns out that none of those solutions can be robustly used with the postselection technique. Thus, we develop the ``weight-preserving flag-state squasher'' to address this issue.
    Additionally, for decoy-state protocols Alice's signal states also live in an infinite-dimensional Hilbert space. We rigorously apply source maps \cite{nahar_imperfect_2023} to reduce the analysis to that of finite-dimensional tagged states \cite{gottesman_security_2004}.
    \item \label{item:oldPSWrong} \underline{Technical flaw in prior work:} The postselection technique prior to our work had a step that was not correct. There was an earlier attempt \cite[Section 3.3.2]{belzig_studying_2020} to fix this misstep in the original paper \cite{christandl_postselection_2009}. However, this also contained a flaw. Our work is the first correct version of this step.
    \item \label{item:rhoAConstraint} \underline{The fixed marginal constraint needed for PM QKD protocols:} The postselection technique prior to our work could not accommodate this fixed marginal constraint which is vital for the security proof of generic PM QKD protocols.
\end{enumerate}
In addition to developing the tools necessary for the \emph{first, complete} security proof, we also improve on the performance drastically. Finally, we also adapt the proof for variable-length protocols \cite{tupkary_security_2023}, which is vital to obtain the best expected key rates.

This paper is organized as follows. In \cref{sec:improvingDeFinetti} we state the improvements to the de Finetti reduction with fixed marginal. In \cref{sec:correctApplicationToQKD} we rigorously apply the de Finetti reductions to lift the security of QKD protocols against IID-collective attacks to that against arbitrary attacks through \cref{cor:liftToCoherent} and \cref{cor:liftToCoherentSymmetries}. In  \cref{sec:PSDecoy}, we explain the usage of the postselection technique for optical protocols by constructing a new version of the flag-state squasher, and discussing the usage of the postselection technique with decoy-state analysis. Many proofs are delegated to the appendices. Finally, in \cref{sec:applicationtothreestate} we apply our results to the time-bin encoded three-state protocol.

\begin{table}[H]
\centering
    \begin{tabular}{ |c|c| } 
        \hline
        Symbol & What it represents  \\
        \hline
        $[k]$ & The set $\{1,2,...,k\}$\\
        \hline
        $A$, $B$, ...  & Registers as well as Hilbert spaces (depending on context)\\
        \hline
        $d_A$, $d_B$, ... & dimensions of spaces $A$, $B$, ...\\
        \hline
        $\matrixset{A}$ & Set of linear operators acting on $A$\\
        \hline
        $\posset{A}$ & Set of positive operators acting on $A$\\
        \hline
        $\stateset{A}$ & Subset of positive operators acting on $A$ with trace 1\\
        \hline
        $\mapset{A,B}$ & Set of linear maps from $\matrixset{A}$ to $\matrixset{B}$\\
        \hline
        $\channelset{A,B}$ & Set of completely positive trace-preserving maps from $\matrixset{A}$ to $\matrixset{B}$\\
        \hline
        $\boundedOpSet{A}$ & Set of bounded operators on $A$\\
        \hline
        $\ket{\rho}$ & Purification of a positive semidefinite matrix $\rho$\\
        \hline
        $\text{Sym}^n\left(\mathbb{C}^x\right)$ & Symmetric subspace of $\left(\mathbb{C}^x\right)^n$\\
        \hline
        $\cost$ & $\dimSym{x}$\\
        \hline
        $\lambdamin{H}$ & Minimum eigenvalue of $H$\\
        \hline
    \end{tabular}
    \caption{Common symbols used in this work}\label{app:Gloss}
\end{table}

\section{Improving the de Finetti reduction} \label{sec:improvingDeFinetti}

Quantum de Finetti theorems \cite{renner_security_2005,christandl_postselection_2009,arnon-friedman_finetti_2015,fawzi_quantum_2015} are useful in reducing the analysis of various quantum information processing tasks to the IID case. In this work, we focus on quantum de Finetti reductions of following form used in Refs.~\cite{christandl_postselection_2009,arnon-friedman_finetti_2015,fawzi_quantum_2015}:
\begin{equation}
    \rho_{A^n B^n} \leq \cost\; \tau_{A^n B^n},
\end{equation}
where $\tau_{A^n B^n}=\int \sigma_{AB}^{\otimes n}\text{d}\sigma_{AB}$ is a normalised density matrix, and $g_{n,x} = \dimSym{x} = \binom{n+x-1}{n}$. (Refer to \cref{sec:plots} for an easily computable upper bound on the dimension of the symmetric subspace.)
We refer to a state of the form $\tau_{A^nB^n}=\int \sigma_{AB}^{\otimes n}\text{d}\sigma_{AB}$ as a ``de Finetti state".
Typically, in reducing the analysis of quantum information processing tasks to the IID case, the following factors come into play:
\begin{enumerate}
    \item The value of $\cost$ : This should be as small as possible, as it appears as a penalty in the reduction to IID states. 
    \item The integral over IID states in $\tau_{A^nB^n}$ : This should be such that the integral is only over states for which the task has been ``analyzed'', e.g.~in the sense that some security property has been proven for all such IID states.
\end{enumerate}

In this section we make several improvements to the value of $\cost$. These improvements are of two types.  We first improve the dimensional scaling ($\cost$) in Lemma 3.1 of Ref.~\cite{fawzi_quantum_2015} for generic states. Second, we also show that the dimensions can be reduced for states that are invariant under certain symmetries, as an extension of Ref.~\cite{arnon-friedman_finetti_2015} to the quantum case.
Proofs of most statements in this section are deferred to \cref{app:deFinettiProfs}.

\subsection{Generic improvement} \label{subsec:genericImprovement}

\begin{restatable}[de Finetti reduction with fixed marginal]{theorem}{genericDeFinetti} \label{thm:genericDeFinetti}
    Let $\promise \in \posset{\mathbb{C}^{d_{A}}}$ and let $\rho_{A^nR^n}$ be any purification of $\left(\promise\right)^{\otimes n}$ supported on the symmetric subspace $\Sym{\C{d_A d_R}}$. Then there exists a probability measure d$\phi$ on the set of purifications $\ketbra{\phi}_{AR}$ of $\promise$, such that
    \begin{align}
        \rho_{A^nR^n} \leq g_{n,d_Ad_R} \int \ketbra{\phi}_{AR}^{\otimes n} \text{d} \phi.
    \end{align}
\end{restatable}
The proof is given in \cref{app:deFinettiProfs}.

Note that crucially, \cref{thm:genericDeFinetti} differs from Lemma 3.1 in Ref.~\cite{fawzi_quantum_2015} in the prefactor $\cost$, where they had $x = \max[d_A^2,d_R^2]$ instead of $x=d_{A}d_{R}$. This leads to a corresponding improvement in the resulting corollary for mixed states. Before stating this corollary, we need to define permutation-invariance of matrices.

\begin{definition}[Permutation-invariance of matrices] \label{def:permutationInvariance}
    Given a matrix $\state \in \matrixset{\left(\C{d_{A} d_{B}}\right)^{\otimes n}}$ and a permutation $\perm \in S_n$ of its subsystems, we denote the action of $\perm$ on $\state$ as $\permutation{d_{A} d_{B}}{n} \state \permutation{d_{A} d_{B}}{n}^\dag$ where $\permutation{d_{A} d_{B}}{n}$ is the standard representation of $\perm$ on $(\C{d_{A} d_{B}})^{\otimes n}$.
    We say that a matrix is \textit{permutation-invariant} if it is invariant under the action of all permutations $\perm \in S_n$.
\end{definition}

\begin{corollary} \label{cor:generalMixedDeFinetti}
    Let $\promise \in \posset{\mathbb{C}^{d_{A}}}$ and $\state$ be any permutation-invariant extension of $\left(\promise\right)^{\otimes n}$. Then there exists a probability measure d$\sigma_{AB}$ on the set of non-negative extensions $\sigma_{AB}$ of $\promise$, such that
    \begin{align}
        \state \leq \cost\int \sigma_{AB}^{\otimes n} \text{d} \sigma_{AB}
    \end{align}
    holds for $x=d_{A}^2d_{B}^2$.
\end{corollary}
\begin{proof}
    We can use Lemma 4.2.2 from Ref.~\cite{renner_security_2005} to construct a purification $\rho_{A^nB^nE^n} \in \Sym{\mathbb{C}^{d_A d_B}\otimes \mathbb{C}^{d_A d_B}}$ of $\state$. From Theorem \ref{thm:genericDeFinetti}, it follows that $$\rho_{A^nB^nE^n} \leq g_{n,d_A^2d_B^2}\int \ketbra{\phi}_{ABE}^{\otimes n} \text{d} \phi,$$ where d$\phi$ is a probability measure over purifications of $\promise$.
    Taking the partial trace over $E^n$ on both sides completes the proof.
\end{proof}

Note that the dramatic improvement over Corollary 3.2 from Ref.~\cite{fawzi_quantum_2015} ($x=d_{A}^2d_{B}^4$ to $x=d_{A}^2d_{B}^2$) is a direct result of \cref{thm:genericDeFinetti}. Additionally, with this improvement we obtain the earlier de Finetti reduction without a fixed marginal \cite[Lemma 2]{christandl_postselection_2009} by considering $A$ to be a trivial system, thus unifying both results. For pedagogical reasons, while applying de Finetti reductions to QKD in \cref{sec:correctApplicationToQKD} we will primarily be using the de Finetti reduction with fixed marginal (\cref{cor:generalMixedDeFinetti}), and not the improvements presented in \cref{subsec:improvementFromSymmetries}. Thus, the remainder of this section can be skipped without affecting the understanding of \cref{sec:correctApplicationToQKD}, though the following results we present are useful for improving finite-size bounds for practical key rate computations.

\subsection{Improvements with symmetries} \label{subsec:improvementFromSymmetries}

In this work, we will use the term ``symmetries'' as a broad term to refer to both block-diagonal structure and invariance under some groups. 
This is due to the following connection between block-diagonal structure and group invariance.
Consider the group $\bigoplus_{i=1}^k\mathcal{U}(1)$ and the representation $\phi:\bigoplus_{i=1}^k\mathcal{U}(1) \rightarrow \mathcal{U}(\bigoplus_{i=1}^k\mathbb{C}^{d_i})$ defined as $\phi(u_1, \ldots u_k) \mapsto \bigoplus_{i=1}^k \bigoplus_{j=1}^{d_i} u_i$. Note that this representation has $k$ irreps of dimensions $\{d_i\}_{i=1}^k$ each. A state being invariant under the action of this representation is equivalent to saying that the state is block-diagonal with $k$ blocks of dimensions $\{d_i\}_{i=1}^k$ each. Hence, we can view block-diagonal structure as essentially corresponding to invariance with respect to some group representation.

The use of de Finetti-like reductions for symmetries was previously mentioned in \cite[footnote 17]{christandl_postselection_2009}, and with more details given in Refs.~\cite{belzig_studying_2020,gross_schur_2021}. We extend this analysis to the de Finetti reduction with fixed marginal, and also show that the reduction in the case of ``IID-symmetries'' can be combined with the reduction in the case of permutational symmetry. A similar analysis to combine permutational symmetry with ``IID-symmetries'' for the de Finetti reduction without a fixed marginal was recently performed by Ref.~\cite{matsuura_tight_2024} independently of our work. Importantly, they also obtain a better cost $\cost$.

Our improvements to \cref{cor:generalMixedDeFinetti} in the presence of symmetries are based on observing that the proof in \cref{subsec:genericImprovement} had the following structure:
\begin{enumerate}
    \item Given a permutation-invariant state with a fixed marginal $\promise^{\otimes n}$, Lemma 4.2.2 from Ref.~\cite{renner_security_2005} is used to obtain a purification in the corresponding symmetric subspace. \label{item:purLemmaStep}
    \item This purification can then be related to a convex mixture of IID states with the same marginal $\promise$, via \cref{thm:genericDeFinetti}. \label{item:deFinettiThmStep}
    \item Tracing out the purifying system gives the de Finetti reduction for mixed states as shown in \cref{cor:generalMixedDeFinetti}.
\end{enumerate}

We show that in the presence of symmetries, steps~\ref{item:purLemmaStep} and~\ref{item:deFinettiThmStep} can be improved.
With this, we prove a generalised version of \cref{thm:genericDeFinetti} that results in better bounds in the presence of symmetries.

\begin{restatable}[de Finetti reduction with symmetries and fixed marginal]{theorem}{promisedDeFinetti}\label{thm:promisedDeFinetti}
   Let $\promise \in \posset{\mathbb{C}^{d_{A}}}$, and let 
     $\rho_{A^nR^n}$ be any purification of $\left(\promise\right)^{\otimes n}$ supported on the symmetric subspace $\Sym{\bigoplus_{i=1}^{k}\C{\blockdim{i}{A}}\otimes \C{\blockdim{i}{R}}}$ where $\bigoplus_{i=1}^k\C{\blockdim{i}{A}}\subseteq\mathbb{C}^{d_A}$ and $\bigoplus_{i=1}^k\C{\blockdim{i}{R}}\subseteq \mathbb{C}^{d_R}$. Then there exists a probability measure d$\phi$ on the set of purifications $\ket{\phi}_{AR}\in \bigoplus_{i=1}^{k}\C{\blockdim{i}{A}}\otimes \C{\blockdim{i}{R}}$ of $\promise$ such that
    \begin{align}
        \rho_{A^nR^n} \leq g_{n,x} \int \ketbra{\phi}_{AR}^{\otimes n} \text{d} \phi,
    \end{align}
    where $x = \sum_{i=1}^k \blockdim{i}{A}\blockdim{i}{R}$.  
\end{restatable} 
The proof is given in \cref{app:deFinettiProfs}.    

We note that in the case where $k=1$, $\blockdim{1}{A}=d_A$ and $\blockdim{1}{R} = d_R$ we recover \cref{thm:genericDeFinetti}. More generally, $\sum_{i=1}^k \blockdim{i}{A}\blockdim{i}{R}\leq d_Ad_R$ leading to a reduced cost $\cost$.
Block-diagonal symmetries are an important special case of generic symmetries that are often seen in optical implementations of quantum information protocols. Thus, we next describe the usage of \cref{thm:promisedDeFinetti} for the case where the state has some block-diagonal symmetry before considering more general symmetries.

\subsubsection{Improvement for IID-block-diagonal states}

\begin{definition}[IID-block-diagonal states] \label{def:iidBLockDiagonal}
    Given a matrix $\state \in \matrixset{(\C{d_{A} d_{B}})^{\otimes n}},$ and a set of orthogonal projections $\{\Pi_i\}_{i=1}^k \subset \matrixset{\C{d_{A} d_{B}}}$, we say that the matrix is \textit{IID-block-diagonal} if
    \begin{align*}
        \state = \sum_{\vec{j}\in [k]^n} \Pi_{\vec{j}} \state \Pi_{\vec{j}}
    \end{align*}
    where $\Pi_{\vec{j}} \coloneqq \Pi_{j_1} \otimes \ldots \otimes \Pi_{j_n}$.
    We denote the rank of projector $\Pi_i$ as $d_i$; it corresponds to the dimension of the $i^{\text{th}}$ block. 
\end{definition}

Just as Lemma 4.2.2 from Ref.~\cite{renner_security_2005} was crucial in the proof of Corollary \ref{cor:generalMixedDeFinetti}, the heart of the improvement that uses block-diagonal structure is the following lemma.
\begin{restatable}{lemma}{blockDiagPurification}  \label{lem:blockDiagPurification}
    Let $\state \in \posset{\left(\C{d_{A} d_{B}}\right)^{\otimes n}}$ be a permutation-invariant and IID-block-diagonal matrix with respect to projections $\{\Pi_{i}\}_{i=1}^k$ of dimension $\{d_i\}_{i=1}^k$. Then there exists a purification of $\state$ supported on $\Sym{\bigoplus_{i=1}^k \C{d_i}\otimes \C{d_i}}$.
\end{restatable}
The proof is given in \cref{app:deFinettiProfs}.
\begin{remark}
    An important special case of Lemma \ref{lem:blockDiagPurification} is when $\state$ is IID-block-diagonal with respect to projections $\{\Pi_{i}^A\otimes\Pi_{j}^B\}_{i,j=1}^{k_A,k_B}$ of dimension $\{\blockdim{i}{A}\blockdim{j}{B}\}_{i,j=1}^{k_A,k_B}$ where $\blockdim{i}{A}$ and $\blockdim{j}{B}$ are the ranks of $\Pi^A_{i}\in \matrixset{\C{d_A}}$ and $\Pi^B_{j}\in \matrixset{\C{d_B}}$ respectively. In that case, there exists a purification of $\state$ on $\Sym{\bigoplus_{i,j=1}^{k_A,k_B} \C{\blockdim{i}{A}}\otimes \C{\blockdim{j}{B}}\otimes \C{\blockdim{i}{A}}\otimes \C{\blockdim{j}{B}}}=\Sym{\bigoplus_{i=1}^{k_A} \C{\blockdim{i}{A}}\otimes \C{\blockdim{i}{A}\sum_{j=1}^{k_B}\blockdim{j}{B}^2}}$.
\end{remark}
Note that this purification of $\rho_{A^nB^n}$ now belongs to $\Sym{\bigoplus_{i=1}^{k_A} \C{\blockdim{i}{A}}\otimes \C{\blockdim{i}{A}\sum_{j=1}^{k_B}\blockdim{j}{B}^2}}$ instead of $\Sym{\C{d_Ad_R}}$ in \cref{thm:genericDeFinetti}.
We thus obtain an improved version of \cref{cor:generalMixedDeFinetti}.

\begin{corollary} \label{cor:blockDiagDeFinetti}
    Let $\promise \in \posset{\mathbb{C}^{d_{A}}}$ and $\state$ be any permutation-invariant and IID-block-diagonal extension of $\left(\promise\right)^{\otimes n}$ with with respect to projections $\{\Pi_{i}^A\otimes\Pi_{j}^B\}_{i,j=1}^{k_A,k_B}$ of dimension $\{\blockdim{i}{A}\blockdim{j}{B}\}_{i,j=1}^{k_A,k_B}$, where $\blockdim{i}{A}$ and $\blockdim{j}{B}$ are the ranks of $\Pi^A_{i}\in \matrixset{\C{d_A}}$ and $\Pi^B_{j}\in \matrixset{\C{d_B}}$ respectively.
    Then there exists some probability measure d$\sigma_{AB}$ over the set of block-diagonal extensions $\sigma_{AB}$ of $\promise$ such that
    \begin{align}
        \state \leq \cost \int \sigma_{AB}^{\otimes n} \ \text{d}\sigma_{AB},
    \end{align}
    where $x=\sum_{i,j=1}^{k_A,k_B} \blockdim{i}{A}^2\blockdim{j}{B}^2$.
\end{corollary}
\begin{proof}
    The proof of this theorem is similar to the proof of \cref{cor:generalMixedDeFinetti}, replacing the use of Lemma 4.2.2 from Ref.~\cite{renner_security_2005} with \cref{lem:blockDiagPurification} and \cref{thm:genericDeFinetti} with \cref{thm:promisedDeFinetti}.
\end{proof}
Although the IID-block-diagonal condition might seem restrictive, we show in \cref{sec:PSDecoy} that optical implementations often naturally result in such IID-block-diagonal structure. Thus, this would greatly tighten the analysis of optical implementations of quantum information protocols.

\subsubsection{Improvement for IID symmetries}

\begin{definition}[IID-group-invariant]
    Let $G$ be a group, and let $\{U_g\}_{g\in G}$ be a unitary representations of $G$ on $\mathbb{C}^{d_A}\otimes\C{d_B}$. We say a matrix $\rho_{A^nB^n}\in\matrixset{(\mathbb{C}^{d_A}\otimes\C{d_B} )^{\otimes n}}$ is \textit{IID-$G$-invariant} if
    \begin{align*}
        U_{\Vec{g}}~\rho~ U_{\Vec{g}}^\dag=\rho
    \end{align*}
    for all $\Vec{g}\in G^n$, where $U_{\Vec{g}}:=\bigotimes_{i=1}^nU_{g_i}$ for $\Vec{g}=(g_1,\ldots,g_n)\in G^n$. 
\end{definition}

Similar to the block-diagonal case, the following lemma gives the improvement in the presence of symmetries.
\begin{restatable}{lemma}{groupPurification}\label{lem:groupPurification}
    Let $G$ be a compact group and let $\{U_g\}_{g\in G}$ be a unitary representation of $G$ on $\mathbb{C}^{d_Ad_B}$ with $k$ irreducible representations with multiplicity $\{m_i\}_{i=1}^k$. If $\state\in\posset{(\mathbb{C}^{d_Ad_B})^{\otimes n}}$ is permutation invariant and IID-$G$-invariant, then there exists a purification of $\state$ on $\Sym{\bigoplus_{i=1}^k\C{m_i}\otimes \C{m_i}}$.
\end{restatable}
The proof is given in \cref{app:deFinettiProfs}.
\begin{remark}
    An important special case of \cref{lem:groupPurification} is when the group $G = G_A \cross G_B$ is a product of compact groups. Let the unitary representations $\{U_{g_A}\}_{g_A\in G_A}\subset \matrixset{\C{d_A}}$ and $\{U_{g_B}\}_{g_B\in G_B}\subset \matrixset{\C{d_B}}$ with $k_A$, $k_B$ irreducible representations with multiplicity $\{m_i^A\}_{i=1}^{k_A}$ and $\{m_i^B\}_{i=1}^{k_B}$ respectively. Then there exists a purification of $\state$ on $\Sym{\bigoplus_{i,j=1}^{k_A,k_B}\C{{m_i^A}}\otimes\C{{m_j^B}}\otimes\C{m_i^A}\otimes\C{m_j^B}}=\Sym{\bigoplus_{i}^{k_A}\C{{m_i^A}}\otimes\C{m_i^A\sum_{j=1}^{k_B}{m_j^B}^2}}$.
\end{remark}
This can now be directly used to prove the de Finetti reduction for mixed states in the presence of symmetries.
\begin{corollary} \label{cor:symmetryDeFinetti}
    Let $\{U_{g_A}^A\}_{g_A\in G_A}$ ($\{U_{g_B}^B\}_{g_B\in G_B}$) be a unitary representation of $G_A$ ($G_B$) on $\mathbb{C}^{d_A}$ ($\C{d_B}$) with $k_A$ ($k_B$) irreducible representations with multiplicity $\{m_i^A\}_{i=1}^{k_A}$ ($\{m_j^B\}_{j=1}^{k_B}$).
    Let $G = G_A \cross G_B$, $\promise \in \posset{\mathbb{C}^{d_{A}}}$ and $\state$ be any permutation-invariant extension of $\left(\promise\right)^{\otimes n}$ that is IID-$G$-invariant. Then there exists some probability measure d$\sigma_{AB}$ over the set of G-invariant extensions $\sigma_{AB}$ of $\promise$ such that
    \begin{align}
        \state \leq \cost \int \sigma_{AB}^{\otimes n} \ \text{d}\sigma_{AB},
    \end{align}
    where $x=\sum_{i,j=1}^{k_A,k_B} {m_i^A}^2{m_j^B}^2$.
\end{corollary}
\begin{proof}
    The proof of this theorem is similar to the proof of \cref{cor:blockDiagDeFinetti}, replacing the use of \cref{lem:blockDiagPurification} with \cref{lem:groupPurification}.
\end{proof}

Of course, these improvements would also apply to the de Finetti reduction without the fixed marginal \cite[Lemma 2]{christandl_postselection_2009} in the case when $A$ is a trivial register.
Moreover, these improvements drastically improve upon the previous de Finetti reduction with fixed marginal \cite[Corollary 3.2]{fawzi_quantum_2015}, thereby increasing the key rates obtained via the postselection technique, as we show later in \cref{sec:applicationtothreestate}.
We now turn to the rigorous application of the de Finetti reduction with fixed marginal to QKD via the postselection technique.

\section{Correct application of de Finetti reductions to QKD} \label{sec:correctApplicationToQKD}

   In this section, we fill in a missing gap in Ref.~\cite{christandl_postselection_2009}, in the reduction of QKD security proofs from arbitrary attacks to IID-collective attacks, first noticed in \cite{belzig_studying_2020}. We also explain how the postselection technique can be applied to prepare-and-measure protocols. For pedagogical reasons, we will present our results for protocols satisfying a permutation-invariance property. For such protocols, we will use the generic de Finetti reduction mentioned in \cref{cor:generalMixedDeFinetti}, without the additional improvements from IID-block-diagonal structure \cref{cor:blockDiagDeFinetti} and IID symmetries \cref{cor:symmetryDeFinetti}.
   Proofs of most statements in this section are deferred to \cref{app:correctApplicationToQKD}.
  
   Given a de Finetti reduction $\state \leq \cost \tau_{A^nB^n}$, Ref.~\cite{christandl_postselection_2009} reduced the security proof of QKD protocols for arbitrary states to that of IID states. This followed in two steps. The first step is a reduction from the security of arbitrary states to the security of the state $\tau_{A^nB^n}$. The second is a reduction from the security for $\tau_{A^nB^n}$ to that of IID states. The second step in their analysis is argued intuitively and is not on sound mathematical grounds. Here, we present a rigorous proof of this step. For the sake of completeness, we explain the first step as well.
   
    Moreover, to use the results of Ref.~\cite{christandl_postselection_2009} one requires an IID security proof against \textit{arbitrary} IID states. This is typically not available for prepare-and-measure (PM) protocols, where the IID security proof considers Alice having a fixed marginal state.
    We address this issue by using the de Finetti reductions with fixed marginal \cite{fawzi_quantum_2015} proved in Sec.~\ref{sec:improvingDeFinetti} instead of the de Finetti reduction in Ref.~\cite{christandl_postselection_2009}.

\subsection{Using de Finetti reductions for QKD}
\label{subsec:conditions}

 We first establish some notation for QKD protocols 
 that produce a key of fixed length upon acceptance. 
 Let $\fullprotMap{l} \in \channelset{A^nB^n, K_AK_B\Cfull}$ be a QKD protocol map, that maps the input state $\rho_{A^nB^n}$ to the output state $\rho_{K_A K_B \Cfull}^{(l)}$. Here $l$ denotes the length of the key produced in the key registers $K_A$,$K_B$ if the protocol accepts, and $\Cfull$ denotes the regiser storing all classical announcements. If the protocol aborts, the key registers $K_A,K_B$ contain the special symbol $\bot$.
        Let $\protMap{l} = \Tr_{K_B} \circ \fullprotMap{l} $ denote the same map but with Bob's key omitted from the output.
        
        The ideal QKD protocol $\fullprotMapId{l}\in \channelset{A^nB^n, K_AK_B\Cfull}$ is one which implements the actual QKD protocol $\fullprotMap{l}$, and then replaces Alice and Bob's key registers with the perfect key of length of $l$ if the protocol accepts (and aborts if the protocol aborts).  Similarly, let $ \protMapId{l}= \Tr_{K_B} \circ \fullprotMapId{l}$ denote the same map but with Bob's key omitted from the output.
        The overall security of a QKD protocol can be described in terms of the maps $\fullprotMap{l}$ and $\fullprotMapId{l}$, as discussed in Ref.~\cite{portmann_cryptographic_2014}; however, as shown in that work, one can break down the security definition into simpler conditions referred to as correctness and secrecy. Correctness is fairly straightforward to prove even in the non-IID case, hence in this work we focus only on secrecy, which can be defined using only the maps $\protMap{l}$ and $\protMapId{l}$, as follows:
\begin{definition}[$\epsSec$-secret PMQKD protocol with fixed marginal $\promise$] \label{def:epsSecPromise}
    A PMQKD protocol $\protMap{l}\in\channelset{A^nB^n,K_A \Cfull} $ is \textit{$\epsSec$-secret with fixed marginal $\promise$} if 
    \begin{equation}
        \begin{aligned} \label{eq:epsSec}
            &\demi \norm{\left(\left(\protMap{l}-\protMapId{l}\right)\otimes \id_{{E}^n}\right) \rho_{A^nB^nE^n} }_1\leq \varepsilon_{\text{sec}}, \\
            &\forall  \rho_{A^nB^nE^n} \text{ such that } \Tr_{B^nE^n}\left( \rho_{A^nB^nE^n} \right) = \left(\promise \right)^{\otimes n},
        \end{aligned}
    \end{equation}
\end{definition}
Since the fixed marginal $\promise$ 
can be constructed from the description of a PMQKD protocol (see \cref{subsec:promisePMQKD}), we will refer to the above definition as $\epsSec$-secrecy of a PMQKD protocol.
Having defined what we mean by secrecy of a QKD protocol, following Ref.~\cite{christandl_postselection_2009} we justify the use of de Finetti reductions for QKD protocols through the following general statements.

We now define what it means for a map to be permutation-invariant. Note that in this definition we also correct a technical error in Ref.~\cite{christandl_postselection_2009} regarding the order in which the maps are applied.
\begin{definition}[Permutation-invariance of maps] \label{def:permInvariantMaps}
    A linear map $\Delta \in \mapset{A^n,B}$ is \textit{permutation-invariant}, if for every $\perm \in S_n $, there exists a $G_{\perm} \in \channelset{B,B}$ such that
    \begin{equation}
        G_\perm \circ \Delta \circ \mathcal{W}_{\perm} = \Delta
    \end{equation}
    where $\mathcal{W}_{\perm} (\cdot) = \permutation{d_{A}}{n} \left(\cdot\right) \permutation{d_{A}}{n}^\dagger$.
\end{definition}
Arguably, the above property might be better described as ``covariance'' rather than ``invariance'', since for instance it does not require that the map is literally ``invariant'' in the sense $\Delta \circ \mathcal{W}_{\perm} = \Delta$. However, for this work we shall follow the existing terminology in the field.
Note that if some channel $\mathcal{F}\in \channelset{A^n,B}$ begins by first applying a uniformly random permutation to its input registers (followed by other operations that do not depend on the choice of permutation), and its output registers include some classical register storing the choice of permutation, then it is a permutation-invariant map according to the above definition, despite not necessarily satisfying $\mathcal{F} \circ \mathcal{W}_{\perm} = \mathcal{F}$. Furthermore, there exists a relatively simple procedure to implement such a random permutation using approximately $n\log(n)$ uniform random bits. (To clarify further: in the context of QKD, this choice of random permutation can be publicly announced and hence these $n\log(n)$ bits can be generated locally by one party and publicly announced; they do not involve a consumption of pre-shared key.) We discuss the details in \cref{app:correctApplicationToQKD}.

We proceed in a manner similar to Ref.~\cite{christandl_postselection_2009}, and prove the following lemma that can be used to relate $\epsSec$-secrecy of arbitrary states to the $\epsSec$-secrecy of a state that is a purification of a de Finetti state.

\begin{restatable}{lemma}{StepOne}
 \label{lemma:endOfStepOne}
    Let $\mathcal{F}, \mathcal{F}^\prime \in \mapset{A^nB^n,K}$ be such that  $\mathcal{F}-\mathcal{F}^\prime$ is a permutation-invariant map. Let  $\rho_{A^nB^nR^{\prime\prime}} \in \posset{A^nB^nR^{\prime\prime}}$ with $\Tr_{B^n R^{\prime \prime }}\left(\rho_{A^nB^nR^{\prime\prime}}\right) = \left(\promise\right)^{\otimes n}$. Then there exists a probability measure d$\sigma_{AB}$ on the set of extensions $\sigma_{AB}$ of $\promise$ such that
    \begin{equation}
        \tracenorm{ \left( \left( \mathcal{F} - \mathcal{F}^\prime \right) \otimes \id_{R^{\prime\prime}} \right) \left( \rho_{A^nB^nR^{\prime\prime}} \right) } \leq \cost \tracenorm{ \left( \left(\mathcal{F} - \mathcal{F}^\prime \right) \otimes \id_{R} \right) \left( \tau_{A^nB^nR} \right) },
    \end{equation}
    where $\tau_{A^nB^nR}$ is any purification of $\tau_{A^nB^n} =  \int \text{d} \sigma_{AB}\  \sigma_{AB}^{\otimes n}$, and $x = d_A^2d_B^2$.
\end{restatable}
The proof is given in \cref{app:correctApplicationToQKD}.

The next step is to relate the security of $\tau_{A^nB^nR}$ to the IID security statements, which we do in the next section.

\subsection{Reducing Security of QKD protocols to the IID case}

 \cref{lemma:endOfStepOne} allows us to reduce the $\epsSec$-secrecy with fixed marginal of a QKD protocol for any arbitrary input state, to the $\epsSec$-secrecy of the protocol when the input state is a purification $\tau_{A^nB^nR}$ of a mixture of IID states $\tau_{A^nB^n}$ with the same fixed marginal.
In this subsection, we will rigorously reduce the $\epsSec$-secrecy of a QKD protocol acting on $\tau_{A^nB^nR}$ to that of a QKD protocol against IID-collective attacks. Note that several techniques for proving the $\epsSec$-secrecy of QKD protocols against IID-collective attacks are known \cite{george_numerical_2021,renner_security_2005}. We will now reduce the $\epsSec$-secrecy of a QKD protocol to one such widely-used IID security proof technique \cite{george_numerical_2021}. We first present the structure of such an IID security proof technique below.

\subsubsection{Structure of IID security proof} \label{subsubsec:structureofIIDsecurityproof}

First to set up some notation, let $ \promiseset= \{ \sigma_{ABE} \in \stateset{ABE} : \Tr_{BE} (\sigma_{ABE}) = \promise \}$ be the set of all states that have a fixed marginal on Alice's subsystem. These are the states for which security must hold.  Suppose the protocol is run with the input state $\sigma^{\otimes n}_{ABE}$. We use $\sigma_{Z^nC^nC_E E^n}$ to denote the state of Alice's raw key $Z^n$, round-by-round announcements $C^n$, error-correction and error-verification announcements $C_E$ and Eve's quantum system $E^n$, just before the privacy amplification step. Moreover, we use $\sigma^{(l)}_{K_A \Cfull E^n}$ be the output of the protocol, where $K_A$ is Alice's $l$-bit key register \footnote{$K_A$ technically also contains the special $\bot$ symbol if the protocol aborts} and $\Cfull = C^nC_EC_P$ where $C_P$ is the classical announcement of the two-universal hash function during privacy amplification.

The $\epsSec$-secrecy of QKD protocols against IID-collective attacks is typically proven by showing that the following statements hold:
\begin{enumerate}
    \item Choose $\epsAT \in [0,1]$, and construct a set $S \subset \promiseset $ such that the set of states not in $S$ but still in $\promiseset$ abort with probability at least $1-\epsAT$, that is,
    \begin{equation} \label{eq:condS}
        \sigma_{ABE}\in \promiseset\setminus S \implies \Pr(\accevent) \leq \epsAT,
    \end{equation}
    where $\accevent$ denotes the event that the protocol does not abort. 
    \item  The hash length $l$, $\epsPA \in [0,1]$ and $\epsbar \in [0,1]$ are chosen to be secure for all states in $S$ \cite{tomamichel_largely_2017}, that is,
    \begin{equation} \label{eq:condLHL}
        \begin{aligned}
            \frac{1}{2}\Pr(\accevent) \tracenorm{ \sigma^{(l)}_{K_A\Cfull E^n | \accevent } - \sigma^{(l), \text{ideal}}_{K_A \Cfull E^n | \accevent}  } &\leq \frac{1}{2} 2^{-\frac{1}{2} \left( \smoothmin{\epsbar} \left(Z^n | C^nC_EE^n\right)_{\sigma \wedge \accevent} - l \right)} + 2 \epsbar \leq			\epsPA + 2\epsbar, \\
            &  \quad \forall \sigma \in S,
        \end{aligned}
    \end{equation}
\end{enumerate}
Note that ~\cref{eq:condS,eq:condLHL} together imply the $\epsSec$-secrecy statement with $\epsSec = \max\{ \epsAT , \epsPA + 2\epsbar \}$
 \begin{equation} \label{eq:condsecrecy}
    \frac{1}{2} \Pr(\accevent) \tracenorm{ \sigma^{(l)}_{K_A\Cfull E^n|\accevent} - \sigma^{(l), \text{ideal}}_{K_A \Cfull E^n|\accevent}  } \leq \epsSec, \quad \forall \sigma \in \promiseset
\end{equation}

\subsubsection{Final reduction}

We can now state the final reduction to IID security proofs in the following theorem.
\begin{restatable}[Postselection Theorem]{theorem}{maintheorem} \label{thm:maintheorem}
    Suppose $\protMap{l} \in \channelset{A^nB^n, K_A \Cfull} $ is such that \cref{eq:condS,eq:condLHL} are satisfied. Let the state $\tau_{A^nB^n}$ be given by
    \begin{align}
        \tau_{A^nB^n} = \int \sigma_{AB}^{\otimes n} \text{d} \sigma_{AB},
    \end{align}
    where d$\sigma_{AB}$ is some probability measure on the set of non-negative extensions $\sigma_{AB}$ of $\promise$ and $\tau_{A^nB^nR}$ be a purification of $\tau_{A^nB^n}$. Let $\protMap{l^\prime}$ be a QKD protocol map identical to $\protMap{l}$, except that it hashes to a length $l^\prime =  l - 2 \log(\cost)$ instead of $l$ upon acceptance, and $x=d^2_Ad_B^2$. Then,
    \begin{equation}
        \frac{1}{2}\tracenorm{\left( \left(\protMap{l^\prime} - \protMapId{l^\prime}  \right)\otimes \id_R \right) (\tau_{A^nB^nR})} \leq \epsPA+2 \epsbar+2 \sqrt{2 \epsAT} .
    \end{equation}
   
\end{restatable}
The proof is given in \cref{app:correctApplicationToQKD}.

Through a series of lemmas and theorems, we have reduced the security proof of a QKD protocol against arbitrary attacks to the security proof of a similar QKD protocol against IID-collective attacks. There are two costs to be paid for this lift. One is a cost paid to the $\epsSec$ as stated in \cref{lemma:endOfStepOne}. The other is a cost paid to the hash length that can be chosen as stated in \cref{thm:maintheorem}. We bring together the entire reduction formally in the following corollary.
\begin{corollary}  \label{cor:liftToCoherent}
     Suppose $\protMap{l} \in \channelset{A^nB^n, K_A \Cfull} $ is a QKD protocol map satisfying \cref{eq:condS,eq:condLHL}, which produces a key of length $l$ upon acceptance.
     Let $\protMap{l^\prime}$ be a QKD protocol map identical to $\protMap{l}$ except that it hashes to a length $l^\prime$ instead of $l$ upon acceptance. 
     Suppose $\protMap{l^\prime} - \protMapId{l^\prime}$ is a permutation-invariant map for all $l^\prime$. Then, $\protMap{l'}$ is $g_{n,x} \left( \epsPA+2\epsbar + 2\sqrt{2 \epsAT} \right)$-secret with fixed marginal $\promise$, for $l^\prime = l -2 \log(\cost)$.
\end{corollary}
\begin{proof} 
Since $\protMap{l^\prime} - \protMapId{l^\prime}$ is permutation-invariant, \cref{lemma:endOfStepOne} states that there exists a probability measure $d\sigma_{AB}$ on the set of extensions $\sigma_{AB}$ of $\promise$ such that 
 \begin{equation} \label{eq:incor2.1}
        \frac{1}{2} \tracenorm{ \left( \left( \protMap{l^\prime} - \protMapId{l^\prime} \right) \otimes \id_{E^n} \right) \left( \rho_{A^nB^nE^n} \right) } \leq  \frac{1}{2} \cost \tracenorm{ \left( \left(\protMap{l^\prime} - \protMapId{l^\prime} \right) \otimes \id_{R} \right) \left( \tau_{A^nB^nR} \right) },
    \end{equation}
    where $\tau_{A^nB^nR}$ is any purification of $\tau_{A^nB^n} =  \int \text{d} \sigma_{AB}\  \sigma_{AB}^{\otimes n}$, and $x=d_A^2d_B^2$. The claim then follows from Theorem \ref{thm:maintheorem}.
\end{proof}

Note that although all our proofs were stated for PMQKD protocols, they are also applicable to entanglement-based QKD protocols by choosing the fixed marginal to be trivial. This is an improvement on Ref.~\cite{christandl_postselection_2009}, whose results are only applicable for entanglement-based QKD protocols. Furthermore, we make rigorous a verbal argument made in Ref.~\cite{christandl_postselection_2009} in Theorem \ref{thm:maintheorem}. In doing so, we notice that the key secrecy parameter is worse than predicted by Ref.~\cite{christandl_postselection_2009}. In particular, Ref.~\cite{christandl_postselection_2009} obtains a secrecy parameter of $\cost \max \{ \epsPA + 2\epsbar, \epsAT \}$ against arbitrary attacks, as compared to $\cost \left( \epsPA+2\epsbar + 2\sqrt{2 \epsAT} \right)$ which we obtain. 
Finally, we highlight that our proof also covers protocols using two-way communication in error correction, as long as the $C_E$ register includes all public communication that takes place in this process, and $Z^n$ is replaced with a register $\breve{Z}$ that represents the string Alice uses as the input to the privacy amplification procedure.

In situations where the QKD protocol satisfies additional symmetries, one can repeat the previous analysis as shown below to obtain the corresponding statement in \cref{cor:liftToCoherentSymmetries}. We first define the notion of IID-group-invariance of maps, analogous to our definition of permutation-invariance of maps.

\begin{definition}[IID-group-invariance of maps] \label{def:groupInvariantMaps}
    Let $H$ be a compact group with unitary representation $W_h$ on $A$. A linear map $\Delta \in \mapset{A^n,B}$ is \textit{IID-$H$-invariant}, if for every $\vec{h} \in H^n $, there exists a $G_{\vec{h}} \in \channelset{B, B}$ such that
    \begin{equation}
        G_{\vec{h}} \circ \Delta \circ \mathcal{W}_{\vec{h}} = \Delta
    \end{equation}
    where $\mathcal{W}_{\vec{h}} (\cdot) = \bigotimes_{i=1}^n W_{h_i} \left(\cdot\right) \bigotimes_{i=1}^n W_{h_i}^\dagger$ for $\Vec{h}\in (h_1, \ldots , h_n) \in H^n$.
\end{definition}

\begin{corollary}  \label{cor:liftToCoherentSymmetries}
     Suppose $\protMap{l} \in \channelset{A^nB^n, K_A \Cfull} $ is a QKD protocol map satisfying \cref{eq:condS,eq:condLHL}, which produces a key of length $l$ upon acceptance.
     Let $\protMap{l^\prime}$ be a QKD protocol map identical to $\protMap{l}$ except that it hashes to a length $l^\prime$ instead of $l$ upon acceptance.
     Suppose for all values of $l^\prime$, $\protMap{l^\prime} - \protMapId{l^\prime}$ is a permutation-invariant and IID-$G$-invariant map for a compact group $G = G_A \times G_B$, where the unitary representation $G_A (G_B)$ has $k_A (k_B)$ irreducible representations with multiplicities $\{m^A_i\}_{i=1}^{k_A}$ $(\{m^B_i \}_{i=1}^{k_B})$. Then $\protMap{l'}$ is $g_{n,x} \left( \epsPA+2\epsbar+2 \sqrt{2 \epsAT} \right)$-secret with fixed marginal $\promise$, for $l^\prime =  l - 2 \log(\cost)$, where $x = \sum_{i,j=1}^{k_A,k_B} (m^A_i)^2 (m^B_j)^2$.
\end{corollary}

For brevity, we do not formally prove \cref{cor:liftToCoherentSymmetries}.
The proof is a simple modification of the steps presented in this work. We use \cref{lemma:psLemma} with permutation-invariance and IID-$G$-invariance of the maps $\protMap{l}$ to restrict our analysis to states which are permutation-invariant and IID-$G$-invariant.
For such states, we use the de Finetti reduction from \cref{cor:symmetryDeFinetti} and restate \cref{thm:maintheorem} with the improved de Finetti reduction. This gives the corresponding improvement in \cref{cor:liftToCoherentSymmetries}.

A special case of \cref{cor:liftToCoherentSymmetries}, is when the QKD protocol is IID-block-diagonal.
\begin{definition}[IID-block-diagonal maps] \label{def:blockDiagMaps}
    Let $\{\Pi_i\}_{i=1}^k$ be a set of orthogonal projectors on $A$. A linear map $\Delta \in \mapset{A^n,B}$ is \textit{IID-block-diagonal}, if
    \begin{equation}
        \sum_{\vec{i}\in[k]^n} \Delta \circ \mathcal{P}_{\vec{i}} = \Delta
    \end{equation}
    where $\mathcal{P}_{\vec{i}} (\cdot) = \bigotimes_{j=1}^n \Pi_{i_j} \left(\cdot\right) \bigotimes_{j=1}^n \Pi_{i_j}$.
\end{definition}

\begin{corollary}  \label{cor:liftToCoherentBlockDiagonal}
     Suppose $\protMap{l} \in \channelset{A^nB^n, K_A \Cfull} $ is a QKD protocol map satisfying \cref{eq:condS,eq:condLHL}, which produces a key of length $l$ upon acceptance.
     Let $\protMap{l^\prime}$ be a QKD protocol map identical to $\protMap{l}$ except that it hashes to a length $l^\prime$ instead of $l$ upon acceptance.
     Suppose for all values of $l^\prime$, $\protMap{l^\prime} - \protMapId{l^\prime}$ is a permutation-invariant and IID-block-diagonal map with respect to projections $\{\Pi_i^A \otimes \Pi_j^B\}_{i,j=1}^{k_A,k_B}$ of dimension $\{d^A_id^B_j \}_{i,j=1}^{k_A,k_B})$. Then $\protMap{l'}$ is $g_{n,x} \left( \epsPA+2\epsbar+2 \sqrt{2 \epsAT} \right)$-secret with fixed marginal $\promise$, for $l^\prime =  l - 2 \log(\cost)$, where $x = \sum_{i,j=1}^{k_A,k_B} (d^A_i)^2 (d^B_j)^2$.
\end{corollary}
The proof is given in \cref{app:correctApplicationToQKD}.

In the next section, we briefly explain the application of the postselection technique to variable-length QKD protocols.

\subsection{Using Postselection technique for variable-length protocols}
\label{subsec:varLift}

So far we have dealt with ``fixed-length'' protocols, which either abort ($\accevent^c$ occurs) or produce a key of fixed-length $l$ ($\accevent$ occurs). In practice, one may wish to implement a variable-length protocol \cite{portmann_cryptographic_2014,hayashi_security_2014,tupkary_security_2023}, where the length of the final key is not fixed and depends on events taking place during the run of the protocol. Such protocols are more practical, since they do not require prior characterization of the honest behavior of the channel, and can 
adapt the length of the final key produced based on observations made during the protocol. A new security proof of variable-length protocols against IID-collective attacks was recently shown in Ref.~\cite{tupkary_security_2023}. In this section, we will apply the postselection technique to lift the security of variable-length protocols of this form to arbitrary attacks.  

We first set up some notation. We let $M$ denote the number of possible lengths of the output key. Thus $K_A$,$K_B$ is now a classical register that stores bit strings up to some maximum length $l_\text{max}$. Analogous to the notation for fixed-length protocols, we use $\protMapVar{l} \in \channelset{A^nB^n, K_A \Cfull}$  to denote the protocol map (excluding Bob's key) for variable-length protocols, which produces a key of length $l_i$ 
in the register $K_A$ upon the  event $\Omega_i$, and makes classical announcements in the register $\Cfull$. Aborting is modelled as the key registers $K_A,K_B$ containing the special symbol $\bot$.
 The ideal protocol map $\protMapIdVar{l} \in \channelset{A^nB^n, K_A \Cfull}$ first implements the real QKD protocol using $\protMapVar{l}$, computes the length of the key produced (i.e.~which event $\Omega_i$ occured), and then replaces the key registers with the perfect keys of the same length (or aborts if the real QKD protocol aborts).

The $\epsSec$-secrecy definition for variable-length protocols is then  given by

\begin{definition}[$\epsSec$-secret variable-length PMQKD protocol with fixed marginal $\promise$] \label{def:VarLengthepsSecPromise}
    A variable-length PMQKD protocol $\protMapVar{l} \in\channelset{A^nB^n,K_A \Cfull} $ is \textit{$\epsSec$-secret with fixed marginal $\promise$} if 
    \begin{equation}
        \begin{aligned} \label{eq:epsSecVar}
            &\demi \norm{\left(\left(\protMapVar{l}-\protMapIdVar{l}\right)\otimes \id_{{E}^n}\right) (\rho_{A^nB^nE^n}) }_1\leq \varepsilon_{\text{sec}},\\
        \end{aligned}
    \end{equation}
    for all $ \rho_{A^nB^nE^n}$  such that $\Tr_{B^nE^n}\left( \rho_{A^nB^nE^n} \right) = \left(\promise \right)^{\otimes n}$.
\end{definition}

The final reductions to IID security proofs for variable-length protocols is stated below as an analogous theorem to \cref{thm:maintheorem}.

\begin{restatable}[Postselection Theorem for Variable-length]{theorem}{maintheoremvar} \label{thm:maintheoremvar}
    Suppose $\protMapVar{l} \in \channelset{A^nB^n, K_A \Cfull} $ is such that the $\epsSec$-secrecy condition (\cref{eq:epsSecVar}) holds for all IID states $\rho_{A^nB^nE^n} = \sigma^{\otimes n}_{ABE}$ satisfying $\Tr_{BE}(\sigma_{ABE}) = \promise$.
    Let the state $\tau_{A^nB^n}$ be given by
    \begin{align}
        \tau_{A^nB^n} = \int \sigma_{AB}^{\otimes n} \text{d} \sigma_{AB},
    \end{align}
    where d$\sigma_{AB}$ is some probability measure on the set of non-negative extensions $\sigma_{AB}$ of $\promise$ and $\tau_{A^nB^nR}$ be a purification of $\tau_{A^nB^n}$. Let $\protMapVar{l^\prime}$ be a variable-length QKD protocol map identical to $\protMapVar{l}$, except that it hashes to a length $l_i^\prime =  l_i - 2 \log(\cost) - 2\log(1/\epstilde)$ instead of length $l_i$, upon the event $\Omega_i$, where $x=d^2_Ad_B^2$.
     Then,
    
    \begin{equation}
        \frac{1}{2} \tracenorm{\left( \left(\protMapVar{l^\prime} - \protMapIdVar{l^\prime}  \right)\otimes \id_R \right) (\tau_{A^nB^nR})} \leq \sqrt{8\epsSec} + \frac{\epstilde}{2} 
    \end{equation}
   \end{restatable}
The proof is given in \cref{app:correctApplicationToQKD}.

Similar to \cref{cor:liftToCoherent}, we then have the following corollary concerning the lift to coherent attacks for variable-length protocols.

\begin{corollary}  \label{cor:liftToCoherentvarx}
    Suppose $\protMapVar{l} \in \channelset{A^nB^n, K_A \Cfull} $ is
    % IID-$\epsSec$-secret, i.e.~
    a map such that
    the $\epsSec$-secrecy condition (\cref{eq:epsSecVar}) holds for all IID states $\rho_{A^nB^nE^n} = \sigma^{\otimes n}_{ABE}$ satisfying $\Tr_{BE}(\sigma_{ABE}) = \promise$. Let $\protMapVar{l^\prime}$ be a variable-length QKD protocol map identical to $\protMapVar{l}$, except that it hashes to a length $l_i^\prime$ instead of $l_i$ upon the event $\Omega_i$.  Suppose $\protMapVar{l^\prime} - \protMapIdVar{l^\prime}$ is a permutation-invariant map for all $l_i^\prime$. Then $\protMapVar{l'}$ is $\cost \left( \sqrt{8 \epsSec} + \frac{\epstilde}{2} \right)$-secret with fixed marginal $\promise$, for $l_i^\prime =  l_i - 2 \log(\cost) - 2 \log(1/\epstilde)$ and $x=d_A^2d_B^2$. 
\end{corollary}
\begin{proof} 
Since $\protMapVar{l^\prime} - \protMapIdVar{l^\prime}$ is permutation-invariant, \cref{lemma:endOfStepOne} states that there exists a probability measure $d\sigma_{AB}$ on the set of extensions $\sigma_{AB}$ of $\promise$ such that 
 
    \begin{equation} 
    \begin{aligned}
        &\tracenorm{ \left( \left( \protMapVar{l^\prime} - \protMapIdVar{l^\prime} \right) \otimes \id_{E^n} \right) \left( \rho_{A^nB^nE^n} \right) }\\
        &\leq \cost \tracenorm{ \left( \left(\protMapVar{l^\prime} - \protMapIdVar{l^\prime} \right) \otimes \id_{R} \right) \left( \tau_{A^nB^nR} \right) },
        \end{aligned}
    \end{equation}
    where $\tau_{A^nB^nR}$ is any purification of $\tau_{A^nB^n} =  \int \text{d} \sigma_{AB}\  \sigma_{AB}^{\otimes n}$ and $x=d_A^2d_B^2$. The claim then follows from Theorem \ref{thm:maintheoremvar}.
\end{proof}

Just as \cref{cor:liftToCoherent} could be modified to yield an improved \cref{cor:liftToCoherentSymmetries} in presence of symmetries, one obtains the following \cref{cor:liftToCoherentvarxSymmetries}

  \begin{corollary}  \label{cor:liftToCoherentvarxSymmetries}
 Suppose $\protMapVar{l} \in \channelset{A^nB^n, K_A \Cfull} $ is 
 % IID-$\epsSec$-secret, i.e.~
 a map such that
 the $\epsSec$-secrecy condition (\cref{eq:epsSecVar}) holds for all IID states $\rho_{A^nB^nE^n} = \sigma^{\otimes n}_{ABE}$ satisfying $\Tr_{BE}(\sigma_{ABE}) = \promise$. Let $\protMapVar{l^\prime}$ be a variable-length QKD protocol map identical to $\protMapVar{l}$, except that it hashes to a length $l_i^\prime$ instead of $l_i$ upon the event $\Omega_i$.  Suppose $\protMapVar{l^\prime} - \protMapIdVar{l^\prime}$ is a permutation-invariant map, and an IID-$G$-invariant for a compact group $G = G_A \times G_B$, where the unitary representation $G_A (G_B)$ has $k_A (k_B)$ irreducible representations with multiplicities $\{m^A_i\}_{i=1}^{k_A}$ $(\{m^B_i \}_{i=1}^{k_B})$. 
  Then $\protMapVar{l'}$ is $g_{n,x} \left( \sqrt{8 \epsSec} + \frac{\epstilde}{2} \right)$-secret with fixed marginal $\promise$, with $l_i^\prime = l_i - 2 \log(\cost) - 2 \log(1/\epstilde)$ and $x = \sum_{i,j=1}^{k_A,k_B} (m^A_i)^2 (m^B_j)^2$.
\end{corollary}

The proof of this corollary is similar to that of \cref{cor:liftToCoherentSymmetries}, replacing the use of \cref{thm:maintheorem} with \cref{thm:maintheoremvar}. We also state a special case of \cref{cor:liftToCoherentvarxSymmetries}.
  \begin{corollary}  \label{cor:liftToCoherentvarxBlockDiagonal}
 Suppose $\protMapVar{l} \in \channelset{A^nB^n, K_A \Cfull} $ is 
 % IID-$\epsSec$-secret, i.e.~
 a map such that
 the $\epsSec$-secrecy condition (\cref{eq:epsSecVar}) holds for all IID states $\rho_{A^nB^nE^n} = \sigma^{\otimes n}_{ABE}$ satisfying $\Tr_{BE}(\sigma_{ABE}) = \promise$. Let $\protMapVar{l^\prime}$ be a variable-length QKD protocol map identical to $\protMapVar{l}$, except that it hashes to a length $l_i^\prime$ instead of $l_i$ upon the event $\Omega_i$.  Suppose $\protMapVar{l^\prime} - \protMapIdVar{l^\prime}$ is a permutation-invariant map, and an IID-block-diagonal map with respect to projections $\{\Pi_i^A \otimes \Pi_j^B\}_{i,j=1}^{k_A,k_B}$ of dimension $\{d^A_id^B_j \}_{i,j=1}^{k_A,k_B})$. 
  Then $\protMapVar{l'}$ is $g_{n,x} \left( \sqrt{8 \epsSec} + \frac{\epstilde}{2} \right)$-secret with fixed marginal $\promise$, with $l_i^\prime = l_i - 2 \log(\cost) - 2 \log(1/\epstilde)$ and $x = \sum_{i,j=1}^{k_A,k_B} (d^A_i)^2 (d^B_j)^2$.
\end{corollary}

Note that unlike \cref{cor:liftToCoherent,cor:liftToCoherentBlockDiagonal,cor:liftToCoherentSymmetries}, the above \cref{cor:liftToCoherentvarx,cor:liftToCoherentvarxBlockDiagonal,cor:liftToCoherentvarxSymmetries} do not have any restrictions on the IID security proofs they lift to coherent attacks. Therefore, they are applicable more generally. Furthermore, a fixed-length protocol is a special case of a variable-length protocols where the number of possible output key lengths ($M$) is 1. Thus, \cref{cor:liftToCoherentvarx,cor:liftToCoherentvarxBlockDiagonal,cor:liftToCoherentvarxSymmetries} can be applied to arbitrary \textit{fixed-length} IID proofs as well. However, the performance of the specialized \cref{cor:liftToCoherent,cor:liftToCoherentBlockDiagonal,cor:liftToCoherentSymmetries} is superior.

Having explained rigorously the application of the postselection technique to QKD protocols we now move on to consider its application to optical protocols.

\section{Postselection technique for optical protocols} \label{sec:PSDecoy}

Decoy-state methods \cite{hwang_quantum_2003, lo_decoy_2005, wang_beating_2005} are essential to perform QKD at large distances in the absence of single photon sources. However, the analysis is performed on infinite-dimensional optical states on both Alice and Bob's spaces.
Thus, the postselection technique cannot be directly applied to it as the dimension enters the correction factor $\cost$. Here, we rigorously show the reductions necessary to apply the postselection technique to decoy-state protocols to finite dimensions through source maps \cite{gottesman_security_2004,nahar_imperfect_2023} and squashing maps \cite{gittsovich_squashing_2014, zhang_security_2021}. In doing so we also develop a new flag-state squasher \cite{zhang_security_2021}, an essential tool to prove security for optical protocols that use realistic detector setups.
We also describe the effect that this reduction plays on the IID decoy-state analysis.
Proofs of most statements in this section are deferred to \cref{app:decoyProofs}.

\subsection{Construction of fixed marginal for PMQKD} \label{subsec:promisePMQKD}

In Definition \ref{def:epsSecPromise}, we defined the $\epsSec$-secrecy of a QKD protocol with fixed marginal $\promise$. In this subsection, we describe the explicit construction of such a fixed marginal for any PMQKD protocol.

A PMQKD protocol where Alice prepares states $\rho_i$ with probability $p(i)$ can be equivalently described by the state preparation
\begin{align} \label{eqRhoMeas}
    \rhoMeas = \sum_{i=1}^{d_{\text{A}}} p(i)\ketbra{i}_A\otimes \rho_i
\end{align}
where the $A$ register represents the classical register where Alice notes the states $\rho_i $ prepared and sent to Bob. Thus, the security definition for such a protocol can be given by the following.
\begin{definition}[$\epsSec$-secret PMQKD protocol]
    A PMQKD protocol $\protMap{l}\in\channelset{A^nB^n,K_A \Cfull} $ is \textit{$\epsSec$-secret} if 
    \begin{align} \label{eq:epsilonSecurityRho}
        \demi \norm{\left(\left(\protMap{l}-\protMapId{l}\right)\otimes \id_{{E}^n}\right) \left[\left(\id_{{A}^n}\otimes \Phi\right)\left[\rhoMeasN\right]\right]}_1\leq \varepsilon_{\text{sec}}
    \end{align}
    for all channels $\Phi\in\channelset{A'^n,B^nE^n}$.
\end{definition}
Here, $\Phi$ can be understood to be Eve's attack on the states Alice sends to Bob.
The IID security proof techniques described in subsection~\ref{subsubsec:structureofIIDsecurityproof} often give non-trivial key lengths only when Alice prepares pure states. Thus, we use the source-replacement scheme \cite{bennett_quantum_1992, curty_entanglement_2004,ferenczi_symmetries_2011} and a shield system \cite{horodecki_general_2009} to construct a fixed marginal involving pure states.
\begin{restatable}[Shield system]{lemma}{lemmaShield} \label{lemma:Shield}
    Let $\protMap{l}$ be a PMQKD protocol with Alice's state preparation described by $\rhoMeas=\sum_{i=1}^{d_{\text{A}}} p(i)\ketbra{i}_A\otimes \rho_i$. Let $\protMapShield{l}$ be another PMQKD protocol identical to $\protMap{l}$ except that Alice's state preparation is given by
    \begin{align} \label{eqShieldSystemRho}
        \rho^\text{prep}_{AA_SA'} = \sum_{i=1}^{d_{A}} p(i)\ketbra{i}_A\otimes \ketbra{\rho_i}_{A_SA'},
    \end{align}
    where $A_S$ (termed the \emph{shield system}) is not sent to Bob and is acted on trivially by Alice. Here, $\ket{\rho_i}$ is related to the signal states Alice prepares $\rho_i = \Tr_{A_S}[\ketbra{\rho_i}]$.
    If the PMQKD protocol $\protMapShield{l}$ is $\epsSec$-secret, then the PMQKD protocol $\protMap{l}$ is $\epsSec$-secret.
\end{restatable}
The proof is given in \cref{app:decoyProofs}.

Note that the purified states $\ket{\rho_i}$ can be chosen to be \textit{any} purification of the signal states $\rho_i$. Thus, the dimension of the shield system $d_{A_S}$ is the largest rank of the signal states.
\begin{restatable}[Source-replacement scheme]{lemma}{lemmaSRS}
    Let $\protMap{l}$ be a PMQKD protocol where Alice's state preparation is given by $\rho^\text{prep}_{AA_SA'} = \sum_{i=1}^{d_{A}} p(i)\ketbra{i}_A\otimes \ketbra{\rho_i}_{A_SA'}$.
    Let $\protMapShieldSR{l}$ be another PMQKD protocol identical to $\protMap{l}$ except Alice's state preparation is given by
    \begin{align} \label{eqSourceReplacementRho}
        \rho_{AA_SA'} = \sum_{i,j=1}^{d_{A}} \sqrt{p(i)p(j)}\ketbra{i}{j}_A\otimes \ketbra{\rho_i}{\rho_j}_{A_SA'},
    \end{align}
    and Alice's register $A$ is measured in the computational basis at the start of the protocol. If the PMQKD protocol $\protMapShieldSR{l}$ is $\epsSec$-secret, then $\protMap{l}$ is an $\epsSec$-secret PMQKD protocol.
\end{restatable}
The proof is given in \cref{app:decoyProofs}.

The fixed marginal used in Definition~\ref{def:epsSecPromise} can thus be constructed as $\promiseShield = \Tr_{A'}\left[\rho_{AA_SA'}\right]$ where $\rho_{AA_SA'}$ is obtained from the source-replacement scheme.
This represents the fact that Eve's channel acts only on the $A'$ register sent to Bob leaving $A$ and $A_S$ unchanged.
Thus, with this fixed marginal, the output of Eve's channel can be replaced with an arbitrary state $\rho_{A^nA_S^nB^nE^n}$ shared by Alice, Bob and Eve with marginal $\promiseShield^{\otimes n}$ on Alice's marginal state.

The usage of \cref{cor:liftToCoherent} to lift a security proof that assumes that Eve's attack $\Phi$ is IID to a security proof against general attacks requires the Hilbert spaces $A_S$ and $B$ to be finite-dimensional. However, for typical optical systems, Bob's detectors are typically infinite-dimensional.
Additionally, for many protocols (such as decoy-state QKD protocols), the shield system $A_S$ is not finite-dimensional.
We describe tools to address this in \cref{subsec:squash} and \cref{subsec:decoyStateLift} respectively.

\subsection{Squashing maps} \label{subsec:squash}

A squashing model is a framework that allows a description of measurements in Hilbert spaces that is smaller than their natural representation. The original propositions \cite{fung_universal_2011, gittsovich_squashing_2014, zhang_security_2021, upadhyaya_dimension_2021} prove the applicability of squashing models by showing that their usage lower bounds the key rate under the assumption of IID-collective attacks. However, this alone is insufficient to apply the squashing model to make Bob's system finite-dimensional and apply the postselection technique.
Thus, the following lemma describes a sufficient condition for the usage of the squashing map for the postselection technique.

\begin{restatable}[Squashing]{lemma}{lemmaSquash} \label{lemma:squash}
    Let $\protMap{l}$ be a QKD protocol where Bob's measurement is described by POVM $\{\Gamma_i\}_{i=1}^{\nmeas} \subset \boundedOpSet{B}$.
    Let $\protMapSquash{l}$ be another QKD protocol identical to $\protMap{l}$ except Bob's measurement is described by POVM $\{F_i\}_{i=1}^{\nmeas} \subset \boundedOpSet{Q}$. If there exists a channel $\Lambda \in \channelset{B,Q}$ such that $\Lambda^\dag\left[F_i\right] = \Gamma_i$ for all $i$, then the $\epsSec$-secrecy of the PMQKD protocol $\protMapSquash{l}$ implies the $\epsSec$-secrecy of the PMQKD protocol $\protMap{l}$.
\end{restatable}
The proof is given in \cref{app:decoyProofs}.

Note that not all squashing models use such a squashing map $\Lambda$. In particular, the universal squashing model \cite{fung_universal_2011} and the dimension reduction method \cite{upadhyaya_dimension_2021} do not proceed through proving the existence of such a map, and thus they cannot be used with the postselection technique via \cref{lemma:squash} \footnote{Since the dimension reduction method is the only reduction to finite dimensions for CVQKD that is currently known, the use of the postselection technique in CVQKD is an open problem.}.

The various multiphoton-to-qubit squashing maps described in Ref.~\cite{gittsovich_squashing_2014} only exist under a very restrictive set of parameter regimes for the QKD protocols which are not robust to device imperfections.
Thus, for the remainder of this manuscript, we turn our attention to the flag-state squasher \cite{zhang_security_2021} which exists for every QKD protocol that uses threshold detectors.

\subsubsection{Weight-preserving flag-state squasher} \label{subsec:WPFSS}

The flag-state squasher exists whenever the POVM elements have a block-diagonal structure
\begin{align*}
    \Gamma_i = \Gamma_{i,m\leq \nFSS} \oplus \Gamma_{i,m > \nFSS}.
\end{align*}
For threshold detectors, $m > \nFSS$ corresponds to the set of photon-numbers $m$ greater than the cutoff $\nFSS$.
The target measurements are given by 
\begin{align}
    \label{eq:oldFSSTargetMeas}F_i = \Gamma_{i,m\leq \nFSS} \oplus \ketbra{i},
\end{align}
where $\{\ket{i}\}_{i=1}^\nmeas$ forms an orthonormal set of vectors termed `flags'. Theorem 1 from Ref.~\cite{zhang_security_2021} guarantees the existence of a squashing map $\Lambda$ for this case.

However, due to the existence of the flags, there exist classical states that live entirely in this subspace that cannot be excluded by acceptance testing on this finite-dimensional space. In other words, by itself, the flag-state squasher would result in a key length of 0. Thus, the usage of the flag-state squasher requires an additional constraint bounding the weight $W$ in the flag space to be useful.

The canonical method of bounding the weight outside the preserved subspace can be found in Ref.~\cite[Section 3.4.2]{li_application_2020} and proceeds as follows. For any event $e$, and all input states it can be shown that
\begin{align} \label{eq:wtOutsideSubspaceGeneric}
    W\leq1-p(m\leq\nFSS) \leq \frac{p(e)- \lambdamin{\Pi_{\nFSS}\Gamma_e\Pi_{\nFSS}}}{\lambdamin{\overline{\Pi}_{\nFSS}\Gamma_e\overline{\Pi}_{\nFSS}}-\lambdamin{\Pi_{\nFSS}\Gamma_e\Pi_{\nFSS}}},
\end{align}
where $\Pi_{\nFSS}$ $(\overline{\Pi}_{\nFSS})$ is the projection on (outside) the space corresponding to $\Gamma_{i,m\leq \nFSS}$.
When working with the infinite-dimensional POVM, some protocol-dependent choices \cite{li_improving_2020,nahar_imperfect_2023} for the event lead to good bounds on the weight $W$. This bound can then be added in as an additional constraint to the finite-dimensional key rate optimisation.

Although this method works well when assuming that Eve performs a IID-collective attack, the use of the postselection technique on the finite-dimensional protocol with the addition of a constraint is not straightforward. Note that attempting to use \cref{eq:wtOutsideSubspaceGeneric} directly with the finite-dimensional target POVM gives only trivial bounds as every event $e$ has $\lambdamin{\overline{\Pi}_{\nFSS}F_e\overline{\Pi}_{\nFSS}}=0$.
This motivates the construction of a modified version of the flag-state squasher - the ``weight-preserving flag-state squasher'' (WPFSS).

\begin{restatable}[Weight-preserving flag-state squasher]{lemma}{lemmaFSS} \label{lemma:FSS}
    Let $\{\Gamma_i\}_{i=1}^\nmeas$ be a POVM where each element is block-diagonal, i.e. $\Gamma_i = \Gamma_{i,m\leq \nFSS} \oplus \Gamma_{i,m > \nFSS}.$ Further, let $\{\ket{i}\}_{i=1}^\nmeas$ be an orthonormal set of vectors for the flag space, and let $0\leq \FSScost \leq \lambdamin{\overline{\Pi}_{\nFSS}\Gamma_1\overline{\Pi}_{\nFSS}},$ where $\overline{\Pi}_\nFSS$ is the projection outside the space corresponding to $\Gamma_{i,m \leq \nFSS}$.
    Then, for the following choice of target measurements
    \begin{align}
     \label{eq:FSSccPOVM}   F_i &= \Gamma_{i,m\leq \nFSS} \oplus (1-\FSScost)\ketbra{i} \quad \forall\  1 < i \leq \nmeas\\
      \label{eq:FSSrestPOVMs}  F_1 &= \Gamma_{1,m\leq \nFSS} \oplus \left( \ketbra{1}+\FSScost\sum_{j=2}^\nmeas \ketbra{j} \right),
    \end{align}
    there exists a channel $\Lambda$ such that $\Lambda^\dag[F_i] = \Gamma_i$ for all $i$.
\end{restatable}
The proof is given in \cref{app:decoyProofs}.

The WPFSS is constructed so that the bounds obtained on the infinite-dimensional states from \cref{eq:wtOutsideSubspaceGeneric} can be directly added to the target POVM elements instead of adding it as an additional constraint. This is because the construction, which is independent of any observed quantity, ensures that $\lambdamin{\overline{\Pi}_{\nFSS}F_1\overline{\Pi}_{\nFSS}}=\FSScost$ is non-zero in contrast to the standard flag-state squasher.
Thus, with this modification, the flag-state squasher can be used with \cref{lemma:squash} to squash Bob's system to finite dimensions.

Note that there is some choice in constructing the WPFSS.
\begin{enumerate}
    \item The POVM element used to estimate the weight $W$ in the flag-space (which we have denoted by $\Gamma_1$ in \cref{lemma:FSS}) can be freely chosen.
    \item $\FSScost$ can be chosen to be any lower bound on the minimum eigenvalue of the component of the POVM element in the flag-space $\lambdamin{\overline{\Pi}_{\nFSS}F_1\overline{\Pi}_{\nFSS}}$.
\end{enumerate}
Thus, it is important to optimise the key length over these choices. The POVM element $\Gamma_1$ is typically \cite{zhang_security_2021,li_improving_2020,nahar_imperfect_2023} chosen based on the specificities of the protocol.
We would generically expect the key rates from the WPFSS to improve as we get closer to a tight bound where $\FSScost$ approaches $\lambdamin{\overline{\Pi}_{\nFSS}F_1\overline{\Pi}_{\nFSS}}$. This can be seen by using \cref{eq:wtOutsideSubspaceGeneric} to estimate the weight in the flag-space when using the WPFSS POVM element $F_1$ - the bound is a monotonically decreasing function of $\lambdamin{\overline{\Pi}_{\nFSS}F_1\overline{\Pi}_{\nFSS}} = \FSScost$.

We also show below that the WPFSS cannot give worse key rates against IID-collective attacks than the standard flag-state squasher. As seen in \cref{eq:FSSccPOVM,eq:FSSrestPOVMs}, the target POVM elements $\{F_i\}_{i=1}^\nmeas$ for the WPFSS are block-diagonal. Thus, \cite[Theorem 1]{zhang_security_2021} can be directly used on this setup to construct a squashing map with target POVM elements described in \cref{eq:oldFSSTargetMeas}. Additionally, \cref{eq:wtOutsideSubspaceGeneric} can also be used to bound the weight inside the flag subspace. Thus, the above observation allows us to use the postselection technique to prove the security against coherent attacks for prior works \cite{zhang_security_2021,li_improving_2020, nahar_imperfect_2023} that use the standard flag-state squasher \footnote{Note that these works find key rates in the asymptotic regime and thus the technical flaw from Ref.~\cite{christandl_postselection_2009} does not pose a problem.}.

Next, we discuss the reduction of $A_S$ to finite-dimensions.

\subsection{Decoy-state lift} \label{subsec:decoyStateLift}

Decoy-state protocols \cite{hwang_quantum_2003,lo_decoy_2005,wang_beating_2005} are often implemented with phase-randomised laser pulses $\rho^\mu$ with varying intensity $\mu$,
\begin{align}
    \rho^\mu = \sum_{m=0}^\infty p(m\vert \mu) \ketbra{m},
\end{align}
where $p(m\vert\mu)$ is a Poisson distribution with mean $\mu$.
Here, the signal information $i$ is typically encoded in an orthogonal mode (for e.g.- polarisation, time-bin, etc.). For simplicity, we assume that this encoding is isometrically implemented resulting in states $\rho_i^\mu = V_i \rho^\mu V_i^\dag$, where $V_i$ is the encoding isometry. 

As this state is full rank and lives in an infinite-dimensional Hilbert space, it would require an infinite-dimensional purifying system $A_S$.
Thus, we aim to use source maps \cite{nahar_imperfect_2023} to construct a convenient virtual protocol with finite-dimensional signal states which can be used with the postselection technique. Although this is a widely used technique, we present a proof here for completeness.
\begin{restatable}[Source maps]{lemma}{lemmaSourceMaps} \label{lemmaSourceMapSecurity}
    Let $\{\rho_{i}^\mu\}\subset\stateset{A''}$ be the set of states prepared by Alice in a PMQKD protocol. Suppose that there exists a \emph{source map} $\Psi\in\channelset{A'',A'}$ relating the real states $\{\rho_{i}^\mu\}$ to a set of virtual states $\{\xi_{i}^\mu\}\subset\stateset{A'}$ such that $\rho_i^\mu = \Psi[\xi_i^\mu]$ for all $i,\mu$. Then $\epsSec$-secrecy for the virtual protocol with $\{\xi_{i}^\mu\}$ implies $\epsSec$-secrecy for the real protocol with $\{\rho_{i}^\mu\}$ instead.
\end{restatable}
The proof is given in \cref{app:decoyProofs}.

We can construct a virtual source, which emits tagged states \cite{gottesman_security_2004} where the space with photon number greater than $\ndecoy$ is tagged as follows,
\begin{align} \label{eq:taggedStates}
    \xi_i^\mu = \sum_{m=0}^\ndecoy p(m\vert \mu) V_i \ketbra{m}V_i^\dag + \left(1-\sum_{m=0}^\ndecoy p(m\vert \mu)\right) \ketbra{i,\mu},
\end{align}
where $\{\ket{i,\mu}\}_{i,\mu}$ form an orthonormal basis for a space orthogonal to the span of $\{\ket{m}\}_{m=0}^\ndecoy$.
If we define $\Psi$ as a map that measures $\ketbra{i,\mu}$ and prepares $\sum_{m=\ndecoy+1}^\infty p(m\vert \mu) V_i \ketbra{m}V_i^\dag$, $\Psi$ is a source map relating the virtual states $\{\xi_i^\mu\}$ to the real states
\begin{align}
    \rho_i^\mu = \sum_{m=0}^\infty p(m\vert \mu) V_i \ketbra{m}V_i^\dag.
\end{align}

Thus, as shown in Lemma \ref{lemmaSourceMapSecurity} it is sufficient to show that the virtual protocol where Alice prepares the states $\{\xi_i^\mu\}$ is an $\epsSec$-secret PMQKD protocol. For such states, the dimension of the shield system $d_{A_S}$ can be obtained from the rank of the tagged states $d_{A_S} = \ndecoy+2$.  Further, as shown in Appendix \ref{app:DecoyShield}, for encoding isometries $V_i$ that preserve the number of photons, the shield system is block-diagonal with all blocks of dimension 1.
Any IID security proof for this can be lifted to a security proof for general attacks as described in \cref{cor:liftToCoherent}.

\subsection{IID decoy-state analysis for tagged sources} \label{subsec:IIDDecoy}

If using the postselection technique, the decoy-state analysis must be performed with the shield system corresponding to tagged states, as constructed in \cref{app:DecoyShield}. As shown in \cref{app:decoyFixedMarginalToGenDecoy}, this is equivalent to the analysis where the source prepares and sends tagged states, which are measured after Eve's interaction.
Thus, we can perform the IID decoy analysis as follows.

As is typically the case, a single intensity $\mu_S$ is chosen for key generation.
Due to the block diagonal structure of the signal states given in Eq. (\ref{eq:taggedStates}), the key rate can be broken up \cite{li_improving_2020} as
\begin{align}
    R = \sum_{m=0}^\ndecoy p(m\vert \mu_S) R_m + (1-\sum_{m=0}^\ndecoy p(m\vert \mu_S)) R_{\text{tag}} \leq \sum_{m=0}^\ndecoy p(m\vert \mu_S) R_m
\end{align}
where each $R_m$ is the key rate of the blocks with $m$ photons, and $R_{\text{tag}}$ is the key rate of the tagged block.

To calculate $R_m$, we compute upper and lower bounds on the $m$-photon statistics using \cite{nahar_imperfect_2023}
\begin{equation} \label{eq:decoyStateSDP}
    \begin{aligned}
        \underset{\Phi}{\text{opt.}}\; &\Tr\left[\Phi\left[V_i\ketbra{m}V_i^\dag\right]\Gamma_j\right]\\
        \text{s.t.}\; &\Tr\left[\Phi\left[\xi_k^\mu\right]\Gamma_l\right] = \gamma_{l\vert k,\mu} \quad \forall l,k,\mu\\
        & \Phi \in \channelset{A',B},
    \end{aligned}
\end{equation}
where $\{\Gamma_i\}$ is Bob's squashed POVM. Here, opt. implies that we perform separate optimisations to find the minimum and maximum value.

While this is the generalised decoy-state optimisation used in Ref.~\cite{nahar_imperfect_2023}, this can be reduced to the standard decoy-state analysis as follows.
We can simplify the constraints as
\begin{align}
    \nonumber \gamma_{l\vert k,\mu} &= \Tr\left[ \Phi\left[\xi_k^\mu\right]\Gamma_l\right]\\
    &= \sum_{m'=0}^\ndecoy p(m'\vert \mu) \Tr\left[ \Phi\left[V_k\ketbra{m'}V_k^\dag\right]\Gamma_l\right]+ (1-\sum_{m'=0}^\ndecoy p(m'\vert \mu)) \Tr\left[ \Phi\left[\ketbra{k,\mu}\right]\Gamma_l\right].
\end{align}
Next, defining $p(\text{det}_l\vert k,m') = \Tr\left[ \Phi\left[V_k\ketbra{m'}V_k^\dag\right]\Gamma_l\right]$, and $p(\text{det}_l\vert k,\mu,\text{tag}) = \Tr\left[ \Phi\left[\ket{k,\mu}\bra{k,\mu}\right]\Gamma_l\right]$ allows us to recast the generalised decoy-state SDPs described in Eq. (\ref{eq:decoyStateSDP}) into linear programs as
\begin{equation}
    \begin{aligned}
        \text{opt. } &p(\text{det}_j\vert i,m)\\
        \text{s.t. } &\sum_{m'=0}^\ndecoy p(m'\vert \mu) p(\text{det}_l\vert k,m')+ (1-\sum_{m'=0}^\ndecoy p(m'\vert \mu)) p(\text{det}_l\vert k,\mu,\text{tag}) = \gamma_{l\vert k,\mu} \quad \forall \; l,k,\mu\\
        & 0\leq p(\text{det}_l\vert k,m') \leq 1\\
        & 0 \leq p(\text{det}_l\vert k,\mu,\text{tag}) \leq 1.
    \end{aligned}
\end{equation}
Finally, this can be recast into the standard form \cite{lo_decoy_2005,wang_beating_2005} by using 0 and 1 as the lower and upper limits of $p(\text{det}_l\vert k,\mu,\text{tag})$ in the linear program constraints.
Note that although a higher photon-number cut-off would lead to better IID key rates due to better decoy-state analysis, it increases the dimension of the shield system leading to worse finite-size performance after the use of the postselection technique.

\section{Application to the Three State Protocol} \label{sec:applicationtothreestate}

So far we have made rigorous the framework to apply the postselection technique to optical prepare-and-measure protocols (including decoy-state protocols). In this section, we first briefly summarise the various results in this manuscript, paying careful attention to the statements needed to easily apply our results. We then illustrate the results by applying them to the time-bin encoded three-state protocol.
This section is formulated so that it can be read without understanding the details of the preceding sections.

\subsection{Recipe for application}

\subsubsection{Generic application}

Recall that the postselection technique reduces the problem of proving security against coherent attacks, to proving security against IID-collective attacks, while paying some cost for the reduction. This is applied via \cref{cor:liftToCoherentvarx,cor:liftToCoherentvarxBlockDiagonal,cor:liftToCoherentvarxSymmetries} to a variable-length protocol \cite{tupkary_security_2023}, and \cref{cor:liftToCoherent,cor:liftToCoherentBlockDiagonal,cor:liftToCoherentSymmetries} to a fixed-length protocol.
Note that the different variable (fixed) length corollaries are essentially the same, differing only in improved costs in the presence of extra structure in the protocol.

Informally, \cref{cor:liftToCoherentvarx,cor:liftToCoherentvarxBlockDiagonal,cor:liftToCoherentvarxSymmetries} state that given a variable-length IID security proof with key lengths $l_i$ and secrecy parameter $\epsSec$, the postselection technique can be used to provide a security proof against coherent attacks with key lengths $l_i-2\log(\cost)-2\log(1/\epstilde)$ and secrecy parameter $\cost (\sqrt{8\epsSec}+\frac{\epstilde}{2})$. Here, $\epstilde$ is a parameter that can be chosen freely and $\cost$ can be thought of as the cost of using the postselection technique that depends on the dimensions of the systems. In the presence of additional protocol structure, this cost can be reduced as detailed in the corollary statements. The fixed-length corollaries (\cref{cor:liftToCoherent,cor:liftToCoherentBlockDiagonal,cor:liftToCoherentSymmetries}) are similar, differing only in the precise numbers.

In practice, this can be used as follows:
\begin{enumerate}
    \item Choose a target secrecy parameter $\epsSec$ based on the application in mind.
    \item \label{item:PScost} Determine the protocol-dependent upper bound on the cost of using the postselection technique as $\cost = \binom{n+x-1}{x-1} \leq \left(\frac{e (n+x-1)}{x-1}\right)^{x-1}$. Here, $x$ depends on the dimensions of the systems, and the structure in the protocol.
    For a generic protocol (in the absence of structure), $x = d_A^2d_B^2$ as described in \cref{cor:liftToCoherent,cor:liftToCoherentvarx} for fixed and variable length protocols respectively. In the presence of block-diagonal structure this can be improved as stated in \cref{cor:liftToCoherentBlockDiagonal,cor:liftToCoherentvarxBlockDiagonal} for fixed and variable length protocols respectively. In the presence of general protocol symmetries this can be improved as stated in \cref{cor:liftToCoherentSymmetries,cor:liftToCoherentvarxSymmetries} for fixed and variable length protocols respectively.
    
    For optical protocols, the subsystems are infinite-dimensional, and so this cost diverges. We comment more on this in \cref{subsec:recipeForOpticalProtocols}.
    \item Pick value of free parameter $\epstilde$ (in principle this value can be optimised over).
    \item Compute key lengths $l_i$ through an IID security proof, with secrecy parameter $\frac{(\epsSec-\cost\epstilde/2)^2}{8\cost^2}$. Note that the details of the IID security proof are out of the scope of this work, although there are multiple existing proofs for fixed \cite{george_numerical_2021} and variable-length protocols \cite{tupkary_security_2023}.
    \item Use key lengths $l_i-2\log(\cost)-2\log(1/\epstilde)$ as the final hash length for the protocol. This protocol is secure against coherent attacks.
\end{enumerate}

\subsubsection{Application to optical protocols} \label{subsec:recipeForOpticalProtocols}

Since optical systems are $\infty$-dimensional, \cref{item:PScost} of the recipe cannot be directly computed. For detection setups with threshold detectors, we can use the weight-preserving flag-state squasher (WPFSS, see \cref{lemma:FSS}) to reduce the dimensions. Note that the existing flag-state squasher \cite{zhang_security_2021} cannot be used. The WPFSS can be thought of as a rigorous way to apply a photon-number cutoff $\nFSS$ to the detection setup. Additionally, the POVM (see \cref{subsec:WPFSS} for details on the POVM construction) has block-diagonal structure. Thus, for a two-mode optical setup \footnote{More generally, the formula is given by $x = d_A^2\left(\sum_{i=0}^\nFSS d_{=i}^2 + \nmeas\right)$, where $d_{=i}$ is the dimension of the subspace with exactly $i$ photons in all modes.} the dimensional-dependent term $x$ in \cref{item:PScost} is given by (\cref{cor:liftToCoherentBlockDiagonal,cor:liftToCoherentvarxBlockDiagonal})
\begin{align} \label{eq:WPFSSOpticalDimension}
    x = d_A^2\left(\sum_{i=0}^\nFSS (i+1)^2 + \nmeas\right),  
\end{align}
where $\nmeas$ is the number of Bob's POVM elements \footnote{Note that technically $\nmeas$ is the number of flags in the WPFSS. In some cases this can be less than the number of Bob's POVM elements.}.

A similar problem arises when applying the postselection technique to decoy-state protocols, where Alice additionally has a shield system $A_S$ as defined in \cref{lemma:Shield}. We have shown that the full optical states can be replaced with finite-dimensional tagged states given in \cref{eq:taggedStates}. Similar to the WPFSS, this is a formal way to apply a photon-number cutoff $\ndecoy$ to Alice's photonic signal states. For most encodings such as polarization, time-bin, etc. (see \cref{subsec:decoyStateLift,app:DecoyShield} for formal details), the dimensional-dependent term $x$ in \cref{item:PScost} is given by
\begin{align} \label{eq:DecoyOpticalDimension}
    x = \nint^2(\ndecoy+2)d_A^2d_B^2,
\end{align}
where $\nint$ is the number of intensities used in the protocol, and $d_A$ is the dimension of Alice's subsytem (without the shield system). The corresponding IID decoy-state analysis for these tagged states is discussed in \cref{subsec:IIDDecoy}.

The above results for decoy-state protocols can be straightforwardly combined with the WPFSS to give
\begin{align} \label{eq:generalOpticalDimension}
    x = \nint^2(\ndecoy+2)d_A^2\left(\sum_{i=0}^\nFSS (i+1)^2 + \nmeas\right).
\end{align}

We now apply this recipe to the three-state protocol, focusing primarily on \cref{item:PScost}.

\subsection{Optical setup} \label{sec:ThreeStateOpticalSetup}

\begin{figure*}[t]
    \centering
    \includegraphics[width = \linewidth]{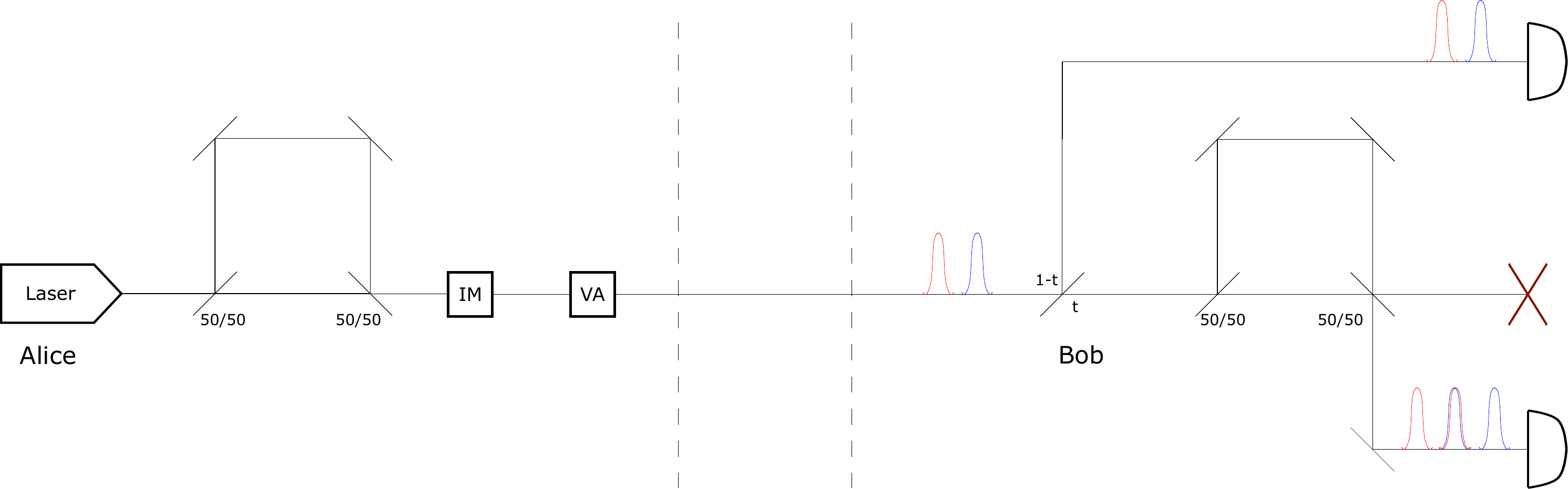}
    \caption{Schematics of the implementation of the three-state protocol as in \cite{boaron_secure_2018}. The numbers below the beam-splitters reflect their transmissivity. IM and VA refer to intensity modulator and variable attenuator respectively.} \label{fig:3StateExpSetup}
\end{figure*}

We consider a protocol where Alice uses a spontaneous parametric down-conversion (SPDC) or quantum dot source to prepare a single-photon in two time-bin modes as follows:
\begin{itemize}
    \item[] $0$: $\ket{0}\otimes\ket{1}$
    \item[] $1$: $\ket{1}\otimes\ket{0}$
    \item[] $+$: $\frac{1}{\sqrt{2}}\left(\ket{0}\otimes\ket{1}+\ket{1}\otimes\ket{0}\right)$.
\end{itemize}
Since the signal states span a two-dimensional vector space, Alice's dimension $d_A$ is 2.

Bob's measurement setup is identical to that considered in \cite{nahar_imperfect_2023, boaron_secure_2018} as depicted in \cref{fig:3StateExpSetup}. Both detectors here are threshold detectors, with 2 and 3 time-bins respectively. So there are $2^5$ possible click patterns. We define the cross-click event cc to be any click pattern that records a click in both detectors, while ignoring all clicks in the middle time slot of the Mach-Zehnder interferometer.
As shown in Appendix C of Ref.~\cite{nahar_imperfect_2023}, we can then find
\begin{align}
    \lambdamin{\overline{\Pi}_{\nFSS} \Gamma_{\text{cc}} \overline{\Pi}_{\nFSS}} \geq 1-t^{N+1}-\left(1-\frac{t}{4}\right)^{N+1}+\left(\frac{3t}{4}\right)^{N+1},
\end{align}
where $t$ is Bob's basis-choice beam-splitting ratio.
This now lets us use the WPFSS described in \cref{lemma:FSS}.

We have some choice of coarse-graining the events when using the flag-state squasher. However, having fewer events leads to a smaller flag space which in turn decreases the dimension (see \cref{eq:WPFSSOpticalDimension}) of the problem improving the usage of the postselection technique.
Thus, we coarse-grain the events to only consider the event in which no detector clicks, the events in which a single detector in a single time-bin clicks, the cross-click event, and all other events. This coarse-graining results in 8 flag states.
For the simulations here, we choose the photon-number cutoff to be $1$. Thus, Bob's system consists of 1 block of dimension 2 (qubit space), and 9 blocks of dimension 1 (vacuum and flags). Thus, \cref{eq:WPFSSOpticalDimension} gives us the dimensional-dependent term $x = 4^2+9\ 2^2 = 52$ to be used in \cref{item:PScost} of the recipe.

We let $\sigma^{\otimes n}_{ABE}$ be the state on which the protocol is run.

\subsection{Classical part}

After the signal transmission and detection, Alice and Bob store their measurement data in their local registers $X^n$ and $Y^n$.
They then choose a random subset of $m$ signals, and announce their measurement outcomes for those rounds in the register $\Cat^m$, to obtain $\Fobs$, the observed frequency of different outcomes.
This is then used for the ``variable-length decision", by computing the length of the key to be produced ($l(\Fobs)$) and the number of bits to be used for error-correction information ($\leak{}(\Fobs)$), as specified below in \cref{subsec:applicationcomputation}. The state of the protocol at this stage is given by $\sigma_{X^n Y^n E^n \Cat^m }$.

For the remaining $\nkey=n-m$ signals, Alice and Bob implement round-by-round announcements $C^{\nkey}$. For this protocol, Bob announces all events without any detections and whether the first (Z-basis) or second (X-basis) detector clicked when he observes a single-click event.
Alice announces the bits in which she encoded the photon in the Z-basis.
All events where Alice prepared the state in the Z-basis, and Bob's first detector clicked are kept and the rest are discarded. Thus, Alice maps her local data $X^{\nkey}$ to the raw key $Z^{\nkey}$, where discarding is modelled as setting $Z=0$. Note that this procedure is equivalent to  \textit{physically} discarding rounds \cite[Lemma 4]{tupkary_security_2023}. 

Alice and Bob then implement error-correction using $\leak{}(\Fobs)$ bits of data, and error-verification by comparing hash values of length $\log(1/\epsEV)$, using the register $C_E$.
The final step is privacy amplification, where Alice and Bob apply a common two-universal hash function, announced in the register $C_P$, to produce a key of $l(\Fobs)$ bits. 
The state of the protocol at this stage is given by $\sigma_{K_A K_B  \Cfull E^n }$, where we write $\Cfull=C^{\nkey} \Cat^m C_E C_P$.

\subsection{IID Key Length Computation} \label{subsec:applicationcomputation}

Recall that in a QKD protocol, when considering IID-collective attacks, one has a fixed but unknown state $\sigma^{\otimes n}_{AB}$, which then gives rise to the random variable $\Fobs$. For a given $\Fobs$, the variable-length protocol must determine $l(\Fobs)$ (length of key to be produced) and $\leak{}(\Fobs)$ (number of bits to be used for error-correction). In this section, we briefly explain how this computation is implemented, using the variable-length framework of Ref.~\cite{tupkary_security_2023}. Note that multiple values of $\Fobs$ will lead to the same value of the final key length $l(\Fobs)$, and therefore constitute the same event $\Omega_i$.

In order to compute the values of $l(\Fobs)$ and $\leak{}(\Fobs)$, the idea is to first construct $\bstat(\Fobs)$, a high-probability lower bound on the R\'enyi entropy of the underlying state in the QKD protocol. In order to do so, 
one first needs to construct a set $V(\Fobs)$ such that it contains the underlying state $\rho_{AB}$ with high probability. That is, 
\begin{equation} \label{eq:VFobs}
    \Pr_{\Fobs} ( \sigma_{AB} \in V(\Fobs) ) \geq 1- \epsAT .
\end{equation}

Given such a $V(\Fobs)$, Section V from \cite{tupkary_security_2023} then specifies $\bstat(\Fobs)$, $l(\Fobs)$ and $\leak{}(\Fobs)$ as follows
\begin{equation} \label{eq:varlengthdecision}
\begin{aligned}
\bstat(\Fobs) &\coloneqq \min_{\substack{\sigma \in V(\Fobs) \\ \Tr_{B}(\sigma_{AB}) = \promise }} \nkey H(Z|CE)_{\sigma} - \nkey (\alpha-1) \log^2(d_Z+1) \\
\leak{}(\Fobs) &\coloneqq \nkey f H(Z|YC)_{\Fobs} \\
l(\Fobs) &\coloneqq \max( \bstat(\Fobs) - \leak{}(\Fobs) - \theta(\epsPA,\epsEV) ,0) \\
\theta(\epsPA,\epsEV) &\coloneqq \frac{\alpha}{\alpha-1} \left ( \log(\frac{1}{4\epsPA}) +\frac{2}{\alpha} \right) + \lceil \log \left( 1/\epsEV \right) \rceil,
\end{aligned}
\end{equation}
where $H$ denotes the conditional von-Neumann entropy, $d_Z$ is the dimension of the $Z$ register, and $1<\alpha<1/\log(d_Z+1)$ is the R\'{e}nyi parameter. The resulting variable-length protocol is shown to be an $\epsEV$-correct \cite[Theorem 2]{tupkary_security_2023} and $(\epsPA+\epsAT)$-secret PMQKD protocol. 
In this work, we choose the optimal $\alpha = 1+\kappa/\sqrt{\nkey}$, where $\kappa = \sqrt{\log(1/\epsPA)} / \log(d_Z+1)$. Moreover, we compute $\leak{}(\Fobs)$ from the observed distribution $\Fobs$, where $f=1.16$ is the efficiency factor. 

One can then use a variety of concentration inequalities to construct $V(\Fobs)$ satisfying \cref{eq:VFobs}  \cite[Lemma 1]{tupkary_security_2023}. In this work, we construct $V(\Fobs)$ in the following lemma using Hoeffding's inequality. We refer the reader to \cref{app:ConstructingNiceSet} for the proof.

\begin{restatable}[Constructing $V(\Fobs)$]{lemma}{lemmaVFobs} \label{lemma:VFobs}

	For any state $\rho$, let $\Fobs \in \mathcal{P}(\Sigma)$ be the frequency vector obtained from measuring the state $m$ times, where $\Sigma$ is the set of possible  outcomes. Let $\Gamma_j$ be the POVM element corresponding to outcome $j$. Define 
			\begin{equation} \label{eq:confinterval}
				\begin{aligned}
					\mu \coloneqq \sqrt{\frac{ \log(2 |\Sigma| / \epsAT ) } {2 m} },
			\end{aligned}
			\end{equation}
			 and the set
			 \begin{equation} \label{eq:Vset}
					V(\Fobs) \coloneqq \{  \sigma \in  S_\circ(AB) \; | \;   | \Tr(\Gamma_j \sigma) - \Fobs_j | \leq \mu, \forall j \in \Sigma \}.
			\end{equation}
			Then, $V(\Fobs)$ contains $\rho$ with probability greater than $1-\epsAT$. That is,
			\begin{equation} 
				\Pr_{\Fobs} \left( \rho \in V(\Fobs) \right) \geq 1-\epsAT.
			\end{equation}
			
\end{restatable}

\subsection{Parameter choices} \label{sec:plots}

For simplicity in our key rate plots, we assume a loss-only channel through an ultralow-loss fiber with an attenuation of 0.16 dB/km. We also assume that Bob's detection setup consists of ideal threshold detectors. Note that this is a simplifying assumption made for the simulation. This can be easily adapted to more realistic scenarios by using the corresponding realistic POVM in our numerics \cite{zhang_security_2021}. We fix Bob's basis-choice beam-splitting ratio to be $t=0.2$, and Alice's probability of preparing a state in the Z-basis to be 0.8. Further $m$ (number of signals used for testing) is taken to be a fixed fraction $m=0.05n$ of the total number of signals. 
In our simulation, we assume that the protocol runs for 3600s using a 3 GHz source. Thus, for the postselection technique plots, the number of signals sent during a run of the protocol is $n=1.08\cross10^{13}$.
We strive to make realistic parameter choices without optimising. A more complete analysis optimising over all parameter choices is accommodated by our proof techniques, but is out of the scope of this work. Additionally, we observe similar qualitative results for other parameter choices, and thus we believe that our conclusions are not dependent on the specific parameters chosen here \footnote{We have made the code for our plots public at \href{https://openqkdsecurity.wordpress.com/repositories-for-publications/}{https://openqkdsecurity.wordpress.com/repositories-for-publications/} for the interested reader to try different parameter choices.}.

We plot the secret key rate per second when the observed frequency of events in the acceptance test $(\Fobs)$ is equal to the frequency expected from honest behavior.
We choose the target security parameter to be $\epsSec^\text{target} = \epsCor^\text{target} = 10^{-12}$. Thus, the resulting protocol is $(\epsSec^\text{target}+\epsCor^\text{target})$-secure. We use $\cost = \binom{n+x-1}{x-1} \leq \left(\frac{e (n+x-1)}{x-1}\right)^{x-1}$ to bound the dimension of the symmetric subspace as described in \cref{item:PScost} of the recipe. This gives better bounds than the often used $\cost \leq (n+1)^{x-1}$.
Furthermore, when seeking to compute key rates for a target secrecy parameter $\epsSec^\text{target}$, for IID key rates we assume that $\epsPA = \epsAT = \epsSec^\text{target}/2$. When using the postselection technique (\cref{cor:liftToCoherentvarxBlockDiagonal}), we assume that $\sqrt{8 \epsSec} = \widetilde{\varepsilon}/2 = \epsSec^\text{target} / 2 \cost$.

Using these parameters, we plot various key lengths against distance in \cref{fig:plots} as follows:
\begin{figure*}[t]
    \centering
    \includegraphics[width = \linewidth]{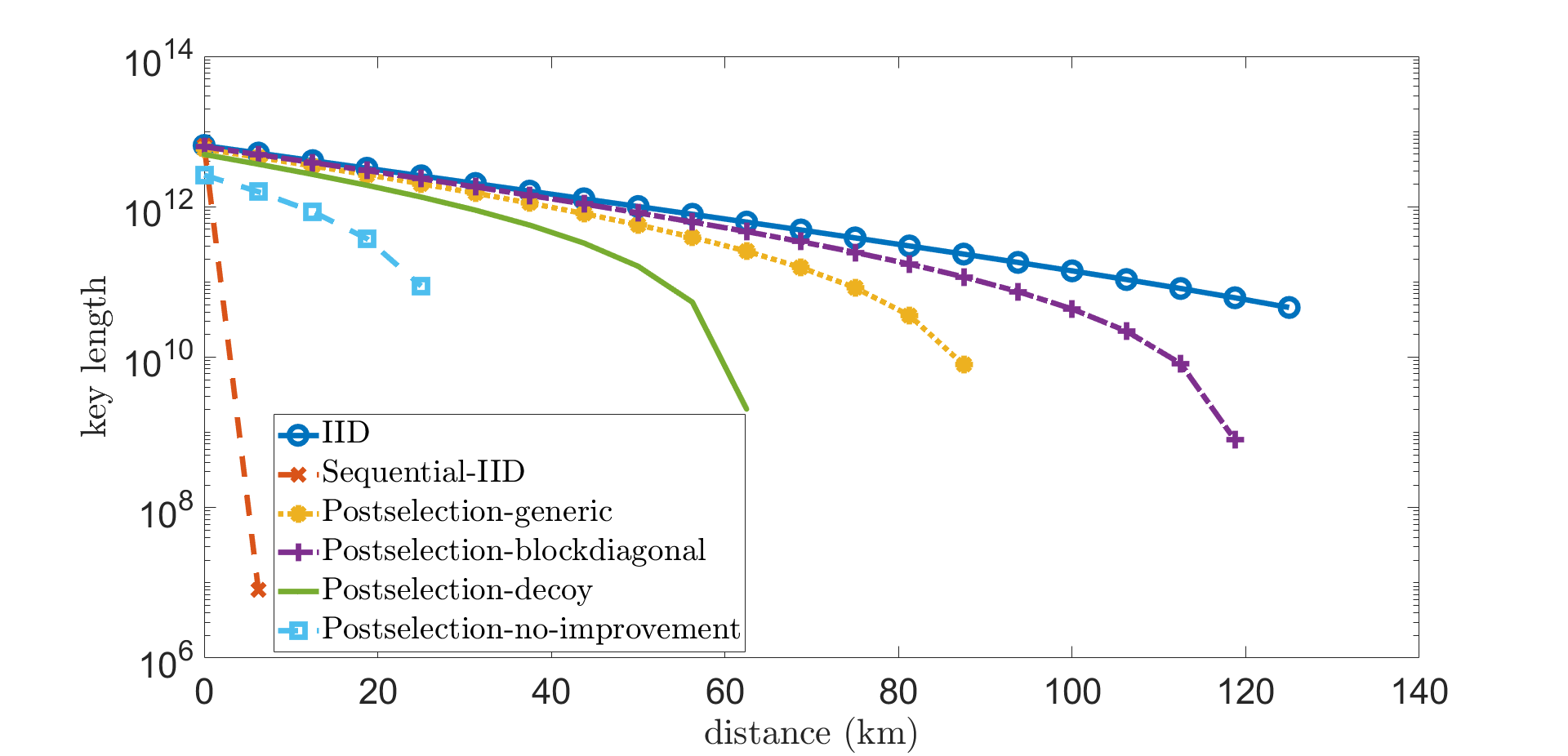}
    \caption{Plots demonstrating the performance of the postselection technique for the three-state protocol.} \label{fig:plots}
\end{figure*}
\begin{enumerate}
    \item \underline{IID}: We plot the key lengths as a function of distance under the assumption that Eve's attack is always IID. This serves as an upper bound for the key length plots.
    
    \item \underline{Sequential IID}: We plot the key length as a function of distance where the repetition rate of the protocol is limited in order to enforce a sequential condition. Here, Alice sends states at long enough intervals that ensure that Eve's attack on a pulse cannot be influenced by pulses that come after.
    As shown in Eq.~(41) of Ref.~\cite{sandfuchs_security_2023}, the sequential condition limits the repetition rate $R_s$ to $\frac{1.5 \cross 10^5}{d}$, where $d$ is the distance between Alice and Bob in $km$, and we have assumed the refractive of the fiber to be 1.5.
    
    Thus, given the protocol duration $T = 3600$s, the maximum number of signals sent with this condition is $n = R_s T = \frac{5.4*10^8}{d}$. We then perform a security analysis assuming that Eve's attack is IID with this value of $n$.
    Note that this value of $n$ decreases rapidly as the distance $d$ becomes non-zero, resulting in low key lengths. This serves as an upper bound on the GEAT key lengths under the sequential attack condition.
    
    \item \label{blockDiagPS} \underline{Postselection with block-diagonal improvement}: We plot the key lengths obtained from applying \cref{cor:liftToCoherentvarxBlockDiagonal}.
    Here, $d_A = 2$, and Bob consists of a block of dimension 2, and nine blocks of dimension 1 resulting in $x = 52$ as detailed in \cref{sec:ThreeStateOpticalSetup}.
    This gives the full secure key length through the use of the postselection technique.

    \item \label{noImpPS} \underline{Postselection with no improvement}:
     We plot the key lengths obtained from applying \cref{cor:liftToCoherentvarx}, except with x = $d_A^2d_B^4$ to demonstrate the impact of our improvements in \cref{cor:generalMixedDeFinetti} on the key rates. Here, $d_A = 2$, and $d_B = 11$ resulting in $x=58564$ as detailed in \cref{sec:ThreeStateOpticalSetup}.
    
    \item \label{genImpPS} \underline{Postselection with generic improvement}: We plot the key lengths obtained from applying \cref{cor:liftToCoherentvarx}. Here, $d_A = 2$, and $d_B = 11$ resulting in $x=484$ as detailed in \cref{sec:ThreeStateOpticalSetup}.
     When compared to \cref{blockDiagPS}, this demonstrates the improvements we have made to the postselection technique by taking into account the block-diagonal structure. Note that even without the block-diagonal structure, this plot already consists of a significant improvement over past work \cite{fawzi_quantum_2015} arising from \cref{cor:generalMixedDeFinetti} as can be seen by comparing it to \cref{noImpPS}.

    \item \underline{Postselection with decoy-state lift}: A more practical alternative to single-photon sources is decoy-state QKD. However, the value of $\cost$ is higher for decoy-state protocols. Thus, we compute key lengths with a larger dimension ($x$) as computed in \cref{eq:generalOpticalDimension}. Here, we use $d_A = 3$ as Alice sends 3 linearly independent states.
    For the simulation, we assume a cut-off $\ndecoy$ of 2 and that $\nint = 3$ intensities were used.
    Bob's dimensions are the same as in \cref{blockDiagPS}: 1 block of dimension 2 and 9 blocks of dimension 1.
    Thus, the dimensional-dependent term is
    \begin{align}
        \nonumber  x & = \nint^2(\ndecoy + 2) d_A^2 (2^2 + 9\times 1^2)\\
        &= 36\times 9 (4+ 9) = 4212.
    \end{align}
    Note that our IID key rate analysis assumes that Alice sends single photons, but we use a higher dimension ($x$) for the postselection lift. We do this, since we expect the  key length obtained from IID decoy analysis along with the postselection lift to be similar.
    
\end{enumerate}

Note that due to Bob's unbalanced basis-choice and the large basis-dependent detection efficiencies, phase-error rate \cite{koashi_simple_2009} and EUR \cite{tomamichel_largely_2017} \emph{cannot} be used (Note that although Ref. \cite{tupkary2024phase} can accommodate small basis efficiency mismatch for active detection setups, their analysis does not apply to the three-state protocol considered here as we use a passive detection setup with large efficiency mismatch induced by the missing detector.).
As shown in \cref{fig:plots}, imposing the sequential condition results in the protocol not producing any key for over 20 km, and thus the GEAT \cite{metger_generalised_2022} would not be useful in this parameter regime. Thus, for realistic QKD protocols, the postselection technique outperforms all other current proof techniques.

\section{Conclusion and Future Work}

In this work, we presented the rigorous and improved application of the postselection technique to practical QKD protocols. In particular, we extended the applicability of the postselection technique to prepare-and-measure protocols and decoy-state protocols. We also develop a new variant of the flag-state squasher which enables the postselection technique to be used with realistic detector setups.
Additionally, we fixed a gap in the original paper \cite{christandl_postselection_2009} to place the postselection technique on rigorous mathematical footing. We have also made several improvements to de Finetti reductions which significantly reduce the penalty imposed on the key rate when using the postselection technique. These improved statements can also be used to improve the performance of other quantum information processing tasks, such as those studied in ~\cite{fawzi_quantum_2015}.

To illustrate our results, we considered a simple implementation \cite{boaron_secure_2018} of the time-bin encoded three-state protocol over a fiber-optic cable.  For this implementation, phase-error rate and EUR-based proof techniques cannot be used.
We find that accounting for the sequential assumption in the GEAT approach by limiting the repetition rate of the protocol leads to significantly worse key rates than our improved postselection technique.
Thus, our results show that the postselection technique currently outperforms all other known security proof techniques for realistic PMQKD protocols.

We believe that many of our results can be improved further. For example, \cref{cor:generalMixedDeFinetti} and \cref{cor:symmetryDeFinetti} can be improved. Since the proof consists of using \cref{thm:genericDeFinetti} and \cref{thm:promisedDeFinetti} respectively, and tracing out the purifying system, any choice of purifying system supported on the symmetric subspace would suffice. Moreover, a mixture of \emph{all} such purifying system would also lead to no change in the proof. However, noting that the infinity norm of this mixture is less than 1 leads to a corresponding improvement to the prefactor in the corollaries.
Another place for improvement is \cref{thm:promisedDeFinetti}. The statement requires any symmetries to act separately on Alice and Bob's spaces. However, it may be possible to remove this requirement, thereby allowing the use of joint symmetries in both Alice and Bob's systems. Finally \cref{thm:maintheorem,thm:maintheoremvar}, which are utilized in applying the postselection technique to QKD protocols, currently impose a square root penalty on the security parameter. There may be a way to improve this penalty.
\section{Author Contributions}

S.N. and Y.Z. developed the proofs in \cref{sec:improvingDeFinetti}. D.T. and E.T. developed the proofs in \cref{sec:correctApplicationToQKD}. S.N. developed the proofs in \cref{sec:PSDecoy}. S.N. and D.T. prepared the plots and code in \cref{sec:applicationtothreestate}.
S.N. contributed the ideas for \cref{sec:improvingDeFinetti,sec:PSDecoy}.
N.L. and E.T. contributed to supervision and general directions for the project.

\section{Acknowledgements}

We thank Arsalan Motamedi for fruitful discussions involving the de Finetti reductions that contributed to our intuition.
We thank Shihong Pan for providing some of the code used in the plots, as well as for some code review.
We thank Florian Kanitschar for pointing out typos in our work.
This work was funded by the NSERC Discovery Grant, and was conducted at the Institute for Quantum Computing, University of Waterloo, which is funded by the Government of Canada through ISED. DT was partially funded by the Mike and Ophelia Laziridis Fellowship.

\appendix

\section{Proof of statements in Sec.~\ref{sec:improvingDeFinetti}}
\label{app:deFinettiProfs}

In this appendix, we complete the proofs of the theorems and lemmas in \Cref{sec:improvingDeFinetti}. To prove \Cref{thm:genericDeFinetti} and \Cref{thm:promisedDeFinetti}, we first need to establish the following lemma.
\begin{lemma}\label{lemma:T}
    Suppose $d_A$ and $d_R$ are two positive integers with $d_A\leq d_R$. Let $\ket{\theta}:=\sum_{i=1}^{d_A}\ket{i}\otimes\ket{i}\in\mathbb{C}^{d_A}\otimes\mathbb{C}^{d_R}$, and let $dU$ be the Haar measure on  $\mathcal{U}(\mathbb{C}^{d_R})$. For every $n\geq 1$, let 
    \begin{align*}
        T_n:=\int_{\mathcal{U}(\mathbb{C}^{d_R})}(\id_{A^n}\otimes U^{\otimes n})\ketbra{\theta}^{\otimes n}(\id_{A^n}\otimes U^{\otimes n})^{\dagger}dU.
    \end{align*}
Then there is an invertible operator $\kappa_n\in \matrixset{(\mathbb{C}^{d_A})^{\otimes n}}$ such that 
\begin{align}
     (\kappa_n\otimes \id_{R^n})^{-1/2}T_n (\kappa_n\otimes \id_{R^n})^{-1/2}=\id_{\Sym{\mathbb{C}^{d_A}\otimes\mathbb{C}^{d_R}} }. \label{eq:kappa}
\end{align}
\end{lemma}
The construction of $\kappa_n$ is similar to the construction of $\kappa_{D_n}$ in the proof of Lemma 3.1 in \cite{fawzi_quantum_2015} which uses Schur-Weyl duality. In this regard, we provide a brief explanation of the Schur-Weyl duality that will be used in the proof. For a more comprehensive treatment, we refer the readers to~\cite{landsberg_tensors_2011}.

Consider a tensor space $(\mathbb{C}^d)^{\otimes n}$, where $d$ and $n$ are positive integers. The unitary group $\mathcal{U}_d$ of $d\times d$ unitaries acts on $(\mathbb{C}^d)^{\otimes n}$ by sending $\ket{\psi_1}\otimes\cdots\otimes\ket{\psi_n}\mapsto U\ket{\psi_1}\otimes\cdots\otimes U\ket{\psi_n}$ for $U\in \mathcal{U}_d$. The symmetric group $S_n$ of permutations on $n$ letters acts on $(\mathbb{C}^d)^{\otimes n}$ by sending $\ket{\psi_1}\otimes\cdots\otimes\ket{\psi_n}\mapsto \ket{\psi_{\pi^{-1}(1)}}\otimes\cdots\otimes\ket{\psi_{\pi^{-1}(n)}}$ for $\pi\in S_n$. Note that these two actions commute and they define representations of $\mathcal{U}_d$ and $S_n$ on $(\mathbb{C}^d)^{\otimes n}$. The Schur-Weyl duality can be viewed as a ``quantitative version" of the double commutant theorem from representation theory. It states that as a $\mathcal{U}_d\times S_n$-module,
\begin{align}
    (\mathbb{C}^d)^{\otimes n}\cong\bigoplus_{\lambda\in \Lambda_{n,d}} W_{d}^\lambda\otimes [\lambda],\label{SWdecomp}
\end{align}
where $\mathcal{U}_d$ acts trivially on $[\lambda]$'s and $S_n$ acts trivially on $W_d^{\lambda}$'s. So in particular, any $U\in\mathcal{U}_d$ acting on $(\mathbb{C}^d)^{\otimes n}$ decomposes as $\bigoplus_{\lambda\in \Lambda_{n,d}} W_{d}^\lambda(U)\otimes \id_{[\lambda]}$, where each $W_{d}^\lambda(U)$ is a unitary acting on the space $W_d^\lambda$. Here $\Lambda_{n,d}:=\{\lambda=(\lambda_1,\ldots,\lambda_d):\lambda_1\geq\lambda_2\geq \cdots \geq \lambda_d\geq 0, \sum_{i=1}^d\lambda_i=n\}$ consists of partitions of $n$ into $d$ parts (which are commonly referred to as Young diagrams). We identify a partition $(\lambda_1,\ldots,\lambda_d,0,\cdots,0)$ with $(\lambda_1,\ldots,\lambda_d)$, so $\Lambda_{n,d}$ is a subset of $\Lambda_{n,d'}$ for $d\leq d'$. In the decomposition (\ref{SWdecomp}), the space $[\lambda]$ is determined by $\lambda$, while the space $W_d^\lambda$ is dependent on both $\lambda$ and $d$.

Now given two spaces $(\mathbb{C}^{d_A})^{\otimes n}$ and $(\mathbb{C}^{d_R})^{\otimes n}$ with $d_A\leq d_R$. The Schur-Weyl duality implies the decompositions
\begin{align*}
        (\mathbb{C}^{d_A})^{\otimes n}\cong \bigoplus_{\lambda\in \Lambda_{n,d_A}} W_{A}^\lambda\otimes [\lambda]_A, \text{ and }(\mathbb{C}^{d_R})^{\otimes n}\cong \bigoplus_{\lambda\in \Lambda_{n,d_R}} W_{R}^\lambda\otimes [\lambda]_R.
    \end{align*}
For every $\lambda\in\Lambda_{n,d_A}\subset \Lambda_{n,d_R}$, we have $[\lambda]_A\cong [\lambda]_R$. The $n$-th symmetric power $\Sym{\mathbb{C}^{d_A}\otimes \mathbb{C}^{d_R}}$ is isomorphic to $\bigoplus_{\lambda\in \Lambda_{n,d_A}}W_A^\lambda\otimes W_R^\lambda$ as a $\mathcal{U}_{d_A}\times \mathcal{U}_{d_R}$-module (see for instance Eq.~(5.27) in Ref.~\cite{harrow_applications_2005}). Indeed, for each $\lambda\in \Lambda_{n,d_A}$ there is a maximally entangled state $\ket{\psi_\lambda}$ such that any vector $\ket{v}$ in $\Sym{\mathbb{C}^{d_A}\otimes \mathbb{C}^{d_R}}$ can be written as $\bigoplus_{\lambda\in \Lambda_{n,d_A}}\ket{\phi_\lambda}\otimes\ket{\psi_\lambda}$ for some $\ket{\phi_\lambda}\in W_A^\lambda\otimes W_R^\lambda$. It follows that $\id_{\Sym{\mathbb{C}^{d_A}\otimes \mathbb{C}^{d_R}}}=\bigoplus_{\lambda\in \Lambda_{n,d_A}}\id_{W^\lambda_A}\otimes \id_{W^\lambda_R}\otimes \ketbra{\psi^{\lambda}}$.

\begin{proof}[Proof of Lemma \ref{lemma:T}]
    We keep using the notation above. Let $\ket{\Psi_\lambda}:=\dim([\lambda])\ket{\psi_\lambda}$ for all $\lambda\in \Lambda_{n,d_A}$. Then
\begin{align*}
    \ket{\theta}^{\otimes n}=\bigoplus_{\lambda\in\Lambda_{n,d_A}}\ket{\Phi^\lambda}\otimes \ket{\Psi^\lambda}
\end{align*}
where $\ket{\Phi^\lambda}=\sum_{j=1}^{\dim(W^\lambda_A)}\ket{w^{\lambda,A}_j}\otimes \ket{w^{\lambda,R}_j}$ for some orthonormal basis $\{\ket{w^{\lambda,A}_j}\}$ of $W^{\lambda}_A$ and orthonormal set $\{\ket{w^{\lambda,R}_j}\}$ in $W^\lambda_R$. For every $\lambda,\lambda'\in\Lambda_{n,d_A}$, $1\leq j,j'\leq \dim(W^\lambda_A)$, and $U\in\mathcal{U}(\mathbb{C}^{d_R})$, by Schur's lemma, we obtain
\begin{align*}
    \int_{\mathcal{U}(\mathbb{C}^{d_R})}W^\lambda_R(U)\ketbra{w^{\lambda,R}_j}{w^{\lambda',R}_{j'}}W^{\lambda'}_R(U)^{\dagger}dU=\begin{cases}
        \frac{1}{\dim(W^\lambda_R)}\id_{W^\lambda_R} & \text{ if } \lambda=\lambda' \text{ and } j=j'\\
        0 & \text{ otherwise}.
    \end{cases}
\end{align*}
It follows that
\begin{align*}
T_n&=\bigoplus_{\lambda,\lambda'\in\Lambda_{n,d_A}}\sum_{\substack{1\leq j\leq \dim(W^{\lambda}_A)\\ 1\leq j'\leq \dim(W^{\lambda'}_A)}}\ketbra{w^{\lambda,A}_j}{w^{\lambda',A}_{j'}}\otimes \left( \int W^\lambda_R(U)\ketbra{w^{\lambda,R}_j}{w^{\lambda',R}_{j'}}W^{\lambda'}_R(U)^{\dagger}dU \right)\otimes \ketbra{\Psi^{\lambda}}{\Psi^{\lambda'}}\\
    &=\bigoplus_{\lambda\in\Lambda_{n,d_A}}\frac{1}{\dim(W^\lambda_R)}\id_{W^\lambda_A}\otimes \id_{W^\lambda_R}\otimes \ketbra{\Psi^{\lambda}}=\bigoplus_{\lambda\in\Lambda_{n,d_A}}\frac{\dim([\lambda])}{\dim(W^\lambda_R)}\id_{W^\lambda_A}\otimes \id_{W^\lambda_R}\otimes \ketbra{\psi^{\lambda}}.
\end{align*}
\Cref{eq:kappa} follows by taking $\kappa_n:=\bigoplus_{\lambda\in\Lambda_{n,d_A}}\frac{\dim([\lambda])}{\dim(W^\lambda_R)}\id_{W^\lambda_A}\otimes \id_{[\lambda]_A}$.
\end{proof}

\genericDeFinetti*
\begin{proof}
    Assume without loss of generality that $d_A\leq d_R$ and $\hat{\sigma}_A$ has full rank. We first observe that
\begin{align*}
   \text{Puri}_R(\hat{\sigma}_A):=&\{\ketbra{\phi}\in \posset{\mathbb{C}^{d_A}\otimes\mathbb{C}^{d_R}}: \Tr_R\left[\ketbra{\phi}\right]=\hat{\sigma}_A\}\\
   =&\{(\promise^{1/2}\otimes U)\ketbra{\theta}(\sigma^{1/2}\otimes U^{\dagger}):U\in\mathcal{U}(\mathbb{C}^{d_R})\}, 
\end{align*}
so the Haar measure $dU$ on $\mathcal{U}(\mathbb{C}^{d_R})$ induces a measure $d\phi$ on the set
of purifications $\text{Puri}_R(\hat{\sigma}_A)$. Let $T_n:=\int_{\mathcal{U}(\mathbb{C}^{d_R})}(\id_{A^n}\otimes U^{\otimes n})\ketbra{\theta}^{\otimes n}(\id_{A^n}\otimes U^{\otimes n})^{\dagger}dU$ be as in \Cref{lemma:T}. Then
\begin{align*}
    \tau_n:=\int_{\text{Puri}_R(\hat{\sigma}_A)} \ketbra{\phi}^{\otimes n}d\phi
    =(\hat{\sigma}_A^{\otimes n}\otimes \id_{R^n})^{1/2}T_n(\hat{\sigma}_A^{\otimes n}\otimes \id_{R^n})^{1/2},
\end{align*}
so we only need to prove that 
\begin{align}
    (\hat{\sigma}_A^{\otimes n}\otimes \id_{R^n})^{-1/2}\rho_{A^nR^n}(\hat{\sigma}_A^{\otimes n}\otimes \id_{R^n})^{-1/2}\leq g_{n, d_Ad_R}T_n.\label{eq:neq1}
\end{align}
By \Cref{lemma:T}, there is an invertible operator $\kappa_n\in \matrixset{(\mathbb{C}^{d_A})^{\otimes n}}$ such that 
\begin{align*}
     (\kappa_n\otimes \id_{R^n})^{-1/2}T_n (\kappa_n\otimes \id_{R^n})^{-1/2}=\id_{\Sym{\mathbb{C}^{d_A}\otimes\mathbb{C}^{d_R}} }.
\end{align*}
Let $\widetilde{\rho}:=(\kappa_n\otimes \id_{R^n})^{-1/2}(\hat{\sigma}_A^{\otimes n}\otimes \id_{R^n})^{-1/2}\rho_{A^nR^n}(\hat{\sigma}_A^{\otimes n}\otimes \id_{R^n})^{-1/2}(\kappa_n\otimes \id_{R^n})^{-1/2}$. To prove \Cref{eq:neq1}, we only need to show that
\begin{align}
    \widetilde{\rho} \leq g_{n,d_Ad_R}\id_{\Sym{\mathbb{C}^{d_Ad_R}}}.\label{eq:rhotilde}
\end{align}
Indeed, since $\rho_{A^nR^n}$ and $\tau_n$ have the same marginal $\hat{\sigma}_A^{\otimes n}$ on $A^n$,  we have
\begin{align}
    \label{eq:infToOneNorm} \norm{ \widetilde{\rho}}_\infty
    \leq& \Tr( \widetilde{\rho})\\
    \nonumber =&\Tr\big( (\kappa_n\otimes \id_{R^n})^{-1/2}(\hat{\sigma}_A^{\otimes n}\otimes \id_{R^n})^{-1/2}\tau_n(\hat{\sigma}_A^{\otimes n}\otimes \id_{R^n})^{-1/2}(\kappa_n\otimes \id_{R^n})^{-1/2} \big)\\
    \nonumber=&\Tr\big( (\kappa_n\otimes \id_{R^n})^{-1/2}T_n(\kappa_n\otimes \id_{R^n})^{-1/2} \big)=\Tr\big(\id_{\Sym{\mathbb{C}^{d_Ad_R}}} \big)\\
    \nonumber =&\dim(\Sym{\mathbb{C}^{d_Ad_R}})=g_{n,d_Ad_R}.
\end{align}
It is clear that the operator $\widetilde{\rho}$ is supported on $\Sym{\mathbb{C}^{d_Ad_R}}$, so Eq.(\ref{eq:rhotilde}) follows.
\end{proof}

\promisedDeFinetti*
\begin{proof}
For convenience, we use $A_i$ and $R_i$ to denote $\mathbb{C}^{d_i^A}$ and $ \mathbb{C}^{d_i^R}$. From \cite[Eq. (6.7.1)]{landsberg_tensors_2011} we know that
    \begin{align*}
        \Sym{\bigoplus_{i=1}^{k}A_i \otimes R_i}=\bigoplus_{n_1+\cdots +n_k = n}\bigotimes_{i=1}^k\text{Sym}^{n_i}\left(A_i\otimes R_i \right).
    \end{align*}
For every $1\leq i\leq k$, let $\ket{\theta_i}:=\sum_{j=1}^{d_A}\ket{j}\otimes\ket{j}\in A_i\otimes R_i$. Then for every purification $\ket{\phi}\in \bigoplus_{i=1}^{k}A_i\otimes R_i$ of $\promise$, there are unitaries $U_1\in\mathcal{U}(R_1),\ldots,U_k\in\mathcal{U}(R_k)$ such that $\ket{\phi}=\bigoplus_{i=1}^k\id_{A_i}\otimes U_i\ket{\theta_i}$. So the Haar measure $dU$ on $\mathcal{U}(R_1)\times \cdots \times \mathcal{U}(R_k) $ induces a measure $d\phi$ on  the set of purifications $\ket{\phi}_{AR}\in \bigoplus_{i=1}^{k}A_i\otimes R_i$ of $\promise$.
It follows that 
\begin{align*}
    \tau:=\int \ketbra{\phi}{\phi}_{AR}^{\otimes n} d\phi = (\hat{\sigma}_A^{\otimes n}\otimes \id_{R^n})^{1/2}T(\hat{\sigma}_A^{\otimes n}\otimes \id_{R^n})^{1/2}, 
\end{align*}
where
\begin{align*}
    T:= \int \left( \Big(\bigoplus_{i=1}^k \id_{A_i}\otimes U_i\ket{\theta_i}\Big)\Big(\bigoplus_{j=1}^k\bra{\theta_j}\id_{A_j}\otimes U_j^{\dag}\Big)\right)^{\otimes n} dU_1\cdots dU_k.
\end{align*}
Since $\int_\mathcal{U} UdU=0$ for any compact unitary group $\mathcal{U}$ and Haar measure $dU$ on $\mathcal{U}$, we have
\begin{align*}
    T= \bigoplus_{n_1+\cdots n_k=n} \bigotimes_{i=1}^k T_{n_i}^{(i)}, \text{ where } T_{n_i}^{(i)}= \int (\id_{A_i^{n_i}}\otimes U_i^{\otimes n_i})\ketbra{\theta_i}^{\otimes {n_i}}(\id_{A_i^{n_i}}\otimes {U_i^\dag}^{\otimes {n_i}})^{\dagger}dU_i.
\end{align*}
By \Cref{lemma:T}, for every $1 \leq i \leq k$ and $1\leq m\leq n $, there is an invertible operator $\kappa_m^{(i)}\in \matrixset{(\mathbb{C}^{A_i})^{\otimes m}}$ such that 
\begin{align*}
     (\kappa_m^{(i)}\otimes \id_{R_i^m})^{-1/2}T_m^{(i)} (\kappa_m^{(i)}\otimes \id_{R_i^m})^{-1/2}=\id_{\text{Sym}^m(A_i\otimes R_i)}.
\end{align*}
Let $\kappa:= \bigoplus_{n_1+\cdots n_k=n} \bigotimes_{i=1}^k \kappa_{n_i}^{(i)}$. Then
\begin{align*}
     (\kappa\otimes \id_{R^n})^{-1/2}T (\kappa\otimes \id_{R^n})^{-1/2}=\bigoplus_{n_1+\cdots n_k=n} \bigotimes_{i=1}^k\id_{\text{Sym}^{n_i}(A_i\otimes R_i)}=\id_{\Sym{\bigoplus_{i=1}^k A_i\otimes R_i}}.
\end{align*}
The rest of the proof follows similarly as in the proof of \Cref{thm:genericDeFinetti}.
\end{proof}

\blockDiagPurification*
\begin{proof}
    Since $\state$ is block-diagonal, its eigenvectors $\{\ket{x}\}_{x\in \mathcal{X}}$ can be picked such that each eigenvector lies in the support of some projection $\Pi_{\vec{i}}$.
    For any eigenvalue $\lambda$, define
    \begin{align*}
        \ket{\Psi^\lambda} \coloneqq \sum_{x\in\mathcal{X}_\lambda}\ket{x}\otimes \overline{\ket{x}}
    \end{align*}
    where $\mathcal{X}_\lambda \coloneqq \{x\in \mathcal{X}\vert \  \state\ket{x}=\lambda\ket{x}\}$ and the complex conjugation is taken with respect to a tensor product basis on $(\mathbb{C}^{d_{A} d_{B}})^{\otimes n}$.
    Since each eigenvector lies in the support of one projection, we have that $\ket{\Psi^\lambda}=\sum_{\vec{i}\in[k]^n}\Pi_{\vec{i}}\otimes \overline{\Pi_{\vec{i}}}\ket{\Psi^\lambda}$. Thus, $\ket{\Psi^\lambda}$ lies in a subspace of $(\mathbb{C}^{d_{A} d_{B}}\otimes \mathbb{C}^{d_{A} d_{B}})^{\otimes n}$. In particular,
    \begin{align} \label{eq:blockDiagPurif}
        \nonumber & \ket{\Psi^\lambda} \in \bigoplus_{\vec{i}\in [k]^n}\left(\bigotimes_{j=1}^n\left(\mathbb{C}^{d_{i_j}}\otimes \mathbb{C}^{d_{i_j}}\right) \right)\\
        \implies & \ket{\Psi^\lambda} \in \left(\bigoplus_{i=1}^k\mathbb{C}^{d_{i}}\otimes \mathbb{C}^{d_{i}} \right)^{\otimes n}.
    \end{align}
    Moreover, as shown in the proof of Lemma 4.2.2 from \cite{renner_security_2005}
    \begin{align} \label{eq:permInvRenner}
        \permutation{d_A^2 d_B^2}{n}\ket{\Psi^\lambda} = \ket{\Psi^\lambda},
    \end{align}
    for all $\perm\in S_n$.
    Combining Eq.(\ref{eq:blockDiagPurif}) with Eq. (\ref{eq:permInvRenner}), we see that
    \begin{align}
        \permutation{\sum_{i=1}^k d_i^2}{n}\ket{\Psi^\lambda} = \ket{\Psi^\lambda},
    \end{align}
    for all $\perm\in S_n$ implying that $\ket{\Psi^\lambda}\in \text{Sym}^n\left(\bigoplus_{i=1}^k \mathbb{C}^{d_i}\otimes \mathbb{C}^{d_i}\right)$.

    Finally, it can be straightforwardly verified that
    \begin{align*}
        \ket{\Psi}\coloneqq \sum_\lambda \sqrt{\lambda}\ket{\Psi^\lambda}
    \end{align*}
    is a purification of $\state$. Since $\ket \Psi$ is a linear combination of states in $\text{Sym}\left(\bigoplus_{i=1}^k \mathbb{C}^{d_i}\otimes \mathbb{C}^{d_i}\right)$, we get the required result.
\end{proof}

\groupPurification*
To prove this, we first need to establish the following lemma:
\begin{lemma}\label{lem:projection}
    Suppose $\pi_1:G\rightarrow\mathcal{U}(\mathbb{C}^d)$ and $\pi_2:G\rightarrow\mathcal{U}(\mathbb{C}^{d'})$ are two irreducible representations of a compact group $G$. Let $\Pi:=\int_G \pi_1(g)\otimes\overline{\pi_2(g)}d\mu(g)$ where $\mu$ is the Haar measure on $G$. Then $\Pi$ is an orthogonal projection and $\Tr(\Pi):=\begin{cases}
      1 & \text{ if } \pi_1\cong \pi_2\\
      0 & \text{ if } \pi_1\not\cong \pi_2
    \end{cases}$. 
\end{lemma}
\begin{proof}
Observe that
\begin{align}
    \pi_1(g)\otimes\overline{\pi_2(g)}\Pi=\int_G\pi_1(gh)\otimes \overline{\pi_2(gh)}d\mu(g)=\int_G\pi_1(gh)\otimes \overline{\pi_2(gh)}d\mu(gh)=\Pi\label{eq:single}
\end{align}
for every $g\in G$. So $\Pi^2=\int_G\pi_1(g)\otimes \overline{\pi_2(g)}\Pi d\mu(g)=\Pi$, and hence $\Pi$ is an orthogonal projection.
Let $\ket{v}\in\mathbb{C}^d\otimes \mathbb{C}^{d'}$ be a vector such that $\Pi\ket{\psi}=\ket{\psi}$. Without loss of generality, we may assume $d\leq d'$. Then there exists a linear map $\Lambda:\mathbb{C}^{d'}\rightarrow\mathbb{C}^d$ such that $\Lambda\otimes \id\ket{\tau_{d'}}=\ket{v}$ where $\ket{\tau_{d'}}:=\tfrac{1}{\sqrt{d'}}\sum_{\ell=1}^{d'}\ket{\ell}\otimes\ket{\ell}$. Then Eq.(\ref{eq:single}) implies
    \begin{align*}
        \pi_1(g)\Lambda\pi_2(g)^{\dagger}\otimes \id\ket{\tau_{d'}}&=\pi_1(g)\Lambda\otimes \overline{\pi_2(g)}\ket{\tau_{d'}}=\pi_1(g)\otimes \overline{\pi_2(g)}\ket{v}\\
        &=\pi_1(g)\otimes \overline{\pi_2(g)}\Pi\ket{v}
        =\Pi\ket{v}=\ket{v}=\Lambda\otimes \id\ket{\tau_{d'}}
    \end{align*}
for all $g\in G$. It follows that $\pi_1(g)\Lambda\pi_2(g)^{\dagger}=\Lambda$ for all $g\in G$. If $\pi_1\not\cong\pi_2$, then Schur's lemma implies $\Lambda=0$, so we must have $\ket{v}=0$, and hence $\Pi=0$. If $\pi_1\cong\pi_2$, let $U$ be the unitary such that $\pi_1(g)=U\pi_2(g)U^\dag$, then by Schur's lemma again, $\Lambda=\lambda U$ for some $\lambda\in\mathbb{C}$, and hence $\Pi$ must be the rank-one projection onto the span of $U\otimes \id\ket{\tau_{d}}$.
\end{proof}

\begin{proof}[Proof of Lemma \ref{lem:groupPurification}] 
For notational convenience, we use $\pi:G\rightarrow \mathcal{U}(\mathbb{C}^d)$ to denote the representation that sends $g\mapsto U_g$ for all $g\in G$ where $d=d_Ad_B$. Without loss of generality, we may assume $\pi=\bigoplus_{i=1}^k \bigoplus_{j=1}^{m_i}\pi_i^{(j)}$ where each $\pi_i^{(j)}$ is an irreducible representations of $G$ on $\mathbb{C}^{d_i}$ such that $\pi_i^{(j)}\cong \pi_i^{(j')}$ for all $i\in [k]$ and $j,j'\in [m_i]$, and $\pi_i^{(j)}\not\cong \pi_{i'}^{(j')}$ for all $i\neq i'$ in $[k]$. Let $\Pi:=\int_G \pi(g)\otimes \overline{\pi(g)}d\mu(g)$. Lemma \ref{lem:projection} implies 
\begin{align*}
    \Pi=\bigoplus_{i,i'=1}^k \bigoplus_{j=1}^{m_i}\bigoplus_{j'=1}^{m_{i'}}\int_G \pi_i^{(j)}(g)\otimes \overline{\pi_{i'}^{(j')}(g)}d\mu(g)=\bigoplus_{i=1}^k \bigoplus_{j,j'=1}^{m_i}\int_G \pi_i^{(j)}(g)\otimes \overline{\pi_{i}^{(j')}(g)}d\mu(g)
\end{align*}
is an orthogonal projection of rank $\sum_{i=1}^k m_i^2$, because every $\int_G \pi_i^{(j)}(g)\otimes \overline{\pi_{i}^{(j')}(g)}d\mu(g)$ is a rank-one projection. Here the conjugation is taken with respect to the standard basis of $\mathbb{C}^{d}=\bigoplus_{i=1}^k\big(\mathbb{C}^{d_i}\big)^{\oplus m_i}$. Now we only need to show that there is purification of $\rho$ on $\Sym{\text{supp}(\Pi)}$. For every eigenvalue $\lambda$ of $\rho$ and $\Vec{g}\in G^n$, since $\rho$ commutes with $\pi({\Vec{g}})$, the eigenspace $\mathcal{H}_\lambda$ of $\lambda$ is invariant under $\pi({\Vec{g}})$. So $\mathcal{H}_\lambda$ must be a direct sum of spaces of the form $\mathbb{C}^{d_{i_1}}\otimes\cdots\otimes \mathbb{C}^{d_{i_n}}$. This implies that every $\mathcal{H}_\lambda$ has a basis $\{\mathcal{X}_\lambda\}$ that is a subset of the standard basis $\mathcal{X}$ for $(\mathbb{C}^d)^{\otimes n}$. Let $\ket{\tau}:=\sum_{\ket{v}\in\mathcal{X}}\ket{v}\otimes \ket{v}$, and let $\ket{\Psi^{\lambda}}:=\sum_{\ket{v}\in\mathcal{X}_\lambda}\ket{v}\otimes\ket{v}$ for every eigenvalue $\lambda$. For every $\Vec{g}\in G^n$, since $\mathcal{H}_\lambda $ is invariant under $\pi({\Vec{g}})$, the orthogonal projection $\Pi_\lambda:=\sum_{\ket{v}\in\mathcal{X}_\lambda}\ketbra{v}$ onto $\mathcal{H}_\lambda$ commutes with $\pi({\Vec{g}})$ and $\overline{\pi({\Vec{g}})}$. It follows that 
\begin{align*}
    \pi({\Vec{g}})\otimes \overline{\pi({\Vec{g}})}\ket{\Psi^\lambda}&= \big(\pi({\Vec{g}})\otimes \overline{\pi({\Vec{g}})}\big)\big(\Pi_\lambda\otimes\Pi_\lambda\big)\ket{\tau}=(\Pi_\lambda\otimes\Pi_\lambda\big)\big(\pi({\Vec{g}})\otimes \overline{\pi({\Vec{g}})}\big)\ket{\tau}\\
&=\big(\Pi_\lambda\otimes\Pi_\lambda\big)\big(\pi({\Vec{g}})\pi({\Vec{g}})^\dag\otimes \id\big)\ket{\tau}=\Pi_\lambda\otimes\Pi_\lambda\ket{\tau}=\ket{\Psi^\lambda}
\end{align*}
for all $\lambda$ and $\Vec{g}$. Then $\ket{\Psi}:=\sum_{\lambda}\sqrt{\lambda}\ket{\Psi^{\lambda}}$ is a purification of $\rho$ satisfying 
\begin{align*}
    \Pi^{\otimes n}\ket{\Psi}&=\big(\int_G\pi(g)\otimes \overline{\pi(g)}d\mu(g)\big)^{\otimes n}\ket{\Psi}=\sum_{\lambda}\sqrt{\lambda}\left(\int_{ G^n}\pi({\Vec{g}})\otimes \overline{\pi({\Vec{g}})}\ket{\Psi^\lambda}d\mu(\Vec{g})\right)\\
&=\sum_{\lambda}\sqrt{\lambda}\ket{\Psi^{\lambda}}=\ket{\Psi}.
\end{align*}
It is clear that $\ket{\Psi}$ must be permutation invariant, so we conclude that $\ket{\Psi}$ is a purification of $\rho$ on $\Sym{\text{supp}(\Pi)}\cong \Sym{\mathbb{C}^{\sum_{i=1}^km_i^2}}$.
\end{proof}

\section{Proof of statements in Sec.~\ref{sec:correctApplicationToQKD}}
\label{app:correctApplicationToQKD}

\newcommand{\permreg}{C}
We begin by reproducing some fairly standard arguments (some of which have been discussed to some extent in Refs.~\cite{christandl_postselection_2009,beaudry_assumptions_2015}) to prove the starting claims in Sec.~\ref{sec:correctApplicationToQKD}, with small modifications in some cases to adapt them to this work.
First, we verify the claim that if a channel $\mathcal{F}$ first applies a uniformly random permutation on its input registers, followed by some operations that do not depend on the choice of permutation, and outputs the permutation choice in a classical register $\permreg$, then $\mathcal{F}$ indeed satisfies \cref{def:permInvariantMaps}.
To do so, observe that the action of such a channel can be written in the form
\begin{align}
\mathcal{F}(\rho) = \frac{1}{n!} \sum_{\perm'} \ketbra{\perm'}_{\permreg} \otimes \mathcal{E} \circ \mathcal{W}_{\perm'}(\rho),
\end{align}
for some channel $\mathcal{E}$. Then for any permutation $\pi$, if we define $G_\perm$ to be a channel that changes the value on $\permreg$ by replacing $\ketbra{\perm'}$ with $\ketbra{\perm' \circ \pi}$, we have the claimed property:
\begin{align}
G_\perm \circ \mathcal{F} \circ \mathcal{W}_{\perm} (\rho) 
&= G_\perm \left(  \frac{1}{n!} \sum_{\perm'} \ketbra{\perm'}_{\permreg} \otimes \mathcal{E} \circ \mathcal{W}_{\perm' \circ \perm}(\rho)  \right) \nonumber\\
&= G_\perm \left( \frac{1}{n!} \sum_{\perm''} \ketbra{\perm'' \circ \perm^{-1}}_{\permreg} \otimes \mathcal{E} \circ \mathcal{W}_{\perm''}(\rho) \right) \nonumber\\
&= \frac{1}{n!} \sum_{\perm''} \ketbra{\perm''}_{\permreg} \otimes \mathcal{E} \circ \mathcal{W}_{\perm''}(\rho) \nonumber\\
&= \mathcal{F} (\rho) 
,
\end{align}
where the second line is just a summation relabelling via $\pi'' \coloneqq \perm' \circ \perm$ (which is a valid bijection on the summation indices) and the third line is by definition of $G_\perm$.

To apply a uniformly random permutation on a string of length $n$ in practice, one can apply the Fisher-Yates shuffle, which is a simple algorithm that can permute the string in-place using approximately $n\log(n)$ random bits. Alternatively, there is a minor variation of that algorithm that can be implemented in a ``streaming'' fashion with respect to the input string, at the cost of requiring enough memory to store the input and output strings simultaneously, as follows. First prepare a ``blank'' output string. Then take the first value in the input string, select a uniformly random position $j\in[n]$ (which uses approximately $\log(n)$ bits), and fill that value in the $j^\text{th}$ position in the output string. Repeat this procedure for each subsequent value in the input string, though in each subsequent step, one should only choose a uniformly random position out of the remaining ``blank'' spaces. This yields a uniformly random permutation of the input string, and uses only about $n\log(n)$ random bits (in fact fewer than that, since later steps require less randomness as the number of blank spaces decreases).

Technically, the above procedure describes permutations on classical strings rather than quantum registers. However, if one were to apply random permutations in QKD protocols in practice, it would indeed usually be performed on the classical output strings rather than the quantum states. This discrepancy can be resolved by noting that if the initial measurement steps in the protocol are described by a channel with an ``IID form'' $\mathcal{M}^{\otimes n}$ (or some other suitable form of ``permutation symmetry''), then the permutation of the output strings commutes with these measurements, and hence we can view it as effectively implementing the permutation directly on the quantum registers, as desired.

We now prove \cref{lemma:endOfStepOne}, by first showing the following lemmas \cref{lemma:psLemma,lemma:tau}. Note that \cref{lemma:psLemma} is also proved in the proof of \cite[Theorem 2]{christandl_postselection_2009}.

\begin{lemma}
    \label{lemma:psLemma}
    Let $\mathcal{F}, \mathcal{F}^\prime \in \mapset{A^nB^n,K}$ be linear maps such that  $\mathcal{F}-\mathcal{F}^\prime$ is a permutation-invariant linear map. Let  $\rho_{A^nB^nR^{\prime \prime}} \in \posset{A^nB^nR^{\prime \prime}}$ be any extension of $\rho_{A^nB^n}$. Then the state $\bar{\rho}_{A^nB^n} = \frac{1}{n!} \sum_{\perm \in S_n}  \mathcal{W}_\perm (\rho_{A^nB^n})$ is permutation-invariant, and for any purification $\bar{\rho}_{A^n B^n R^\prime}$ of that state, we have
    \begin{equation}
        \tracenorm{ \left( \left( \mathcal{F} - \mathcal{F}^\prime \right) \otimes \id_{R^{\prime\prime}}\right) \left( \rho_{A^nB^nR^{\prime\prime}} \right) } \leq	\tracenorm{ \left( \left(\mathcal{F} - \mathcal{F}^\prime \right) \otimes \id_{R^\prime}\right) \left( \bar{\rho}_{A^nB^nR^\prime} \right) }. 
    \end{equation}
\end{lemma}

\begin{proof}
    Construct 
    \begin{equation}
    	\bar{\rho}_{A^nB^nR^{\prime\prime} \widetilde{R}} = \frac{1}{n!} \sum_{\perm \in S_n} \left( \mathcal{W}_\perm \otimes I_{R^{\prime\prime}} \right) \left( \rho_{A^nB^n R^{\prime\prime}} \right) \otimes \ketbra{\perm}{\perm}_{\widetilde{R}}
    \end{equation}
    as an extension of $\bar{\rho}_{A^nB^n}$. Therefore, there exists $\Phi \in \channelset{ R^{\prime} , R^{\prime \prime} \widetilde{R}}$ such that $ \left( \id_{A^nB^n} \otimes \Phi \right) \left( \bar{\rho}_{A^nB^nR^\prime} \right) = \bar{\rho}_{A^nB^n R^{\prime \prime} \widetilde{R}}$. Since trace norm cannot increase under CPTNI maps, we have
    \begin{equation} \label{eq:dataprocessing2}
    	\tracenorm{ \left( \left(\mathcal{F} - \mathcal{F}^\prime \right) \otimes \id_{ R^{\prime \prime} \widetilde{R}}\right) \left( \bar{\rho}_{A^nB^n R^{\prime \prime} \widetilde{R} } \right) } \leq	\tracenorm{ \left( \left(\mathcal{F} - \mathcal{F}^\prime \right) \otimes \id_{R^\prime}\right) \left( \bar{\rho}_{A^nB^nR^\prime} \right) } .
    \end{equation}
    Next, making use of the permutation-invariance of $\mathcal{F}-\mathcal{F}^\prime$, we have
    \begin{equation} \label{eq:perminvariancetracenorm}
    	\begin{aligned} 
    			\tracenorm{ \left( \left(\mathcal{F} - \mathcal{F}^\prime \right) \otimes \id_{R^{\prime\prime} }\right) \left( \rho_{A^nB^nR^{\prime\prime}} \right) }  & = \frac{1}{n!} \sum_{\perm \in S_n}	\tracenorm{ \left( G_\perm \circ \left( \mathcal{F} - \mathcal{F}^\prime \right) \circ \mathcal{W}_\perm \otimes \id_{R^{\prime\prime} }\right) \left( \rho_{A^nB^nR^{\prime\prime}} \right) } \\
    			& \leq \frac{1}{n!} \sum_{\perm \in S_n}	\tracenorm{ \left(  \left( \mathcal{F} - \mathcal{F}^\prime \right) \circ \mathcal{W}_\perm \otimes \id_{R^{\prime \prime} }\right) \left( \rho_{A^nB^nR^{\prime \prime}} \right) } \\
    			& = \tracenorm{ \left(  \left( \mathcal{F} - \mathcal{F}^\prime \right) \otimes \id_{ R^{\prime \prime} \widetilde{R} }\right) \left( \bar{\rho}_{A^nB^n R^{\prime \prime} \widetilde{R}} \right) } ,
    	\end{aligned}
    \end{equation}
    where the inequality again follows from the fact that CPTNI maps cannot increase trace norm, and the final inequality follows from the fact that the states $\left(  \left( \mathcal{F} - \mathcal{F}^\prime \right) \circ \mathcal{W}_\perm \otimes \id_{R^{\prime \prime} }\right) \left( \rho_{A^nB^nR^{\prime \prime}} \right)$ are orthogonal for different $\perm$.
    Putting Eqs.~\eqref{eq:dataprocessing2} and \eqref{eq:perminvariancetracenorm} together, we get the desired result.
\end{proof}

\begin{lemma}
    \label{lemma:tau} 
    Let $\rho_{A^nB^n} \in \posset{A^nB^n}$ and $\tau_{A^nB^n} \in \posset{A^nB^n}$ be such that $\rho_{A^nB^n}  \leq \cost \tau_{A^n B^n}$ for some $\cost \in \mathbb{R}_{+}$. Let $\rho_{A^nB^nR^\prime}$ be any extension of $\rho_{A^nB^n}$, and let $\tau_{A^nB^nR}$ be any purification of $\tau_{A^nB^n}$. Then for any two maps $\mathcal{F}, \mathcal{F}^\prime \in \mapset{A^nB^n,K}$, 
    \begin{equation}
        \tracenorm{ \left( \left( \mathcal{F} - \mathcal{F}^\prime \right) \otimes \id_{R^\prime} \right) \left( \rho_{A^nB^nR^\prime} \right) } \leq \cost \tracenorm{ \left( \left(\mathcal{F} - \mathcal{F}^\prime \right) \otimes \id_{R} \right) \left( \tau_{A^nB^nR} \right) }	
    \end{equation}	
\end{lemma}

\begin{proof}
     Since $\rho_{A^nB^n}  \leq \cost  \tau_{A^n B^n}$, there exists a $\omega_{A^nB^n} \in \stateset{A^nB^n}$ such that $\rho_{A^nB^n} + (\cost -1) \omega_{A^nB^n} = \cost \tau_{A^nB^n}$. One can then construct an extension of of $\tau_{A^nB^n}$ as,
    	\begin{equation}
    		\tau_{A^nB^n R^\prime M } = \frac{1}{\cost} \rho_{A^nB^n R^\prime} \otimes \ketbra{0}{0}_M + \left( 1- \frac{1}{\cost} \right) \omega_{A^nB^nR^\prime} \otimes \ketbra{1}{1}_M.
    	\end{equation}
        Since the map $(\mathcal{F}-\mathcal{F}^\prime) \otimes \id_{R^\prime M}$ acts identically on $M$, the two terms above are orthogonal before and after the action of the map. Therefore,
    	\begin{equation} \label{eq:triangle}
    		\begin{aligned}
    		\tracenorm{ \left( \left(\mathcal{F} - \mathcal{F}^\prime \right) \otimes \id_{R^\prime M}\right) \left( \tau_{A^nB^nR^\prime M} \right) } &= 	\frac{1}{\cost} \tracenorm{ \left( \left(\mathcal{F} - \mathcal{F}^\prime \right) \otimes \id_{R^\prime M}\right) \left( \rho_{A^nB^nR^\prime} \otimes \ketbra{0}{0}  \right) }\\
    		&+ \left( 1- \frac{1}{\cost} \right)	\tracenorm{ \left( \left(\mathcal{F} - \mathcal{F}^\prime \right) \otimes \id_{R^\prime M}\right) \left( \omega_{A^nB^nR^\prime } \otimes \ketbra{1}{1} \right) } \\
    		&\geq 	\frac{1}{\cost} \tracenorm{ \left( \left(\mathcal{F} - \mathcal{F}^\prime \right) \otimes \id_{R^\prime}\right) \left( \rho_{A^nB^nR^\prime}  \right) }
    		\end{aligned}
    	\end{equation}
    	Finally, for any purification $\tau_{A^nB^nR}$ of $\tau_{A^nB^n}$, and extension $\tau_{A^nB^nR^\prime M}$ of $\tau_{A^nB^n}$, there exists a CPTP map $\Phi \in \channelset{ R ,R^\prime M}$ such that $ \left( \id_{A^nB^n} \otimes \Phi \right) \left( \tau_{A^nB^nR} \right) = \tau_{A^nB^nR^\prime M}$. Therefore, 
    	\begin{equation} \label{eq:dataprocessing}
    			\tracenorm{ \left( \left(F - F^\prime \right) \otimes \id_{R}\right) \left( \tau_{A^nB^nR} \right) } \geq 	\tracenorm{ \left( \left(F - F^\prime \right) \otimes \id_{R^\prime M}\right) \left( \tau_{A^nB^nR^\prime M} \right) } 
    	\end{equation}
        follows from the fact that CPTP maps cannot increase trace norm. 
    	Putting Eq.~\eqref{eq:triangle} and \eqref{eq:dataprocessing} together, the desired result is obtained.
 \end{proof}

Lemma \ref{lemma:psLemma} lets us assume without loss of generality that the input state to a permutation-invariant QKD protocol is permutation-invariant on $A^nB^n$. Thus, the input states to such protocols satisfy the de Finetti reductions (\cref{cor:generalMixedDeFinetti}) described in Section \ref{sec:improvingDeFinetti}. Using \cref{lemma:tau} on this de Finetti reduction allows us to prove the following lemma after combining \cref{lemma:psLemma,lemma:tau,cor:generalMixedDeFinetti}.

\StepOne*

We now turn to the main technical proof for this section, which addresses the flaw in Ref.~\cite{christandl_postselection_2009}. Essentially, the gap in the argument was the claim that one can compensate for the purifying register $V$ in the state $\tau_{Z^nC^nC_EE^nV}$ (defined below) by simply subtracting $\log\dim(V)$ from the key length. That claim would have been true if we had a bound on the smooth min-entropy for \emph{every} IID state contributing to the mixture $\tau_{Z^nC^nC_EE^n}$ and chose the hash length accordingly. However, the structure of typical IID security proofs (as we described in \cref{subsubsec:structureofIIDsecurityproof}) does not straightforwardly yield such a bound, as it does not consider the smooth min-entropy of states outside the set $S$. We fix this by applying a different argument to lower bound the smooth min-entropy of $\tau_{Z^nC^nC_EE^n \wedge \accevent}$, at the cost of a worse smoothing parameter, though we leave for future work the question of whether an alternative argument might prove the original claim in Ref.~\cite{christandl_postselection_2009}.

\maintheorem*

\begin{proof}
Suppose the protocol is run with the input state $\tau_{A^nB^nR}$. Then, the output states of the real and ideal protocols can be written as
 \begin{equation} 
 \begin{aligned}
    \left( \protMap{l^\prime} \otimes \id_R \right)(\tau_{A^nB^nR}) &= \Pr(\Omega_{\text{acc}}) \tau^{(l^\prime)}_{K_A \Cfull R | \Omega_\text{acc}} + (1-\Pr(\Omega_{\text{acc}}) ) \tau^{(\bot)}_{K_A \Cfull R | \Omega^c_\text{acc}} \\
 \left( \protMapId{l^\prime} \otimes \id_R \right)(\tau_{A^nB^nR}) &= \Pr(\Omega_{\text{acc}}) \tau^{(l^\prime,\text{ideal})}_{K_A \Cfull R | \Omega_\text{acc}} + (1-\Pr(\Omega_{\text{acc}}) ) \tau^{(\bot,\text{ideal})}_{K_A \Cfull R | \Omega^c_\text{acc}}
        \end{aligned}
    \end{equation}
where we separate the output states conditioned on the event $\Omega_\text{acc}$ (protocol accepts) and its complement $\Omega_\text{acc}^c$.
Since the output states of the real and ideal protocol are the same when the protocol aborts, we have
\begin{equation} 
    			\tracenorm{  \left( \left( \protMap{l^\prime}  - \protMapId{l^\prime}\right) \otimes \id_R \right) \left( \tau_{A^nB^nR} \right) } = \Pr(\Omega_\text{acc})\tracenorm{\tau^{(l^\prime)}_{K_A \Cfull R | \Omega_\text{acc}} -\tau^{(l^\prime,\text{ideal})}_{K_A \Cfull R | \Omega_\text{acc}} }.
    	\end{equation}
Let the state of the protocol just before privacy amplification be given by $\tau_{Z^nC^nC_ER}$, where $Z^n$ denotes Alice's raw key register, $C^n$ denotes round-by-round announcements, $C_E$ denotes error-correction and error-verification announcements.

We will first obtain a lower bound on the smooth min-entropy of the state $\tau_{Z^nC^nC_EE^n \wedge \accevent}$, where $\tau_{A^nB^nE^n}$ is an IID extension of $\tau_{A^nB^n}$.
We will then obtain a lower bound on the smooth min-entropy of $\tau_{Z^nC^nC_ER \wedge \accevent}$.
We will then show that the choice of $l^\prime$ is such that an $\epsSec$-secret key of $l^\prime$ bits can be safely extracted from the state $\tau_{Z^nC^nC_ER \wedge \accevent}$.

\textbf{Obtaining a lower bound on } $\smoothmin{\epsbar}(Z^n|C^nC_EE^n)_{\tau \wedge \Omega_\text{acc}}$\textbf{: } We will utilize the IID security proof statements \cref{eq:condS,eq:condLHL} for this part of the argument. First, we set up some notation to split the set of IID states into two convenient parts. Let $T^\prime = \{\sigma_{AB} : \Tr_{B}(\sigma_{AB}) = \promise, \Pr(\accevent)_{\sigma} \leq \epsAT )\}$ be the set of states that the protocol accepts with probability less than $\epsAT$ . Let $T = \{\sigma_{AB} : \Tr_{B}(\sigma_{AB}) = \promise, \Pr(\accevent)_{\sigma} > \epsAT)\}$. 
        Recall from Eq.~\eqref{eq:condS} of the IID security proof statement, that $S$ is the set of states such that $\sigma \notin S \implies \Pr(\accevent)_\sigma \leq \epsilon_{AT}$.
        Clearly, $T \subseteq S$. We can then write $\tau_{A^nB^n}$ and its extension $\tau_{A^nB^nE^n}$ as
    \begin{equation} \label{eq:tausplit}
        \begin{aligned}
        \tau_{A^nB^n} &= \int_{\sigma \in T} d\sigma \sigma^{\otimes n}_{AB} + \int_{\sigma \in T^\prime} d\sigma \sigma^{\otimes n}_{AB} = \tau^{(1)}_{A^nB^n}+\tau^{(2)}_{A^nB^n} \\
            \tau_{A^nB^nE^n} &= \int_{\sigma \in T} d\sigma \sigma^{\otimes n}_{ABE} + \int_{\sigma \in T^\prime} d\sigma \sigma^{\otimes n}_{ABE} = \tau^{(1)}_{A^nB^nE^n}+\tau^{(2)}_{A^nB^nE^n}
        \end{aligned}
    \end{equation}
    where $\sigma_{ABE}$ is a purification of $\sigma_{AB}$, and 
    where we used the fact that $\tau_{A^nB^n}$ is a mixture of IID states which are extensions of $\promise$, so we can split the mixture into two components $\tau^{(1)},\tau^{(2)}$ defined by integrals over $T$ and $T'$ respectively.

    We also have $\tau_{ Z^nC^nC_EE^n \wedge \accevent} = \tau^{(1)}_{ Z^nC^nC_EE^n \wedge \accevent} + \tau^{(2)}_{ Z^nC^nC_EE^n \wedge \accevent}$ by linearity.  Since the IID states in $\tau^{(2)}_{A^nB^n}$ abort with high probability, we expect $\tau_{ Z^nC^nC_EE^n \wedge \accevent}$ to be ``close'' to $\tau^{(1)}_{ Z^nC^nC_EE^n \wedge \accevent}$. To bound the distance between these states, we have
    \begin{equation}
    \begin{aligned}
    	\tracenorm{\tau_{Z^nC^nC_EE^n \wedge \accevent} - \tau^{(1)}_{Z^nC^nC_EE^n \wedge \accevent}}  &= \tracenorm{\tau^{(2)}_{Z^nC^nC_EE^n \wedge \accevent}} \\
     &= \tracenorm{ \int_{\sigma \in T^\prime}   d\sigma \Pr(\accevent)_{\sigma}  \sigma_{Z^nC^nC_EE^n | \accevent}}\\
     &     \leq \int_{\sigma \in T^\prime} d\sigma \Pr(\accevent)_{\sigma} \leq  \epsAT.
     \end{aligned}
    \end{equation}
Thus, the generalized trace distance \cite{tomamichel_quantum_2015} between the two states satisfies  $\Delta(\tau_{Z^nC^nE^n \wedge \accevent} , \tau^{(1)}_{Z^nC^nC_EE^n \wedge \accevent} ) = \frac{1}{2}\tracenorm{ \tau^{(2)}_{Z^nC^nC_EE^n \wedge \accevent}} + \frac{1}{2} | \text{Tr}(\tau^{(2)}_{Z^nC^nC_EE^n \wedge \accevent}) | \leq  \epsilon_{AT}$. Therefore, the purified distance can be bounded as 
     \begin{equation}
    P(\tau_{Z^nC^nC_EE^n \wedge \accevent} , \tau^{(1)}_{Z^nC^nC_EE^n \wedge \accevent} ) \leq   \sqrt{2\Delta(\tau_{Z^nC^NE^n \wedge \Omega} , \tau^{(1)}_{Z^nC^nC_EE^n \wedge \accevent} )} = \sqrt{2 \epsilon_{AT}}
    \end{equation}
    using \cite[Lemma 3.5]{tomamichel_quantum_2015}. This allows us to relate the smooth min entropies of the two states as follows.
    
    Let $\smoothmin{\epsbar} (Z^n | C^nC_E E^n)_{\tau^{(1)}_{\wedge \accevent}} = H_{\text{min}}(Z^n | C E^n)_{\rho^{(1)}}$ with $P(\rho^{(1)}_{Z^nC^nC_EE^n}, \tau^{(1)}_{Z^nC^nC_EE^n \wedge \accevent}) \leq \epsbar$. By the triangle inequality for purified distance, we have $P(\rho^{(1)}_{Z^nC^nC_EE^n},\tau_{Z^nCE^n\wedge \Omega_\text{acc}}) \leq \epsbar + \sqrt{2 \epsAT}$. Therefore, we obtain
    \begin{equation} \label{eq:boundingsmoothedmin}
    \begin{aligned}
        \smoothmin{\epsbar+\sqrt{2\epsAT}} (Z^n | C^nC_EE^n)_{\tau \wedge \accevent}  &\geq H_\text{min} (Z^n | C^nC_E E^n)_{\rho^{(1)}}\\
        &=\smoothmin{\epsbar} (Z^n| C^nC_E E^n)_{\tau^{(1)} \wedge \accevent} \\
        &\geq   \min_{\sigma \in T} 	\smoothmin{\epsbar}(Z^n | C^nC_E E^n)_{\sigma \wedge \accevent} \\
           &\geq   \min_{\sigma \in S} 	\smoothmin{\epsbar}(Z^n | C^n C_E E^n)_{\sigma \wedge \accevent}
      \end{aligned}
    \end{equation}
    where we used $P(\tau_{X_ACE^n \wedge \Omega} , \tau^{(1)}_{X_ACE^n \wedge \Omega} ) \leq  \sqrt{2 \Delta} = \sqrt{2 \epsilon_{AT}}$ to obtain the first inequality, Lemma \ref{lemma:infsmoothedmin} to obtain the second inequality, and $T \subset S$ to get the third. Notice that $\min_{\sigma \in S} 	\smoothmin{\epsbar}(Z^n | C^n C_E E^n)_{\sigma \wedge \accevent}$ is the same expression that appears in the IID key length expression \cref{eq:condLHL}.
    
    \textbf{Obtaining a lower bound on } $\smoothmin{\epsbar}(Z^n|C^nC_ER)_{\tau \wedge \Omega_\text{acc}}$\textbf{: }
    Since we require $\tau_{A^nB^nR}$ to be a purification, we can purify $\tau_{A^nB^nE^n}$ to $\tau_{A^nB^nE^nV}$. Note that since each $\sigma^{\otimes n}_{ABE}$ in Eq.~\ref{eq:tausplit} is supported on $\Sym{\mathbb{C}^{d^2_Ad^2_B} }$,  $\tau_{A^nB^nE^n}$ is supported on $\Sym{\mathbb{C}^{d^2_Ad^2_B} }$. Thus, the dimension of the purifying register $V$ is bounded by  $\dim(V) \leq \dim(\Sym{\mathbb{C}^{d^2_Ad^2_B} })  = \cost$ with $x=d_A^2d_B^2$. Then, using \cite[Eq. (8)]{winkler_impossibility_2011}, we have 
    \begin{equation}
        \label{eq:splittingoffV}
        \smoothmin{ \epsbar + \sqrt{2 \epsAT}} (Z^n | C^nC_EE^nV)_{\tau \wedge \accevent} \geq \smoothmin{\epsbar + \sqrt{2 \epsAT}}  (Z^n | C^nC_EE^n)_{\tau \wedge \accevent} - 2 \log(\cost) 
    \end{equation}

    \textbf{Reducing the hash length:} Let $l^\prime = l - 2 \log{\cost}$. Applying the leftover hashing lemma \cite{tomamichel_quantum_2015} to $\tau_{Z^nC^nC_EE^nV}$, we obtain
    \begin{equation}
    	\begin{aligned}
    		\frac{1}{2} \Pr(\accevent)_{\tau} \tracenorm{ \tau^{(l^\prime)}_{K_A C^nC_EE^n V | \accevent}  - \tau^{(l^\prime), \text{ideal}}_{K_A C^nC_EE^n V | \accevent} } &\leq 	\frac{1}{2}  2^{-\frac{1}{2} \left( \smoothmin{\epsbar +  \sqrt{2 \epsAT} }(Z^n | C^n C_E E^n V)_{\tau \wedge \accevent} - (l - 2 \log(\cost) ) \right)}\\
      &+ 2 (\epsbar +  \sqrt{2 \epsAT}) \\
    		& \leq 	\frac{1}{2} 2^{- \frac{1}{2} \left(H_\text{min}^{\epsbar +  \sqrt{2 \epsAT} }(Z^n | C^n C_E E^n )_{\tau \wedge \Omega} - l \right)} + 2 (\epsbar +  \sqrt{2 \epsAT} ) \\
    		& \leq 	\frac{1}{2}2^{-\frac{1}{2} (\min_{\sigma\in S} 	\smoothmin{\epsbar}(Z^n | C^n C_E E^n)_{\sigma \wedge \accevent} - l )} + 2 (\epsbar +  \sqrt{2 \epsAT} )   \\
    		&\leq \epsPA+2\epsbar+  2 \sqrt{2 \epsAT}
    	\end{aligned}
    \end{equation}
    where we used Eq.~\eqref{eq:splittingoffV} for the second inequality, and Eq.~\eqref{eq:boundingsmoothedmin} for the third inequality, and Eq.~\eqref{eq:condLHL} for the final inequality.
    
    Identifying $E^nV$ with $R$, we obtain the required statement
    
    \begin{equation}
        \tracenorm{ \left( \left(\mathcal{E}^{(l^\prime)}_\text{QKD} - \mathcal{E}^{(l^\prime),\text{ideal}}_\text{QKD} \right) \otimes \id_{R} \right) \left( \tau_{A^nB^nR} \right) } \leq  \left( \epsilon_{PA}+ \epsbar + 2 \sqrt{2 \epsilon_{\text{AT}}} \right)
    \end{equation}
\end{proof}

The following lemma was used in the proof above.

\begin{lemma} \label{lemma:infsmoothedmin}
    Let $\rho_{AB} = \int_{\sigma \in S} d\sigma \sigma_{AB}$ for some set of states $S$. Then,
    \begin{equation}
        H^{\epsbar}_\text{min}(A|B)_\rho \geq \inf_{\sigma \in S} H^{\epsbar}_\text{min} (A|B)_\sigma
    \end{equation}
\end{lemma}
\begin{proof}
From \cite{vesely_libor_notes_nodate}, $\rho_{AB}$ can always be written as a finite sum of states in $S$, i.e.~$\rho_{AB} = \sum_z \Pr(z) \sigma^{(z)}_{AB}$, where $\sigma^{(z)}_{AB} \in S$. Next, we define  $\rho_{ABZ} \coloneqq \sum_z \Pr(z) \sigma^{(z)}_{AB} \otimes \ket{z}\bra{z}$,  and let $H_\text{min}(A|B)_{\widetilde{\sigma}^{(z)} } = H^{\epsbar}_\text{min}(A|B)_{\sigma^{(z)} }$, with $P(\widetilde{\sigma}^{(z)}, \sigma^{(z)} ) \leq \epsbar$.  Define $\widetilde{\sigma}_{ABZ} \coloneqq \sum_z \Pr(z) \widetilde{\sigma}^{(z)}_{AB} \otimes \ket{z}\bra{z}$. Then, from \cite[Eq.~3.59]{tomamichel_quantum_2015}, we obtain
    \begin{equation}
        P \left( \sum_z \Pr(z) \sigma^{(z)}_{AB} \otimes \ket{z}\bra{z}, \sum_z \Pr(z) \widetilde{\sigma}^{(z)}_{AB} \otimes \ket{z}\bra{z} \right) \leq \max_z P \left( \sigma^{(z)}_{AB}, \widetilde{\sigma}^{(z)}_{AB} \right) \leq \epsbar
    \end{equation}

    Therefore,
    \begin{equation}
        \begin{aligned}
            \smoothmin{\epsbar}(A|BZ)_\rho \geq H_\text{min}(A|BZ)_{\tilde{\sigma}} \geq \inf_z H_\text{min} (A|B)_{\tilde{\sigma}^{(z)}} = \inf_z \smoothmin{\epsbar}(A|B)_{\rho^{(z)}}
        \end{aligned}
    \end{equation}
    where we used \cite[Eq.~6.25]{tomamichel_quantum_2015} for the second inequality. The required claim then follows from the fact that $\smoothmin{\epsbar}(A|B)\geq \smoothmin{\epsbar}(A|BY)$.

\end{proof}

We next prove the result for variable-length protocols. Our proof is based on an argument that was used in~\cite{beaudry_assumptions_2015} to remove the random permutation at the start of the protocol (we highlight however that in order for the argument in that work to be valid, the error-correction code must satisfy a permutation-invariance property that does not necessarily hold for error-correction procedures used in practice). In principle we could have applied a similar argument in our analysis of fixed-length protocols above; however, the approach we used there results in a somewhat better secrecy parameter.

\maintheoremvar*
\begin{proof}
Recall that in the variable-length protocol, multiple events may occur. Either the protocol aborts and does not produce any key ($\accevent^c$), or it accepts and produces a key of length $l_i$ (when event $\Omega_i$ occurs). Thus, we write the output states for $\protMapVar{l^\prime},\protMapIdVar{l^\prime}$, for the input state $\tau_{A^nB^nR}$, as
\begin{equation}
    \begin{aligned}
         \left(\protMapVar{l^\prime} \otimes \id_R \right) (\tau_{A^nB^nR}) = \sum_{i=1}^M \Pr(\Omega_i) \tau^{(l^\prime_i)}_{K_A \Cfull R | \Omega_i} +\Pr(\accevent^c) \tau^{(\bot)}_{K_A \Cfull R | \accevent^c}  \\
             \left(\protMapIdVar{l^\prime} \otimes \id_R \right) (\tau_{A^nB^nR}) = \sum_{i=1}^M \Pr(\Omega_i) \tau^{(l^\prime_i, \text{ideal})}_{K_A \Cfull R | \Omega_i} + \Pr(\accevent^c) \tau^{(\bot,\text{ideal})}_{K_A \Cfull R | \accevent^c}
    \end{aligned}
\end{equation}
where recall that the protocol produces a key of length $l_i^\prime$ when the event $\Omega_i$ occurs. Similarly, we write the output states for $\protMapVar{l},\protMapIdVar{l}$, for the input state $\tau_{A^nB^nE^n}$ as

\begin{equation}
    \begin{aligned}
         \left(\protMapVar{l} \otimes \id_{E^n} \right) (\tau_{A^nB^nE^n}) = \sum_{i=1}^M \Pr(\Omega_i) \tau^{(l_i)}_{K_A \Cfull E^n | \Omega_i}+\Pr(\accevent^c) \tau^{(\bot)}_{K_A \Cfull E^n | \accevent^c} \\
             \left(\protMapIdVar{l} \otimes \id_{E^n} \right) (\tau_{A^nB^nE^n}) = \sum_{i=1}^M \Pr(\Omega_i) \tau^{(l_i, \text{ideal})}_{K_A \Cfull E^n | \Omega_i} +\Pr(\accevent^c) \tau^{(\bot,\text{ideal})}_{K_A \Cfull E^n | \accevent^c}
    \end{aligned}
\end{equation}
where \begin{align} \label{eq:temptau}
        \tau_{A^nB^nE^n} = \int \sigma_{ABE}^{\otimes n} \text{d} \sigma.
    \end{align}

Recall that since each event $\Omega_i$ leads to a different length of the final key,  the states $\tau^{(l_i)}_{K_A \Cfull E^n | \Omega_i}$ have orthogonal supports. Moreover, the output state conditioned on abort is identical for the real and ideal protocols. Therefore, from the $\epsSec$-secrecy statement (\cref{eq:epsSecVar}) for IID states, we obtain

\begin{equation}
    \begin{aligned}
      \frac{1}{2}  \tracenorm{   \left(\left(\protMapVar{l} - \protMapIdVar{l} \right)\otimes \id_{E^n} \right)    (\tau_{A^nB^nE^n})       } & =  \frac{1}{2} \sum_{i=1}^M \Pr(\Omega_i) \tracenorm{\tau^{(l_i)}_{K_A \Cfull E^n | \Omega_i} - \tau^{(l, \text{ideal})}_{K_A \Cfull E^n | \Omega_i}}  \\
        &= \sum_{i=1}^M  \lambda_i \leq \epsSec,
    \end{aligned}
\end{equation}

where we define
\begin{equation}
	 	\lambda_i \coloneqq  \frac{1}{2}	\Pr(\Omega_i ) \tracenorm{ \tau^{(l_i)}_{K_A \Cfull E^n| \Omega_i}  - \tau^{(l_i, \text{ideal})}_{K_A  \Cfull E^n| \Omega_i } }. 
	 	\end{equation}

Without loss of generality, we can assume $\Pr(\Omega_i ) > \lambda_i$ (if equality occurs, then the required bound in \cref{eq:varproofend} for the proof follows trivially). The converse bound for privacy amplification \cite[Theorem 7.7]{tomamichel_quantum_2015} allows us to bound the smooth min-entropy of the state prior to privacy amplification, as follows : 
	  \begin{equation} \label{eq:conversePA}
		  	H^{\sqrt{2 \lambda^\prime_i - \lambda_i^{\prime 2} }}_\text{min} (Z^n | E^n C ^n C_E)_{\tau | \Omega_i } \geq l_i, \quad \lambda^\prime_i = \frac{\lambda_i}{\Pr(\Omega_i ) }
	 \end{equation}
  
    Since we require $\tau_{A^nB^nR}$ to be a purification, we purify $\tau_{A^nB^nE^n}$ in the same manner as in the proof of \cref{thm:maintheorem}. Since each $\sigma^{\otimes n}_{ABE}$ in Eq.~\ref{eq:temptau} belongs to $\Sym{\mathbb{C}^{d^2_Ad^2_B} }$,  $\tau_{A^nB^nE^n}$ is supported on $\Sym{\mathbb{C}^{d^2_Ad^2_B} }$. Thus, the dimension of the purifying register $V$ is bounded by  $\dim(V) \leq \dim(\Sym{\mathbb{C}^{d^2_Ad^2_B} })  \leq \cost$ with $x=d_A^2d_B^2$. Then, using \cite[Eq. (8)]{winkler_impossibility_2011}, we have
	\begin{equation} \label{eq:afterconversepa}
		\begin{aligned}
		H^{\sqrt{2 \lambda^\prime_i - \lambda_i^{\prime 2} }}_\text{min} (Z^n | E^n V C ^n C_E )_{\tau | \Omega_i } &\geq 
		H^{\sqrt{2 \lambda^\prime_i - \lambda_i^{\prime 2} }}_\text{min} (Z^n | E^n C ^n C_E)_{\tau | \Omega_i } - 2 \log( \dim(V))\\
		& \geq l_i - 2 \log(\cost)
		\end{aligned}
	\end{equation}
	
	 Therefore, consider the modified protocol $\protMapVar{l^\prime}$ that hashes to $l^\prime_i = l_i -  2\log(\cost) - 2 \log(1/ \epstilde)$ instead of $l_i$, upon the event $\Omega_i$, where $x=d_A^2d_B^2$. In this case, using the leftover hasing lemma \cite[Proposition 9]{tomamichel_largely_2017} for smooth min-entropy, we have
	 \begin{equation}
	 	\begin{aligned}
		\tracenorm{ \tau^{(l^\prime_i)}_{K_A  \Cfull E^nV | \Omega_i }  - \tau^{(l^\prime_i,\text{ideal})}_{K_A  \Cfull E^n V| \Omega_i } } &\leq 2^{-\frac{1}{2} \left(   	H^{\sqrt{2 \lambda^\prime_i - \lambda_i^{\prime 2} }}_\text{min} (Z^n | E^n V C^n C_E )_{\tau | \Omega_i  }  - l^\prime_i \right)} + 4 \sqrt{2 \lambda^\prime_i - \lambda_i^{\prime 2}} \\
		&\leq 2^{ \log(\epstilde)} + 4 \sqrt{2 \lambda^\prime_i - \lambda_i^{\prime 2}} \\
		& =\epstilde +  4 \sqrt{2 \lambda^\prime_i - \lambda_i^{\prime 2}} \\
		\end{aligned}
	\end{equation}
 Therefore, bringing all the terms together, we obtain

	 \begin{equation} \label{eq:varproofend}
	\begin{aligned}
		&\frac{1}{2} \tracenorm{ \left( \left( \protMapVar{l^\prime} -\protMapIdVar{l^\prime} \right) \otimes \id_{E^nV} \right) \left( \tau_{A^nB^nE^nV} \right)} \\
  &=\sum_{i}  \frac{1}{2}	\Pr(\Omega_i ) \tracenorm{ \tau^{(l^\prime_i)}_{K_A  \Cfull E^n V | \Omega_i }  - \tau^{(l^\prime_i, \text{ideal})}_{K_A  \Cfull E^n V | \Omega_i } } \\
	& \leq  	\sum_{i}  \frac{1}{2}	\Pr(\Omega_i)  \left( \epstilde+  4 \sqrt{2 \lambda^\prime_i - \lambda_i^{\prime 2}}  \right) \\
	&\leq 	\sum_{i}  \frac{1}{2}	\Pr(\Omega_i  )  \left( \epstilde +  4 \sqrt{2 \lambda^\prime_i}  \right) \\
	&= \sum_{i}  \frac{1}{2} 	\Pr(\Omega_i ) \epstilde + \sqrt{8} \sum_i \Pr(\Omega_i  ) \sqrt{	 \lambda^\prime_i} \\
	&\leq \frac{\epstilde}{2} + \sqrt{8} \sqrt{	 \sum_i \Pr(\Omega_i  )  \lambda^\prime_i} \\
	&\leq  \frac{\epstilde }{2} + \sqrt{8} \sqrt{\epsSec}
	\end{aligned}
\end{equation}
where we used concavity of the square root function in the penultimate inequality.
	
		\end{proof}

\section{Proof of statements in \cref{sec:PSDecoy}}
\label{app:decoyProofs}

\lemmaShield*
\begin{proof}
    As Eve's channel does not act on the shield system, $\epsSec$-secrecy of the protocol with the shield system is given by
    \begin{align}
        \demi \norm{\left(\left(\protMapShield{l}-\protMapShieldId{l}\right)\otimes \id_{{E}^n}\right) \left[\left(\id_{{A}^nA_S^n}\otimes \Phi\right)\left[\rho^\text{prep}_{A^nA_S^nA'^n}\right]\right]}_1\leq \varepsilon_{\text{sec}}
    \end{align}
    for all channels $\Phi$.
    Further, since the QKD protocol acts trivially on the shield system, $\protMap{l} =\protMapShield{l} \circ \Tr_{A_S^n}$, and $\protMapShieldId{l} =\protMapId{l} \circ \Tr_{A_S^n}$.
    Combining these equations gives us the required result.
\end{proof}

\lemmaSRS*
\begin{proof}
    Let $\Psi_{\text{meas}}$ be the channel that measures Alice's systems $A^n$ in the computational basis. Since the first step of the QKD protocol is measuring Alice and Bob's system, $\protMap{l} = \protMapShieldSR{l} \circ \Psi_{\text{meas}}$ and $\protMapId{l} = \protMapShieldSRId{l} \circ \Psi_{\text{meas}}$. Finally, noting that $\Psi_{\text{meas}}[\rho_{AA_SA'}] = \rho^\text{prep}_{A^nA_S^nA'^n}$ gives us the required result.
\end{proof}

\lemmaSquash*
\begin{proof}
    The first step of the QKD protocol is to measure Bob's received state, i.e. a quantum to classical channel $$ \state \rightarrow \sum_{\vec{i}\in [\nmeas]^n}\Tr_{B^n}\left[\mathbb{I}_{A^n}\otimes\Gamma_{\Vec{i}}\state\right]\otimes\ketbra{\vec{i}},$$ where $\Gamma_{\vec{i}} = \bigotimes_{j=1}^n \Gamma_{i_j}$. Since $\Gamma_i = \Lambda^\dag\left[F_i\right],$ we have that
    \begin{align*}
        \Tr_{B^n}\left[\mathbb{I}_{A^n}\otimes\Gamma_{\Vec{i}}\state\right] &= \Tr_{B^n}\left[\mathbb{I}_{A^n}\otimes{\Lambda^\dag}^{\otimes n}\left[F_{\Vec{i}}\right]\state\right]\\
        &= \Tr_{B^n}\left[\left(\mathbb{I}_{A^n}\otimes F_{\vec{i}}\right) \id_{A^n}\otimes\Lambda^{\otimes n}\left[\state\right]\right].
    \end{align*}
    Thus, $\protMap{l} = \protMapSquash{l}\circ \Lambda^{\otimes n}$ and the security condition can be reduced to
    \begin{align}
        &\norm{\left(\left(\protMap{l}-\protMapId{l}\right)\otimes \id_{E^n}\right)\left[\state\right]}_1 \quad \forall\state\in\stateset{A^nB^n}\\
        =&\norm{\left(\left(\protMapSquash{l}-\protMapSquashId{l}\right)\otimes \id_{E^n}\right)\left[\left(\id_{A^n}\otimes\Lambda^{\otimes n}\right)[\state]\right]}_1 \quad \forall\state\in\stateset{A^nB^n}\\
        \leq&\norm{\left(\left(\protMapSquash{l}-\protMapSquashId{l}\right)\otimes \id_{E^n}\right)\left[\stateSquash\right]}_1 \quad \forall\stateSquash\in\stateset{A^nQ^n} \label{eq:sqashSecurityBound}\\
        \leq& \epsSec,
    \end{align}
    where \cref{eq:sqashSecurityBound} holds as $\left(\id_{A^n}\otimes\Lambda^{\otimes n}\right)[\state] \in \stateset{A^nQ^n}$ for all $\state\in\stateset{A^nB^n}$.
    This completes the proof.
\end{proof}

\lemmaFSS*
\begin{proof}
    First, note that $\FSScost\leq \lambdamin{\overline{\Pi}_\nFSS \Gamma_1 \overline{\Pi}_\nFSS}$ implies that $\FSScost\leq 1$. Thus, $\{F_i\}_{i=1}^\nmeas$ is a POVM. We denote the Hilbert space on which the original measurements $\Gamma_i$ act to be $B$, and the Hilbert space where the target measurements $F_i$ act to be $Q$.
    Now define the map
    \begin{align}
        \nonumber \Lambda[\rho] = &\Pi_\nFSS \rho \Pi_\nFSS + \left(1- \frac{1- \Tr\left[\Gamma_1\overline{\Pi}_\nFSS \rho \overline{\Pi}_\nFSS\right]}{1-\FSScost}\right)\ketbra{1}\\
        &+ \sum_{i=2}^\nmeas \frac{1}{1-\FSScost}\Tr\left[\Gamma_i \overline{\Pi}_\nFSS \rho \overline{\Pi}_\nFSS\right] \ketbra{i},
    \end{align}
    for all $\rho \in \stateset{B}$. It is easy to verify that this map is trace-preserving through explicit computation. Further, $\FSScost\leq \lambdamin{\overline{\Pi}_\nFSS \Gamma_1 \overline{\Pi}_\nFSS}$ implies that $\Tr\left[\Gamma_1\overline{\Pi}_\nFSS \rho \overline{\Pi}_\nFSS\right] \geq \FSScost$. Thus, it can be easily verified that the map is positive.
    
    For complete positivity, we show that it can be constructed by the composition of channels.
    The first channel is a measurement with Krauss operators $\Pi_\nFSS$ and $\overline{\Pi}_\nFSS$. The next channel leaves the outcome corresponding to $\Pi_\nFSS$ as is, and is a prepare-and-measure channel on the outcome corresponding to $\overline{\Pi}_\nFSS$. The measurements are given by the POVM $\{\frac{1}{1-\FSScost}\Gamma_i\}_{i=2}^\nmeas \cup \{\frac{\Gamma_1-\overline{\Pi}_\nFSS}{1-\FSScost}\}$, and the state prepared is the classical state corresponding to each measurement result.

    Finally, through explicit computation, we can verify that $\Tr\left[\Gamma_i \rho\right]= \Tr\left[F_i\Lambda[\rho]\right]$ for all $i$, and $\rho \in \stateset{B}$. Thus, $\Lambda^\dag\left[F_i\right] = \Gamma_i$ for all $i$.
\end{proof}

\lemmaSourceMaps*
\begin{proof}
        The existence of the source map implies that $\rhoMeas = \left(\id_{A^n}\otimes \Psi\right)\left[\tauMeas\right]$ where $\rhoMeas$ and $\tauMeas$ are Alice's state preparations as described in Eq. (\ref{eqRhoMeas}).
        Thus, the secrecy condition
        \begin{align} \label{eqEpsilonSecurityTauMeas}
            \norm{\left(\left(\protMap{l}-\protMapId{l}\right)\otimes \id_{E^n}\right)\left[\left(\id_{A^nA_S^n}\otimes\Phi'\right)[\tauMeasN]\right]}_1\leq\epsSec \quad \forall \Phi'\in\channelset{A''^n,B^nE^n}
        \end{align}
        trivially implies that $\norm{\left(\left(\protMap{l}-\protMapId{l}\right)\otimes \id_{E^n}\right)\left[\left(\id_{A^nA_S^n}\otimes\Phi'\right)[\tauMeasN]\right]}_1\leq\epsSec$ for any subset of channels. In particular, consider the subset
        \begin{align} \label{eqSourceMapChannels}
            \mathcal{C}_{\Psi} \coloneqq \{ \Phi' \vert \Phi' = \Phi \circ \Psi^{\otimes N},\  \Phi \in \channelset{{A'}^n,B^nE^n}\}.
        \end{align}
        Thus, it follows that
        \begin{align}
            \epsSec\geq&\norm{\left((\protMap{l}-\protMapId{l})\otimes \id_{E^n}\right)\left[\left(\id_{A^n}\otimes(\Phi\circ\Psi^{\otimes N})\right)[\tauMeasN]\right]}_1 \quad \forall \Phi\in \channelset{{A'}^n,B^nE^n}\\
            = &\norm{\left((\protMap{l}-\protMapId{l})\otimes \id_{E^n}\right)\left[\left(\id_{A^n}\otimes\Phi\right)\left[\rhoMeasN\right]\right]}_1 \quad \forall\Phi\in \channelset{{A'}^n,B^nE^n}.
        \end{align}
    \end{proof}

\section{Decoy-state constructions}
\subsection{Construction of shield system for decoy-state protocol} \label{app:DecoyShield}

In this appendix, we will construct the shield system for a decoy-state protocol. This will enable us to calculate the dimensions of Alice's systems $AA_S$ to use with \cref{cor:liftToCoherentvarxBlockDiagonal}.

In a decoy-state protocol, the tagged source described by \cref{eq:taggedStates} can be purified as
\begin{align} \label{eq:tagStatePurification}
    \ket{\xi_i^\mu}_{A_SA'} = \sum_{m=0}^\ndecoy \sqrt{p(m\vert \mu)} \ket{m}\otimes V_i \ket{m} + \sqrt{1-\sum_{m=0}^\ndecoy p(m\vert \mu)} \ket{\text{tag}}\otimes \ket{i,\mu},
\end{align}
where $\ket{\text{tag}}$ is a state orthogonal to $\{\ket{m}\}_{m=0}^{\ndecoy}$. Let there be $\nint$ intensities used in the decoy-state analysis.
Then in the source replacement scheme described in \cref{eqSourceReplacementRho}, we can obtain the fixed marginal
\begin{align}
    \nonumber \promiseShield &= \Tr_{A'}\left[\rho_{AA_SA'}\right]\\
    \nonumber &= \sum_{\substack{i,j=1\\\mu,\nu}}^{d_{A},\nint} \sqrt{p(i,\mu)p(j,\nu)}\ketbra{i,\mu}{j,\nu}_A\otimes \Tr_{A'}\left[\ketbra{\xi_i^\mu}{\xi_j^\nu}_{A_SA'}\right]\\
    \nonumber &= \sum_{\substack{i,j=1\\\mu,\nu}}^{d_{A},\nint} \sqrt{p(i,\mu)p(j,\nu)}\ketbra{i,\mu}{j,\nu}_A\otimes\\
    &\phantom{        }\left(\sum_{m=0}^\ndecoy \sqrt{p(m\vert \mu)p(m\vert \nu)}\bra{m}V_j^\dag V_i \ket{m} \ketbra{m} + \sqrt{\left(1-\sum_{m=0}^\ndecoy p(m\vert \mu)\right)\left(1-\sum_{m=0}^\ndecoy p(m\vert \nu)\right)} \ketbra{\text{tag}}\right), \label{eq:decoyPromisedState}
\end{align}
where the last equality used the fact that the encoding isometries $V_i$ preserve the photon number i.e. $\bra{m}V_j^\dag V_i\ket{m'} = 0$ if $m\neq m'$. Thus, the fixed marginal is block-diagonal in the photon number.
This has $\ndecoy+2$ diagonal blocks, each of dimension $\nint d_A$.

\subsection{Decoy-state analysis with fixed marginal} \label{app:decoyFixedMarginalToGenDecoy}

We have constructed the shield system in \cref{app:DecoyShield}. Thus, after the use of the postselection technique, the IID analysis is performed on some state $\sigma_{AA_SB}$ with a fixed marginal $\promiseShield$ given in \cref{eq:decoyPromisedState}. However, the analysis in \cref{subsec:IIDDecoy} is interpretated as Alice preparing some set of states $\{\xi_k^\mu\}_{k,\mu}$ which she sends through the channel representing Eve's (IID) attack $\Phi$ and being measured by POVM $\{\Gamma_l\}_l$. In this appendix we describe how to translate between the two pictures. More specifically, we show that the former implies the latter.

In particular, we show that the constraints used in the IID analysis on the state $\sigma_{AA_SB}$ with fixed marginal $\promiseShield$ imply \cref{eq:decoyStateSDP}. Let $\sigma_{AA_SB}$ be any state such that
\begin{equation}
    \begin{aligned} \label{eq:IIDDecoyWithPromiseStartingPoint}
        \Tr\left[\left(\ketbra{k,\mu}\otimes \mathbb{I}_{A_S}\otimes \Gamma_l\right)\; \sigma_{AA_SB}\right] &= p(i,\mu)\gamma_{l\vert k,\mu} \quad \forall l,k,\mu,\\
        \Tr_B[\sigma_{AA_SB}] &= \promiseShield.
    \end{aligned}
\end{equation}
Any extension $\sigma_{AA_SB}$ of $\promiseShield$ can be written as 
\begin{equation*}
    \begin{aligned}
        \id_{AA_S}\otimes \Phi\left(\ketbra{\psi}_{AA_SA'}\right)
    \end{aligned}
\end{equation*}
for some channel $\Phi \in \channelset{A',B}$, where
\begin{align}
    \ket{\psi}_{AA_SA'} &= \sum_{\substack{i,j=1\\\mu}}^{d_{A},\nint} \sqrt{p(i,\mu)}\ket{i,\mu}_A \otimes \ket{\xi_i^\mu}_{A_SA'}
\end{align}
is a purification of $\promiseShield$ and $\ket{\xi_i^\mu}_{A_SA'}$ is as defined in \cref{eq:tagStatePurification}. The rest follows \cref{eq:tagStatePurification}, noting that
\begin{align}
    \Tr\left[\left(\ketbra{k,\mu}\otimes \mathbb{I}_{A_S}\otimes \Gamma_l\right)\; \sigma_{AA_SB}\right] = p(i,\mu)\Tr\left[\Phi\left[\xi_k^\mu\right]\Gamma_l\right].
\end{align}

\section{Proof of statements in \cref{sec:applicationtothreestate}}
\label{app:ConstructingNiceSet}

\lemmaVFobs*

\begin{proof}
    	Consider the observed frequency of the $j$th outcome, which happens in each round with probability $\Tr(\Gamma_j \rho) = p_j$, where $\Gamma_j$ is the POVM element associated with outcome $j$. Given that one observed $\Fobs_j m$ events after sampling $m$ times, Hoeffdings inequality for $p_j$ gives us
  
				\begin{equation}
					\Pr_{\Fobs} \left(  p_j \in [\Fobs_j - \mu,\Fobs_j+\mu]  \right) \geq 1-\frac{\epsAT}{|\Sigma^\prime|} \quad \forall j \in \Sigma
				\end{equation}
    
			Combining these expressions for all $j \in \Sigma^\prime$, we obtain
				\begin{equation}
					\begin{aligned}
							\Pr_{\Fobs} \left( p_j \notin [\Fobs_j - \mu,\Fobs_j+\mu] \right) &\leq \frac{\epsAT}{|\Sigma|}, \\
								\Pr_{\Fobs} \left( \bigcup_{ j \in \Sigma} p_j \notin [\Fobs_j - \mu,\Fobs_j+\mu] \right ) &\leq \sum_{j \in \Sigma} \frac{\epsAT}{|\Sigma|} = \epsAT, \\
									\Pr_{\Fobs} \left( \bigcap_{ j \in \Sigma} p_j \in [\Fobs_j - \mu,\Fobs_j+\mu] \right) &\geq 1-\epsAT, \\
										\Pr_{\Fobs} \left( \rho \in V(\Fobs) \right) &\geq 1-\epsAT,
					\end{aligned}
					\end{equation}
				where we used the union bound for probabilities to obtain the second inequality.
				
\end{proof}

\bibliography{references}
\bibliographystyle{apsrev4-1}

\end{document}